\documentclass{aastex631}

\shorttitle{Testing the accuracy of SED modeling techniques}
\shortauthors{Faucher \& Blanton}

\graphicspath{{./}{figures/}}
\newcommand\HII{$\textrm{H}\scriptstyle\mathrm{II}$}
\begin{document}

\title{Testing the accuracy of SED modeling techniques
using the NIHAO-SKIRT-Catalog}

\author{Nicholas Faucher}
\affiliation{Center for Cosmology and Particle Physics, Department of Physics, New York University,
726 Broadway, New York, New York 10003, USA}

\author{Michael R. Blanton}
\affiliation{Center for Cosmology and Particle Physics, Department of Physics, New York University,
726 Broadway, New York, New York 10003, USA}

\begin{abstract}
\noindent We use simulated galaxy observations 
from the NIHAO-SKIRT-Catalog to test the accuracy of Spectral Energy Distribution
(SED) modeling techniques. SED modeling is an essential tool for inferring star-formation 
histories from nearby galaxy observations, but is fraught with difficulty due to our 
incomplete understanding of stellar populations, chemical enrichment processes, and the 
nonlinear, geometry-dependent effects of dust on our observations. The NIHAO-SKIRT-Catalog
uses hydrodynamic simulations and radiative transfer to produce SEDs from the ultraviolet 
(UV) through the infrared (IR), accounting for the effects of dust. We use the commonly
used Prospector software to perform inference on these SEDs, and compare the inferred
stellar masses and star-formation rates (SFRs) to the known values in the simulation. We match 
the stellar population models to isolate the effects of differences in the star-formation 
history, the chemical evolution history, and the dust. We find that the combined effect of model mismatches for high mass ($> 10^{9.5} M_{\odot}$) galaxies leads to inferred SFRs that are on average underestimated by a factor of 2 when fit to UV through IR photometry, and a factor of 3 when fit to UV through optical photometry. These biases lead to significant inaccuracies in the resulting sSFR-mass relations, with UV through optical fits showing particularly strong deviations from the true relation of the simulated galaxies. In the context of massive existing and upcoming 
photometric surveys, these results highlight that star-formation history inference 
from photometry remains imprecise and inaccurate, and that there is a pressing need
for more realistic testing of existing techniques.
\end{abstract}

\section{Introduction} 
\label{Introduction}

\noindent Understanding our place in the universe requires
untangling the history of galaxies over cosmic time.
This paper utilizes simulated observations of galaxies 
accounting for dust absorption, scattering, and emission \citep{faucher23a}
to test the spectral energy distribution (SED) modeling
techniques that many investigators use to infer the growth and evolution of galaxies from observations. 

Galaxy evolution research progresses along two paths---the archaeological record of individual galaxies at redshift zero and high redshift studies of how populations of galaxies change over time. This paper concentrates on understanding the evolutionary histories of galaxies at redshift zero. However, inferring galaxy histories in the local Universe through SED modeling presents formidable challenges and uncertainties stemming from the difficulty of accurately modeling star formation, stellar evolution, stellar structure,  stellar atmospheres, chemical evolution, and dust \citep{conroy10a, conroy13a, hayward15a, lower20a}. In this paper we will use mock observations of simulated galaxies from the Numerical Investigation of a Hundred Astrophysical Objects (NIHAO; \citealt{wang15a}) project to probe these uncertainties, especially those related to dust.

A particularly informative property of a galaxy is 
its  star formation rate (SFR), which tells us how 
rapidly the galaxy is converting gas into stars. Because young stellar populations emit much more strongly 
in the ultraviolet (UV) than do older populations, the 
UV luminosity emitted by stars is closely related to 
a galaxy's SFR. However, dust preferentially absorbs and scatters UV light relative to higher wavelengths and tends to be located near newly formed stellar populations \citep{salim20a}. Thus, the observed UV luminosity is typically substantially less than the emitted luminosity. The combination of absorption and scattering due to a dusty medium between
the observer and a source is known as extinction,
and its wavelength dependence as the extinction curve.
The extinction curve depends on the distribution of dust
grain sizes and chemical composition.
The dusty medium and the light sources are 
intermixed, so in a real system some sources experience more extinction than others. In addition, flux can be scattered by dust into the line of sight.
In the presence of all of these effects, the ratio of the total observed  to emitted flux is known as the attenuation. The attenuation's dependence on wavelength can differ from the extinction curve of the dusty medium.

The most commonly used and comprehensive approach to determine a galaxy's SFR is through SED modeling, a process in which observed broadband photometry and/or spectra are modeled under a specific set of assumptions about the underlying stellar populations, the form and flexibility of the star-formation histories (SFHs) and chemical evolution histories (CEHs), and the nature of the attenuation and emission by dust \citep{conroy13a}. For instance, \cite{moustakas_etal13} use SED modeling with UV through
NIR broadband photometry, exponentially declining SFHs with stochastic bursts, Chabrier IMF \citep{chabrier03a}, and a two-component power law dust attenuation model \citep{charlot00a} to study the quenching of star-forming galaxies and the evolution of the stellar mass function out to $z\sim 1$. \cite{salim16a} use 
UV through Optical broadband photometry to determine the masses, SFRs, and dust attenuation properties of $\sim 700,000$ galaxies with the SED modeling package CIGALE \citep{noll09a, boquien19a}. Using UV through Optical broadband photometry of 366 nearby galaxies with spectral indices from the Calar Alto Legacy Integral Field Area (CALIFA; \citealt{sanchez12a, husemann13a, garcia15a, sanchez16a}) survey, \cite{fernandez18a} use SED modeling to infer SFHs and study the cosmic evolution of the SFR density ($\rho_{\rm SFR}$), the sSFR, the main sequence of star formation (MSSF), and the stellar mass density ($\rho_{\ast}$). \cite{leja19a} fit the photometry of 58,461 galaxies from the 3D-HST catalogs \citep{skelton14a} using the Prospector-$\alpha$ model \citep{leja17a} and find lower SFR inferences ranging from $\sim 0.1-1$ dex compared to those made by \cite{whitaker14a} from a simple UV+IR luminosity relation, leading to a lower observed cosmic SFR density by $\sim 0.2$ dex. By comparing local inferred SFHs to the stellar mass functions from look-back studies \citep{tomczak14a}, they find that the Prospector-$\alpha$ model predicts SFHs that are likely too old in the $10^{10} - 10^{11} M_{\odot}$ stellar mass range. \cite{popesso23a} compile a wide variety of literature studies that use both SED modeling and various SFR indicators to investigate the evolution of MSSF across cosmic time. They find that at a given redshift, sSFR tends to be constant at low stellar masses and suppressed at higher masses, with the corresponding turnover mass getting larger with increasing redshift. They argue that the suppression of star-formation at high masses is caused by the decreasing availability of cold gas at lower redshifts. 
In short, there are numerous scientific results that fundamentally depend on
accurate inference of stellar masses and SFRs from SED modeling.

Correcting for the attenuation is a significant source of 
uncertainty in inferring the SFR of a galaxy. From galaxy
to galaxy, and within galaxies, there is variation in the 
amount of dust, in the dust extinction curves, and 
in the geometry of dust relative to the stars.
All of these differences can cause a substantial variation in 
the global attenuation curves.
Of course, we have no real way of knowing the true attenuation 
curve of a galaxy, since we don't have access to the emitted 
light before interacting with dust. This lack of a ground 
truth makes it impossible to test the accuracy of SED modeling 
inferences from observations alone. In order to test our ability to infer physical properties of 
galaxies from observations, \cite{faucher23a} provides an open source catalog of realistic photometry from the NIHAO galaxy simulation suite, called the NIHAO-SKIRT-Catalog, for which the ground truth is known, and from which we will test our inferences. 

In contrast to the assumptions of most SED modeling packages, including
the Prospector package used in this work, the results of the radiative 
transfer calculations in the NIHAO-SKIRT-Catalog show that we should
not expect real galaxies to exhibit energy  balance between 
dust  attenuation and emission along all lines of sight. 
This discrepancy between SED models and the likely real nature of 
galaxies makes correcting for dust attenuation an even greater challenge 
in  the inferences of masses and SFRs. Determining the extent to which 
this particular model mismatch affects our ability to infer 
physical parameters of galaxies is one of the main goals of this work. 

Related work by \cite{hayward15a} uses the 3D Monte Carlo dust radiative transfer code SUNRISE to generate mock observations of a simulated isolated disk galaxy with total stellar mass $5.6 \times 10^{10} \, M_{\odot}$, and a simulated merger of two of the same disk galaxies at 10-Myr intervals, each at 7 different viewing angles. In addition to Starburst99 SEDs for the general population of stars, they implement sub-grid recipes that include emission from the \HII{} and photodissociation regions (PDRs) surrounding young star clusters (star particles younger than 10 Myrs). Using MAGPHYS as their SED modeling code with a two-component power law dust attenuation model, they claim that most physical parameters are recovered reasonably well with the exception of total dust mass, which is systematically underestimated. We note that the simulated galaxies used in this work \citep{cox_etal06} do not model the cold gas phase of the ISM, leading to artificially smooth dust density structures and attenuation curves that are more similar to the underlying extinction curves. We expect this to lessen the impact of model mismatches related to dust on their SED modeling inferences. 

Along these same lines, \cite{lower20a} generates mock observations of $\sim 1600$ redshift zero galaxies from the SIMBA cosmological simulation using the 3D radiative transfer code POWDERDAY to test the accuracy of inferences from the SED modeling code Prospector. However, the main analysis of their work doesn't actually use the star-to-dust geometry of the simulations, but rather employs a uniform dust screen with a fixed optical depth to all stars and an additional uniform dust screen with the same optical depth applied to young stellar populations. The same fixed dust model is used in the Prospector fits, which means that their tests are not comparing Prospector's 
model to a more realistic dust distribution. At the end of their paper, they do perform a limited analysis where they consider the 
effect of the star-to-dust geometry, now using the flexible Kriek and Conroy dust model, but without the implementation of subgrid recipes to account for \HII\ and PDRs, which are below the resolution of the simulations. While the fits using non-parametric SFHs are only marginally affected by the resolved diffuse dust, those using delayed-$\tau$ SFHs are significantly less accurate. 

\cite{lower22a} also uses mock observations of galaxies from the SIMBA cosmological simulations using POWDERDAY, this time actually using the star-to-dust geometry of the simulations to calculate the dust attenuation, in order to test Prospector's ability to recover the simulated attenuation curves. However, they limit their analysis to the attenuation and emission from the diffuse ISM, ignoring contributions from the birth clouds surrounding young stellar populations in both the simulations and the fits. This makes it difficult to extrapolate their results to determine how well their fitting methods would perform on galaxies in the real Universe. Furthermore, the main results of the work are done with no error bars on the mock observations, and they only include a short appendix related to how well the model performs under three unrealistic uncertainty choices: One with a constant signal-to-noise ratio ($S/N$) of 5, one with mixed $S/N$ in each band ranging from 5 to 25, and one in which the flux in each band is perturbed with a $S/N$ of 10. We note that none of these methods capture the trend found in real observations for IR bands to have significantly larger uncertainties than UV through optical bands.  

\cite{trcka20a} generates mock observations of $\sim 7000$ galaxies from the EAGLE simulations using the radiative transfer software SKIRT. CIGALE UV-IR fits of these mock observations using a two-component dust attenuation curve without a UV bump, fixed stellar metallicity, a Salpeter IMF \citep{salpeter64}, and a delayed-tau SFH with a variable star-formation burst, give stellar mass and SFR inferences that are consistently overestimated, but generally do not deviate by more than $\sim 0.2$ dex. They claim that these differences are mainly caused by the assumed SFHs and the use of different IMFs between the fits and the radiative transfer calculation. Similarly to the simulations used by \cite{hayward15a}, the EAGLE simulations do not model the cold gas phase of the ISM \citep{schaye15a}, which reduces spatial fluctuations in the dust density and therefore reduces the resulting diversity of the resulting attenuation curves. We expect this to lessen the impact of model mismatches related to dust on their SED modeling inferences. 

In this work, we use the NIHAO-SKIRT-Catalog \citep{faucher23b} to test the accuracy of inferences from the SED modeling code Prospector. Section \ref{Methods} describes our methodology for using mock observations to infer physical parameters of the simulated galaxies. 
Section \ref{Results} shows the results of our analysis. 
Section \ref{Discussion} explores how inference inaccuracies affect the resulting sSFR-mass relation. 
Section \ref{Summary} summarizes our findings.  

\section{Methods} 
\label{Methods}

\noindent In this section we describe the general methodology used in this work. The first 
three subsections describe the details of how we generated realistic mock observations.
Section \ref{Simulations} describes the process of generating mock observations from simulated 
galaxies, Section \ref{Mock Photometry Uncertainties} describes how we assign realistic 
photometric uncertainties to our mock observations, and Section \ref{Mock Photometry Distances} 
covers how we decide appropriate distances for each simulated galaxy based on total stellar mass. 
The last three subsections describe how we perform inference on the mock observations. 
Section \ref{Stellar Population Synthesis} describes stellar population modeling choices we 
use for our inferences, Section \ref{Dynamic Nested Sampling} describes our sampling 
method for exploring our model parameter spaces, and Section \ref{Credible Intervals} 
explains how we use the distributions of sampled parameters to report credible intervals 
on estimated physical parameters. 

\subsection{Simulations} \label{Simulations}

\noindent The mock observations used in this work are from \cite{faucher23a} and we briefly describe them here. The underlying hydrodynamical galaxy simulations are from the Numerical Investigation of Hundred Astrophysical Objects
(NIHAO) project \citep{wang15a}, which have proven to be successful in reproducing the observed relation between stellar mass and halo mass \citep{wang15a}, the disk
gas mass and disk size relation 
\citep{Maccio2016}, the
Tully–Fisher relation \citep{Dutton2017}, as well as a large variety of scaling relations based on the MaNGA sample \citep{Arora2023}. 

To generate mock observations from these simulations, \cite{faucher23a} uses the Stellar Kinematics Including
Radiative Transfer (SKIRT; \citealt{camps20a}) 3D Monte
Carlo radiative transfer code, with FSPS \citep{conroy09a, conroy10a, foreman-mackey14a} using the MIST isochrones \citep{paxton11a, paxton13a, paxton15a, choi16a, dotter16a} and the MILES spectral library \citep{sanchez06a} assuming a Chabrier stellar initial mass function (IMF) \citep{chabrier03a} as the underlying simple stellar population (SSP) model. 
SKIRT calculates the dust attenuation and emission from the diffuse interstellar medium (ISM) locally using The Heterogeneous dust Evolution Model for Interstellar Solids (THEMIS; \citealt{jones17a}) dust model. Due to the fact that photodisassociation regions (PDRs) are unresolved in the simulated galaxies, we use subgrid recipes for young stellar populations which already include the attenuation and emission from dust \citep{faucher23a}. 

\begin{figure}
\begin{center}
\includegraphics[width=0.45\textwidth]{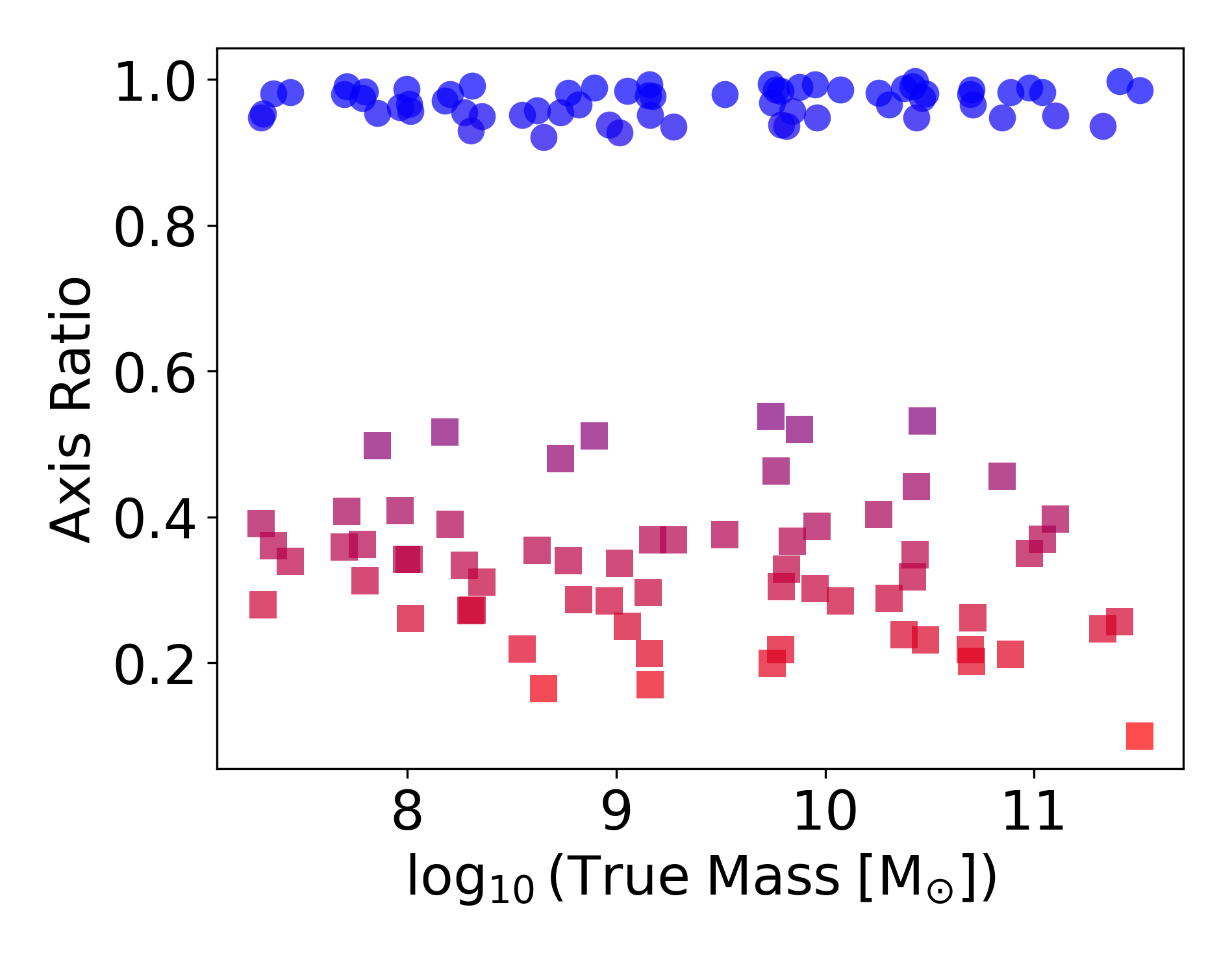}
\includegraphics[width=0.45\textwidth]{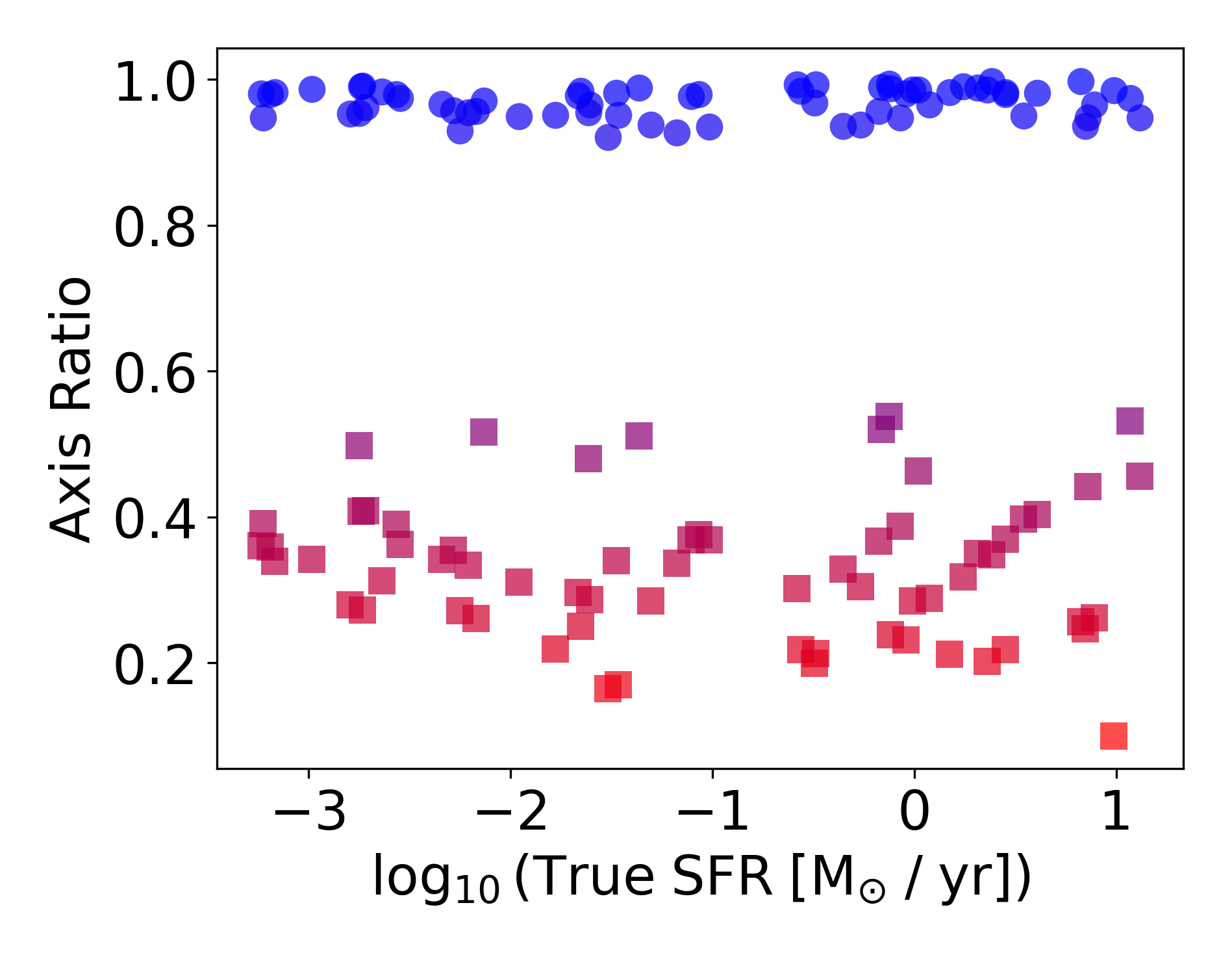}
\end{center}
\caption{\label{fig:axisRatioColors} \small 
Simulated galaxy observed axis ratios as a function of total stellar mass (left) and SFR (right) for face-on and edge-on orientations. Each point is colored by its axis ratio from 0.1 (red) to 1 (blue). This color scheme will be used throughout the rest of this work.}
\end{figure}

For our analysis, we will use 65 galaxies from the 
NIHAO-SKIRT-Catalog, each observed under one roughly face-on and one 
roughly edge-on orientation. We also include versions of each 
orientation that exclude the effects of dust (and are therefore
identical in their SEDs between the orientations). Throughout this work, 
we will find it 
useful to separate our analysis between the face-on and edge-on 
orientations of the simulated galaxies. When plotting inferred 
parameters of individual fits, we will also color points by their axis 
ratios (see \cite{faucher23a} for a detailed description of axis ratio 
calculations). Figure \ref{fig:axisRatioColors} shows the axis ratios values as a function of total stellar masses and SFRs using this color scheme for face-on and edge-on orientations.

A feature of the NIHAO simulations is that the degree of 
dust attenuation is a strong function of stellar mass, and
in particular below $10^{9.5} M_\odot$ there is minimal attenuation.
We will therefore use that threshold below in our analysis 
to separate the low mass and high mass galaxies, which behave 
qualitatively differently from each other. In this sample, as
in the real universe, the stellar mass is strongly correlated with
SFR, so that the lowest SFR galaxies are largely the same ones as the
lowest stellar mass galaxies, and therefore have little dust attenuation.

\subsection{Mock Photometry Uncertainties} \label{Mock Photometry Uncertainties}

\noindent Accurately testing SED modeling requires having relative flux
errors that are realistic, i.e. similar to an observed sample. In our
case, we match the uncertainties in our sample to those of the DustPedia sample \citep{davies17a}. The DustPedia sample is close to a 
best case scenario for SED modeling, with its panchromatic data on
nearby galaxies. 

In order to assign realistic uncertainties to the mock observations, we fit photometry from the DustPedia \citep{davies17a} catalog with a simple function:
\begin{equation}
\sigma =  \, \sqrt{c \, + \, \alpha \, f_{\nu} + \left( \sigma_{\rm calib} \, f_{\nu} \right)^2},
\label{eq:sigma}
\end{equation}
\noindent where $c$ represents the background noise, $\alpha$ represents the Poisson noise, and $\sigma_{\rm calib}$ is the calibration uncertainty given in Table 1 of \cite{clark18a}. Table \ref{table:1} of this work shows our calculated $c$ and $\alpha$ values for each filter.

These fits are performed using a least squares method from {\tt scipy.optimize} \citep{virtanen20a} between $\log_{10}(\sigma)$ and $\log_{10}(f_{\nu})$, where $f_{\nu}$ is the flux in each photometric band in units of Janskys. In cases where flux values are consistent with zero up to 1$\sigma$, for the figures in this 
paper we will plot $\log_{10}(f_{\nu})$ error bars as extending all the way to the bottom edge of the plot. 

\begin{table}[h!]
\begin{center}
\begin{tabular}{|l|l|l|}
\hline
Filter Name & $c \; [\rm Jy^2]$ & $\alpha \; [\rm Jy]$ \\
\hline
\texttt{FUV} & 1.4566$\times 10^{-10}$ & 4.3922$\times 10^{-6}$ \\
\texttt{NUV} & 4.588$\times 10^{-10}$ & 2.4238$\times 10^{-6}$ \\
\texttt{u} & 1.0962$\times 10^{-8}$ & 9.6788$\times 10^{-5}$ \\
\texttt{g} & 1.6421$\times 10^{-8}$ & 2.2165$\times 10^{-5}$ \\
\texttt{r} & 2.5926$\times 10^{-8}$ & 2.8862$\times 10^{-5}$ \\
\texttt{i} & 1.2963$\times 10^{-7}$ & 6.7169$\times 10^{-5}$ \\
\texttt{z} & 9.0480$\times 10^{-7}$ & 2.2377$\times 10^{-4}$ \\
\texttt{J} & 1.0103$\times 10^{-5}$ & 1.4023$\times 10^{-3}$ \\
\texttt{H} & 1.1989$\times 10^{-5}$ & 8.485$\times 10^{-3}$ \\
\texttt{K} & 2.2551$\times 10^{-5}$ & 3.6033$\times 10^{-3}$ \\
\texttt{W1} & 3.0633$\times 10^{-7}$ & 5.4390$\times 10^{-5}$ \\
\texttt{W2} & 4.3207$\times 10^{-7}$ & 1.2953$\times 10^{-4}$ \\
\texttt{W3} & 3.0377$\times 10^{-6}$ & 7.0860$\times 10^{-4}$ \\
\texttt{W4} & 3.5885$\times 10^{-5}$ & 2.4209$\times 10^{-3}$ \\
\texttt{Irac1} & 6.4139$\times 10^{-7}$ & 4.1288$\times 10^{-5}$ \\
\texttt{Irac2} & 4.2164$\times 10^{-7}$ & 5.8425$\times 10^{-5}$ \\
\texttt{Irac3} & 6.3356$\times 10^{-6}$ & 4.8108$\times 10^{-4}$ \\
\texttt{Irac4} & 3.1927$\times 10^{-6}$ & 1.4007$\times 10^{-4}$ \\
\texttt{Mips24} & 2.561$\times 10^{-5}$ & 9.4577$\times 10^{-4}$ \\
\texttt{Mips70} & 7.9693$\times 10^{-3}$ & 2.0658$\times 10^{-2}$ \\
\texttt{Mips160} & 1.5751$\times 10^{-2}$ & 1.1444$\times 10^{-1}$ \\
\texttt{Pacs70} & 1.927$\times 10^{-1}$ & 1.2964$\times 10^{-1}$ \\
\texttt{Pacs100} & 1.0821$\times 10^{-1}$ & 1.7251$\times 10^{-1}$ \\
\texttt{Pacs160} & 1.3729$\times 10^{-1}$ & 1.1827$\times 10^{-1}$ \\
\texttt{Spire250} & 1.0465$\times 10^{-2}$ & 1.2453$\times 10^{-2}$ \\
\texttt{Spire350} & 6.0181$\times 10^{-3}$ & 1.1519$\times 10^{-2}$ \\
\hline
\end{tabular}
\caption{Table of mock photometry uncertainty parameters from Equation \ref{eq:sigma} calculated from the DustPedia sample. }
\label{table:1}
\end{center}
\end{table}

Assigning uncertainties to the fluxes
is necessary to accurately explore the degeneracies of the SED modeling
and to determine what systematic errors are important relative to statistical
errors.  
We do not actually add random errors to the fluxes that we 
will fit.

\subsection{Mock Photometry Distances} \label{Mock Photometry Distances}

\noindent To obtain a simulated sample with similar properties to DustPedia, in
addition to a similar flux error distribution, we need a similar distance 
distribution. In DustPedia, like many samples, the lower luminosity
galaxies are closer, and to get similar fractional flux errors our simulated
sample needs to reflect the observations in this way. 
Since the mock observations from \cite{faucher23a} are all 
modeled at 100 Mpc, we therefore need to  
adjust their distances to match observed samples and flux errors. 

\begin{figure}
\begin{center}
\includegraphics[width=0.4\textwidth]{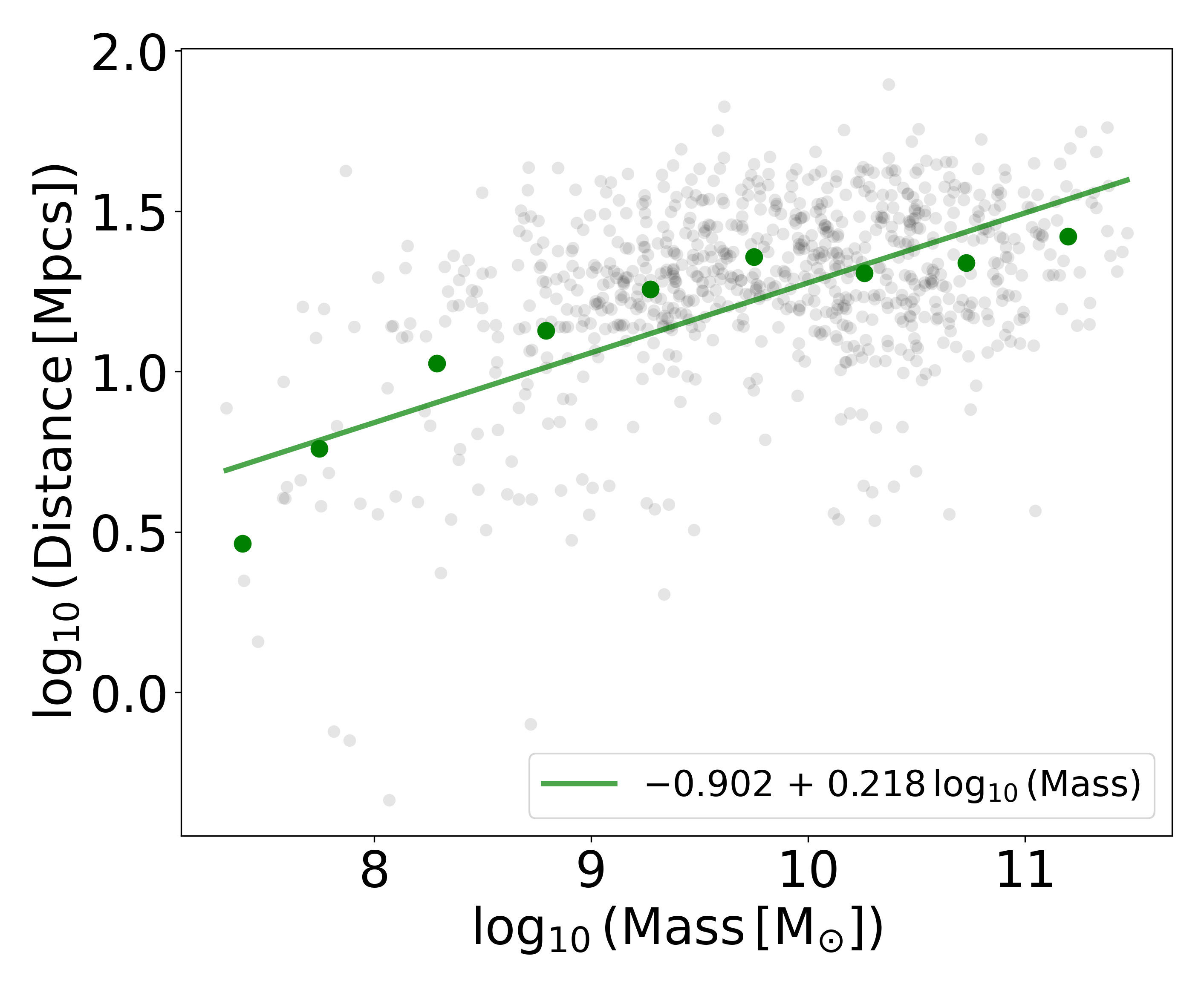}
\end{center}
\caption{\label{fig:distances} \small 
Distribution of DustPedia galaxies' inferred masses and distances (grey), along with the average binned $\log_{10}(\rm mass)$, and $\log_{10}(\rm distance)$ in green, and the line fitted to the average binned points. The equation for the fitted line is provided in the legend.}
\end{figure}

We use inferred masses and distances of galaxies in the DustPedia catalog, restricted to those which fall within the range of masses of our simulated galaxy sample, to assign distances to our simulated galaxies. Since the DustPedia sample includes more high mass galaxies, we bin the observed galaxies by $\log_{10}(M_{\ast})$ and find the mean $\log_{10}(M_{\ast})$ and $\log_{10}(d)$ within each bin, where $d$ is the luminosity distance to the galaxy. We then fit a line to the resulting set of points. Figure \ref{fig:distances} shows the resulting fitted line, the points used to fit the line in green, and the original DustPedia data in light grey. We then use the fitted line to assign our simulated galaxies a realistic observed distance based on its mass, and adjust the mock observations according to the inverse square law. 

\begin{figure}
\gridline{\fig{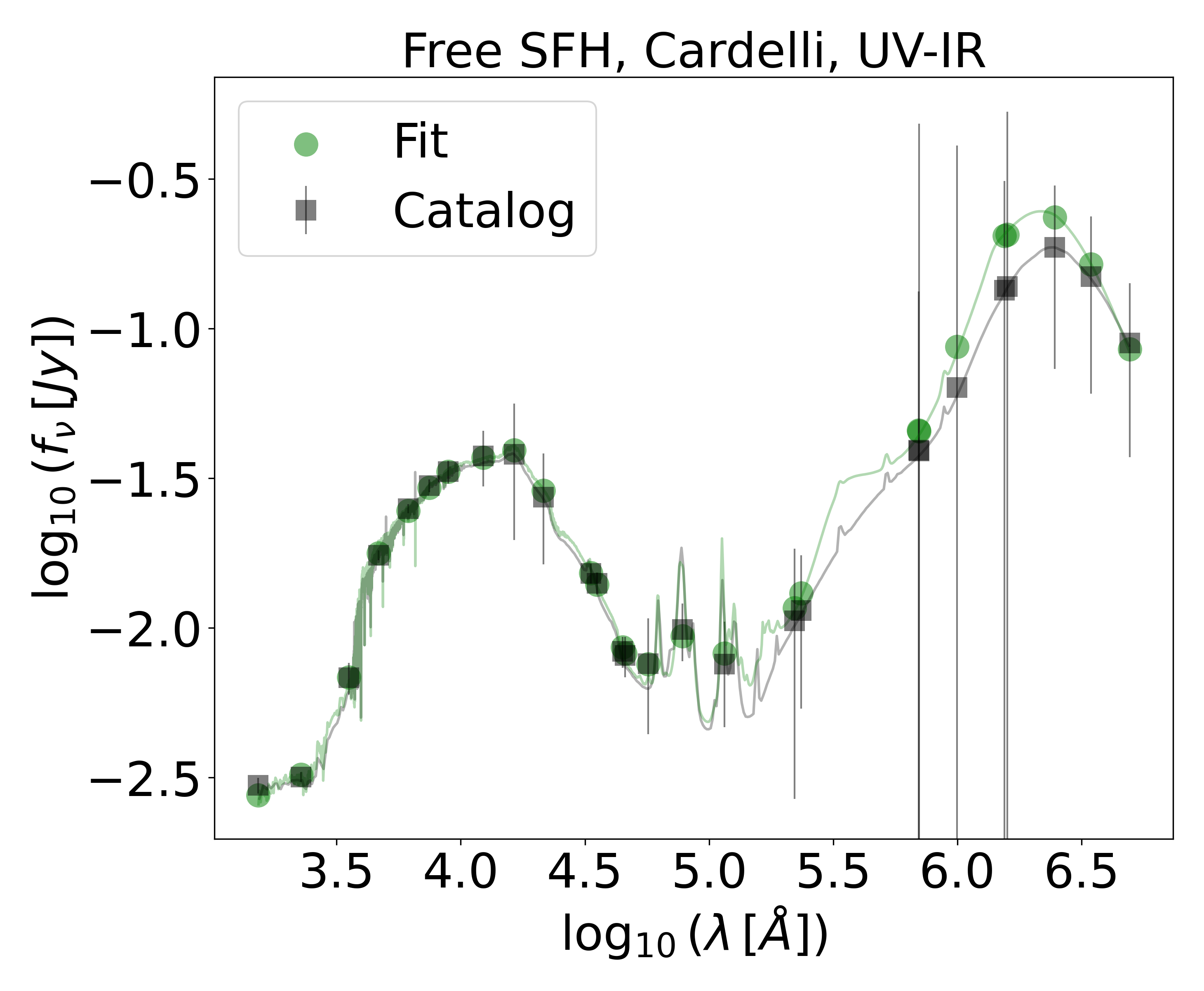}{0.5\textwidth}{}
\fig{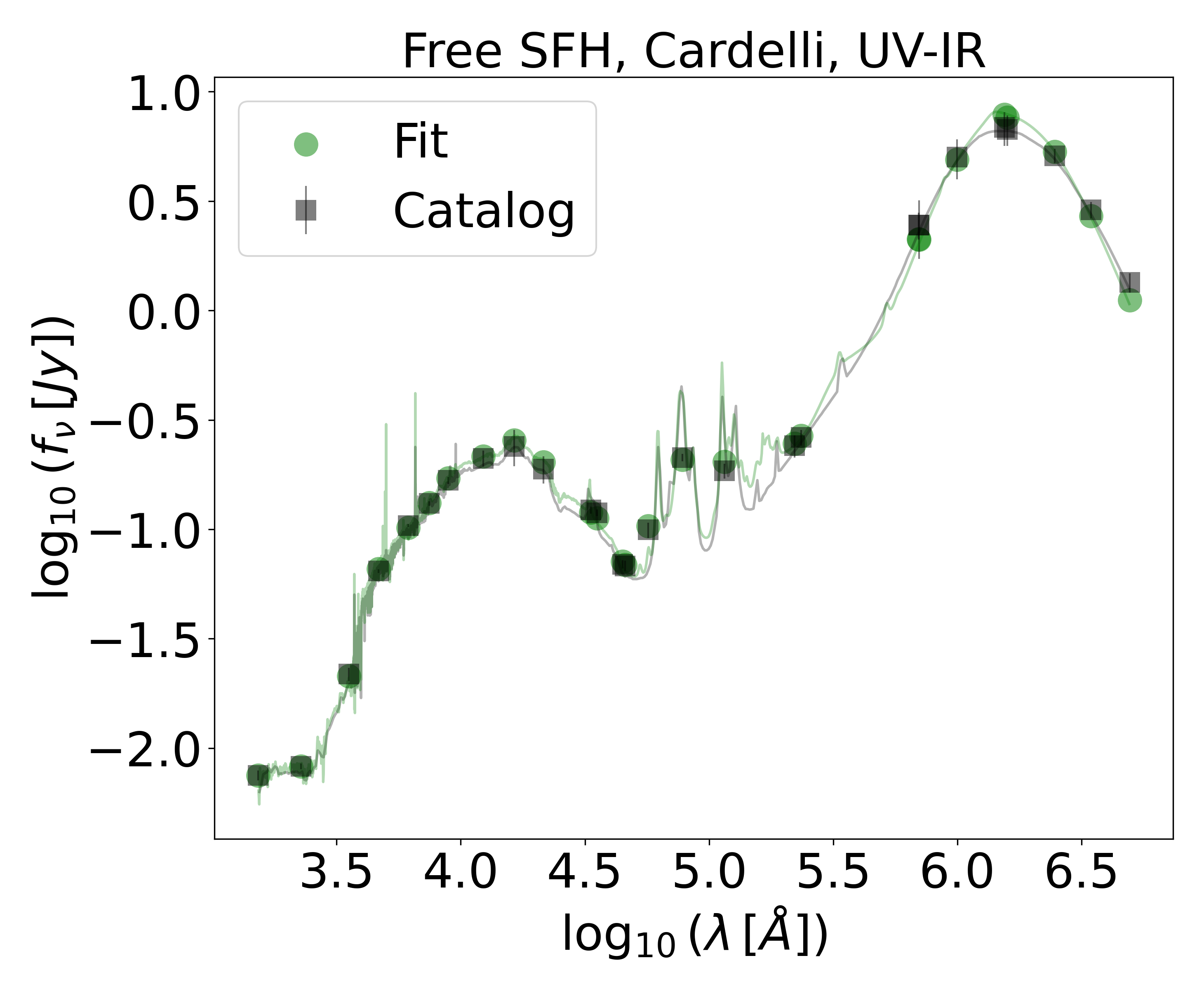}{0.5\textwidth}{}}
\gridline{\fig{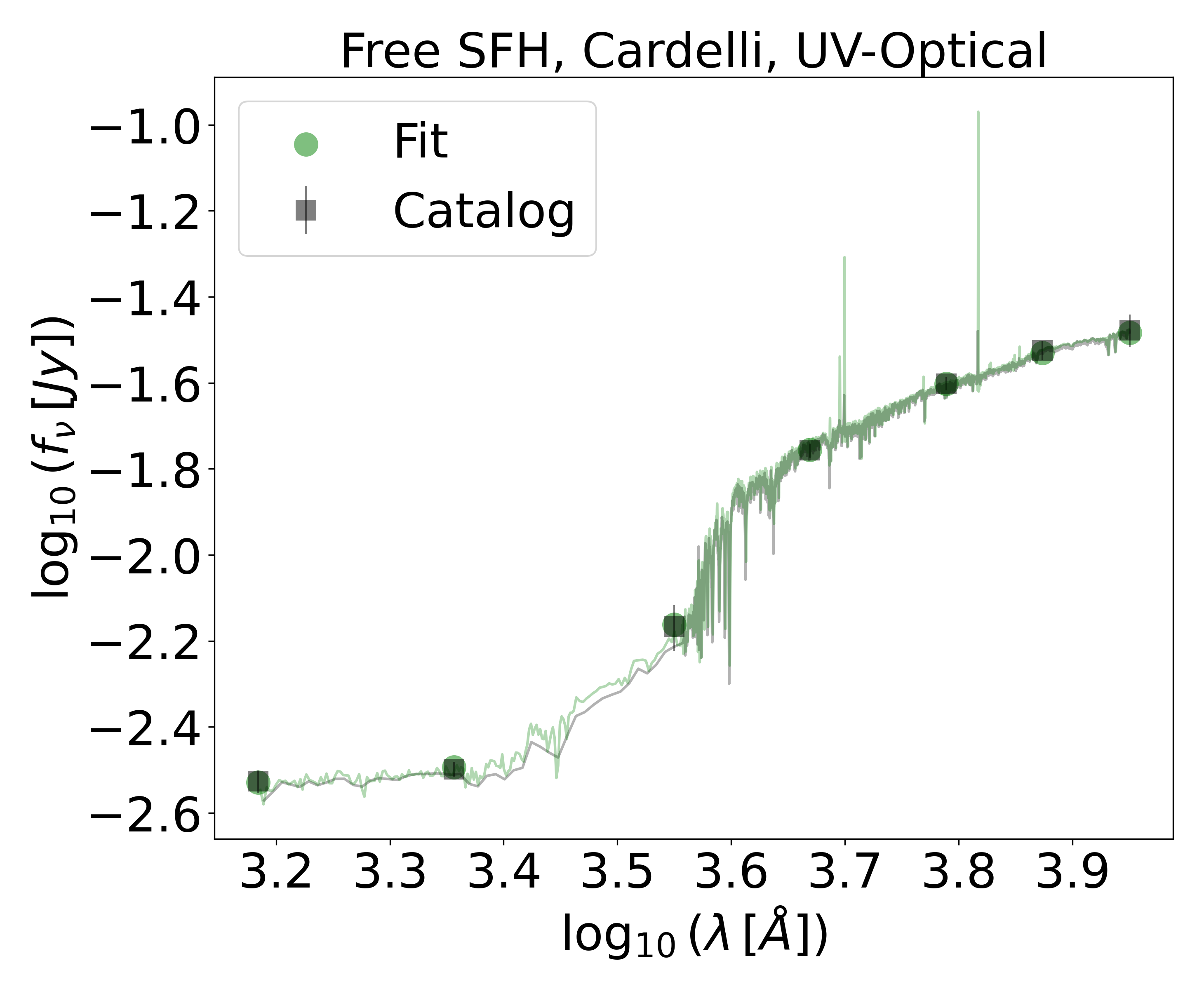}{0.5\textwidth}{}
\fig{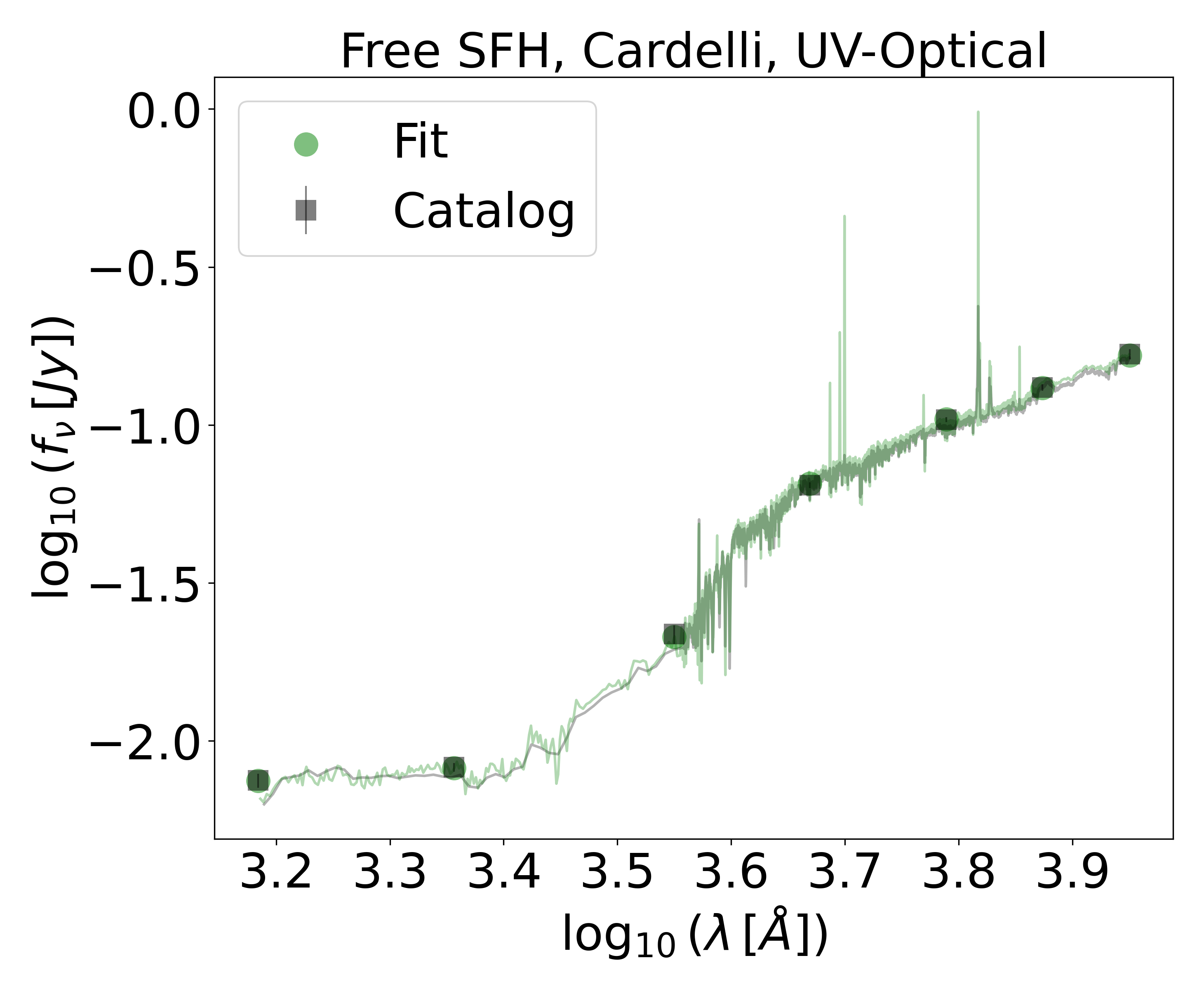}{0.5\textwidth}{}}
\caption{\label{fig:fits} Fits of the face-on orientation of a low (left) and high (right) stellar 
mass galaxy using the Cardelli dust attenuation model, and UV-IR (top) and UV-Optical (bottom) bands with all SFH parameters free. Mock photometry is shown in black and fitted photometry in green. Error bars that extend to the bottom of the plot represent fluxes that are consistent with zero up to one sigma.}
\end{figure}

To illustrate typical resulting fluxes and uncertainties, Figure \ref{fig:fits} shows spectra
(grey line) and photometry (black points and errors) for several examples. The left and right panels
correspond to the face-on orientation of a low mass and a high mass galaxy, respectively.
Error bars that extend to the bottom of each plot represent fluxes that are consistent with 
zero up  to $1\sigma$. 
Also shown in green are the best fits using the methods described in Section 
\ref{Stellar Population Synthesis}. These particular fits use the Cardelli dust model with a Free SFH.

\subsection{Stellar Population Synthesis and SED Modeling} \label{Stellar Population Synthesis}

\noindent The spectral fitting in this work is performed using the Bayesian SED modeling framework Prospector \citep{leja17a}, which also uses FSPS as its underlying SSP model. In order to narrow the potential causes of discrepancies between the catalog mock photometry and the model photometry, we choose to also use the MIST isochrones \citep{paxton11a, paxton13a, paxton15a, choi16a, dotter16a} the MILES spectral library \citep{sanchez06a}, and a Chabrier IMF \citep{chabrier03a} in our SED modeling procedure.

We explore two different sets of broadband photometry to include in our fits. The first includes 
only UV through Optical (hereafter, ``UV-Optical'') whereas the second includes UV through far IR 
(hereafter, ``UV-IR''). Referring back to Table \ref{table:1}, UV-Optical filters range from FUV to z, and UV-IR filters represent all filters in the table. If it were the case that both our 
mock observations and fitting procedure were based on the same underlying model (IMFs, isochrones, stellar libraries, SFHs, CEHs, and dust models), the inclusion of more bands should always lead to more accurate and precise inferences. However, as noted in \cite{faucher23a}, the energy balance between dust absorption and emission along a particular line of sight can be violated by up to a factor of three in the 
mock observations, in contrast to the strict assumption of energy balance made in Prospector. 
Because of this assumption, the inclusion of IR bands can provide a misleading constraint on the 
amount of dust absorption in the fits, and may actually lead to less accurate inferences. 

Another cause of discrepancies between the mock observations and the fits is the shape of
the attenuation curve. In the simulations, the shape of the attenuation curve is primarily
a result of the underlying extinction curve from THEMIS and the star-dust geometry of each 
galaxy at each viewing orientation. In Prospector, we explore four different commonly used 
dust attenuation models: Calzetti \citep{calzetti00a}, Power Law \citep{charlot00a}, Kriek and Conroy (K\&C; \citealt{Kriek06}), and Cardelli \citep{cardelli89a}. In addition, we apply a dust-free 
version of Prospector to our dust-free simulated SEDs as a baseline test. We note that, while all of 
the attenuation curves of the simulated galaxies in this work contain a UV-bump around 
$2175$ \AA, only the Kriek and Conroy and Cardelli dust attenuation models include this 
feature. Furthermore, only the Cardelli model has the flexibility to independently vary 
the strength of the UV-bump without changing its other parameters. 

Of particular relevance 
to this work, \cite{faucher24a} perform a detailed test of the flexibility of these 
commonly used dust attenuation models against the attenuation curves from the NIHAO-SKIRT-Catalog. 
To briefly summarize these results, although none of the dust models are able to fully capture 
the shapes of the catalog attenuation curves, Cardelli performs the best, Kriek and 
Conroy and Power Law perform similarly, and Calzetti performs the worst. 
\cite{faucher24a} consider the two cases in which the dust models can and cannot allow a 
variable fraction of the stellar light to reach the observer unattenuated. In contrast, in 
the current work we follow the more standard practice of limiting our analysis to the case 
where all stellar light receives attenuation, while still allowing for variable attenuation 
between young and old stellar populations in the two-component models (all except Calzetti). 
In this case, the number of parameters associated with the dust attenuation models are:
one for Calzetti, four for Power Law, four for Kriek and Conroy, and five for Cardelli.

While in the simulations the shape of the dust emission spectrum results from the temperature distribution of the dust grains, which are heated stochastically by the local radiation field \citep{camps15a}, for SED modeling in this work we use the dust emission model of \cite{draine07a}. This model has three free parameters which determine the shape of the dust emission spectrum, while the total emitted energy is fixed to the energy attenuated by dust. We will see in later sections that this assumption of energy balance between dust attenuation and emission can have a strong effect on our ability to make inferences from UV-IR data. 

To model the star-formation histories (SFHs) of the galaxies, we use the non-parametric method described in 
\cite{leja17a}. This method discretizes the SFH into 6 bins, using
a Dirichlet prior to regularize the SFH, biasing it towards being smooth. 
Additionally, as is common practice in SED modeling, all stellar populations in the Prospector are assigned a 
single metallicity, that is, there is no chemical evolution history (CEH). Given that we purposefully 
match the underlying SSP models and IMFs, these differences between SFHs and CEHs, along with the 
differences between dust attenuation and emission models, are the primary drivers of discrepancies 
between the true and inferred physical parameters.

In order to disentangle how these various model mismatches affect physical parameter inferences, 
we perform four separate tests in which we selectively fix model parameters in the fits 
to their true values. From most to least constrained, the methods are as follows: 
\begin{enumerate}
    \item  Fixed SFH: All SFH bins and the total stellar masses are fixed. 
    \item Fixed specific SFH (sSFH): All SFH bins are fixed but total stellar mass is free. 
    \item Fixed mass: total stellar mass is fixed, all SFH bins are free. 
    \item Free SFH; all parameters related to the SFH are free. 
\end{enumerate}
In all four methods, we fix the luminosity distance and redshift to their true values. The 
idea behind performing these tests is that as we remove SFH related information from the fits, 
we can see how Prospector uses the additional flexibility to compensate for model mismatches. 

In all cases, we attempt to find the model parameters that maximize the posterior probability density, which is proportional to the likelihood (probability density of the data given the model) multiplied by the prior probability density, which represents our prior beliefs about the model parameters before seeing the data. For a given set of model parameters, we assume that the $f_\nu$ residuals are drawn from a Gaussian distribution with standard deviations given by Equation \ref{eq:sigma}. 
This assumption leads to a log-likelihood function of the form 
\begin{equation}
\ln \left( P \left( D \, | \, \theta \right) \right) = -\frac{1}{2} \, \sum^{N}_{i=1} \left( \left( \frac{\delta_{i}^{2}}{\sigma_{i}^{2}} \right) + \ln \left( 2 \, \pi \sigma_{i}^2 \right) \right)
\label{eq:likelihood}
\end{equation}

\noindent where $\delta_{i}$ is the $i^{th}$ observed flux minus the $i^{th}$ model flux, $\sigma_{i}$ is the $i^{th}$ flux uncertainty given by equation \ref{eq:sigma}, and $N$ is the total number of bands used in the fit. 

\subsection{Dynamic Nested Sampling} \label{Dynamic Nested Sampling}

\noindent For all fits in this work, we use DYNESTY \citep{speagle20a}, a dynamic nesting sampling package for estimating Bayesian posteriors and evidences. Relative to traditional Markov Chain Monte Carlo (MCMC) methods, dynamic nested sampling is particularly well suited for efficiently sampling multimodal distributions by adaptively sampling parameters based on posterior structure.

This method relies heavily on the ability to efficiently draw samples from a constrained prior. While this can be difficult for arbitrary priors, it is much simpler in the case of uniform distributions. For this reason, a prior transform function is used to create a uniform multidimensional unit cube prior distribution to sample from. While there are several methods that can be used to constrain the prior, we choose to use multiple, possibly overlapping ellipsoids, which are able to model complex, multimodal distributions.

To lay a baseline set of samples, we first sample $50 \times n$ live points from this transformed prior, where $n$ is the dimensionality of the model. Then we take the sample with the minimum likelihood, add it to a list of dead points, and use this minimum likelihood value to constrain the prior and take another sample. This process repeats until the estimated upper bound of the evidence is less than 2\% of the current estimated evidence, where the estimated remaining evidence is given by the maximum likelihood sample multiplied by the remaining prior volume. Once this criteria is met, the remaining live points are added to the list of dead points in order from minimum to maximum likelihood. This process is referred to as static nested sampling.

In order to make the rest of the sampling dynamic, we make the choice to fully prioritize estimating the posterior over the evidence. In this case, the importance weight of a particular sample, which weights its contribution to the approximated posterior, is given by the average likelihood of that point and the one with the next lowest likelihood, multiplied by the difference in the constrained prior volume between those two points. Next, another $50 \times n$ points are sampled from the prior constrained to have importance weights greater than or equal to 80\% of the maximum importance weight from the set of dead points. The minimum likelihood sample is added to the list of dead points and another point is sampled repeatedly until the likelihood of the last dead point is larger than any of the previous samples. This processes is repeated until the KL-divergence between the sampled posterior distribution and its bootstrap is less than 2\%.  

Following the recommendation of \cite{speagle20a}, we choose to use uniform sampling when the dimensionality of the model is less than 10, and random walks sampling when the dimensionality is between 10 and 20. When using random walks, each walker must take at least 30 steps before proposing a new live point.

\subsection{Credible Intervals} \label{Credible Intervals}

\noindent Due to model mismatches, the peak of the posterior distribution 
does not in general correspond to the true parameters, even though
we have not added any noise to our data.

We need a way of assessing how far away our the peak of the
posterior is from the true parameters relative to the method's 
statistical uncertainties. To demonstrate how we do so, we here
describe our  procedure for calculating credible intervals of 
inferred parameters.  A $1\sigma$ credible interval represents a region in the marginalized posterior probability distribution of each parameter that contains 68\% of the posterior probability. In other words, if we draw a fair sample from the posterior distribution, that sample will have a 68\% chance of being within a $1\sigma$ credible interval. 

In contrast to MCMC, the distribution of samples from dynamic nested sampling is not a direct approximation 
of the posterior distribution. Instead, each sample is assigned an importance weight, as described in Section \ref{Dynamic Nested Sampling}, that determines the relative contribution of each sample to the posterior. When using a sampling method whose samples directly approximate the posterior, the $1\sigma$ credible interval for an inferred parameter is typically reported as a parameter range that includes 68\% of the samples. In the case where weighted samples approximate the posterior, $1\sigma$ credible intervals must be reported as the parameter ranges that include 68\% of the total weight. In both cases, there is ambiguity in how this range is determined. 


The method we use in this work, called the equal-tailed interval, is the interval that contains 
the central 68\% of the weight; that is, 16\% of the weight is below the interval and 16\% of 
the weight is above the interval. We calculate this interval using the cumulative distribution 
function (CDF) of the weighted samples sorted by their parameter values and finding the values 
at which the fractional CDF is $0.16$ and $0.84$. These values define the lower and upper parameter 
values of the credible interval, respectively. This method allows us to account for asymmetries in 
the marginalized parameter posterior distributions with asymmetric error bars around the maximum 
posterior probability parameter values. A peculiarity of this method is that it is possible for the 
maximum posterior sample to be outside of the $1\sigma$ credible interval. This can happen when 
the estimated marginalized posterior distributions are strongly skewed to one side. For plots 
that contain parameter inferences, we will plot those that are below the credible interval as 
only having error bars above, and plot those that are above the credible interval as only having 
error bars below. 

\section{Results} 
\label{Results}

\begin{figure}
\gridline{\fig{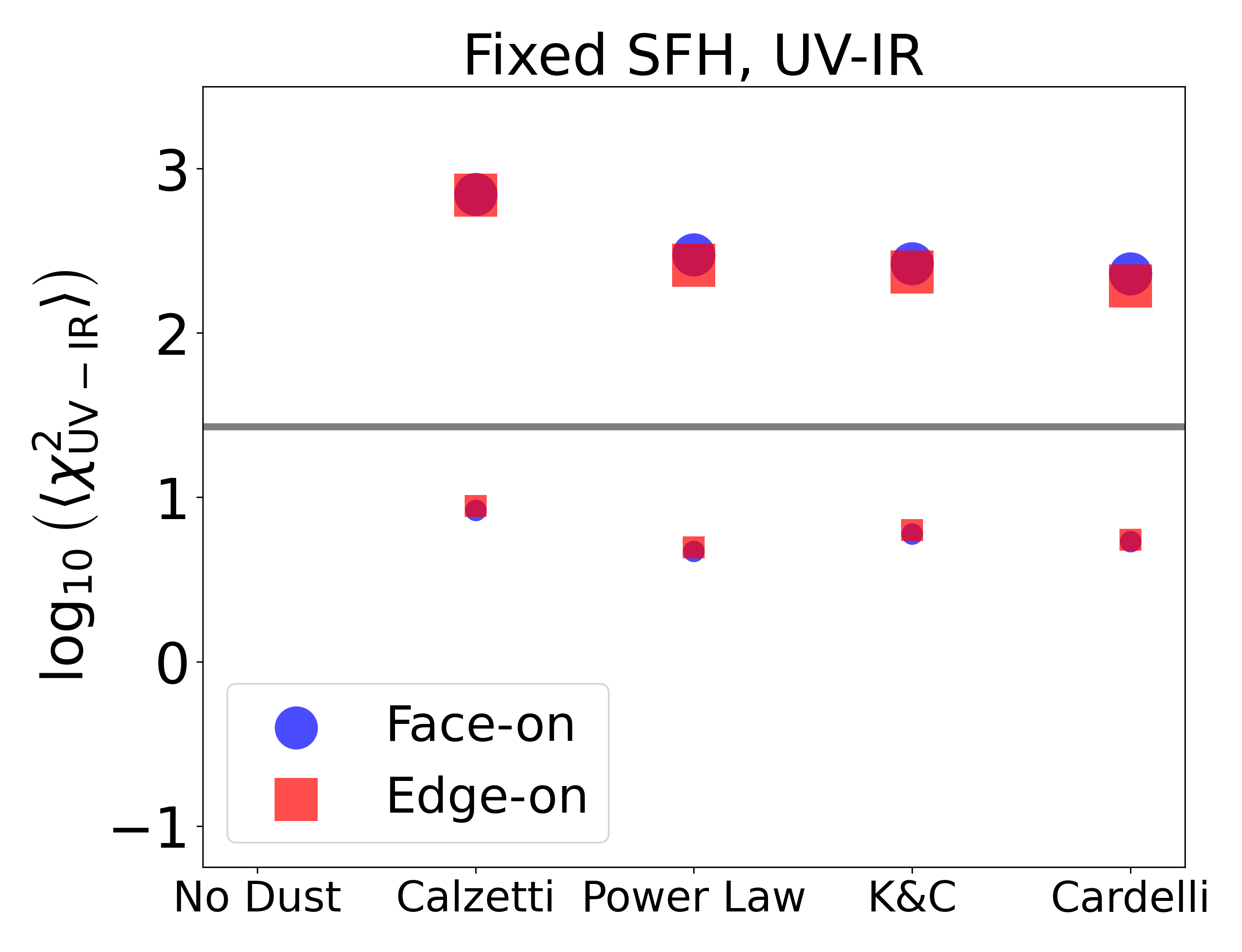}{0.25\textwidth}{}    \fig{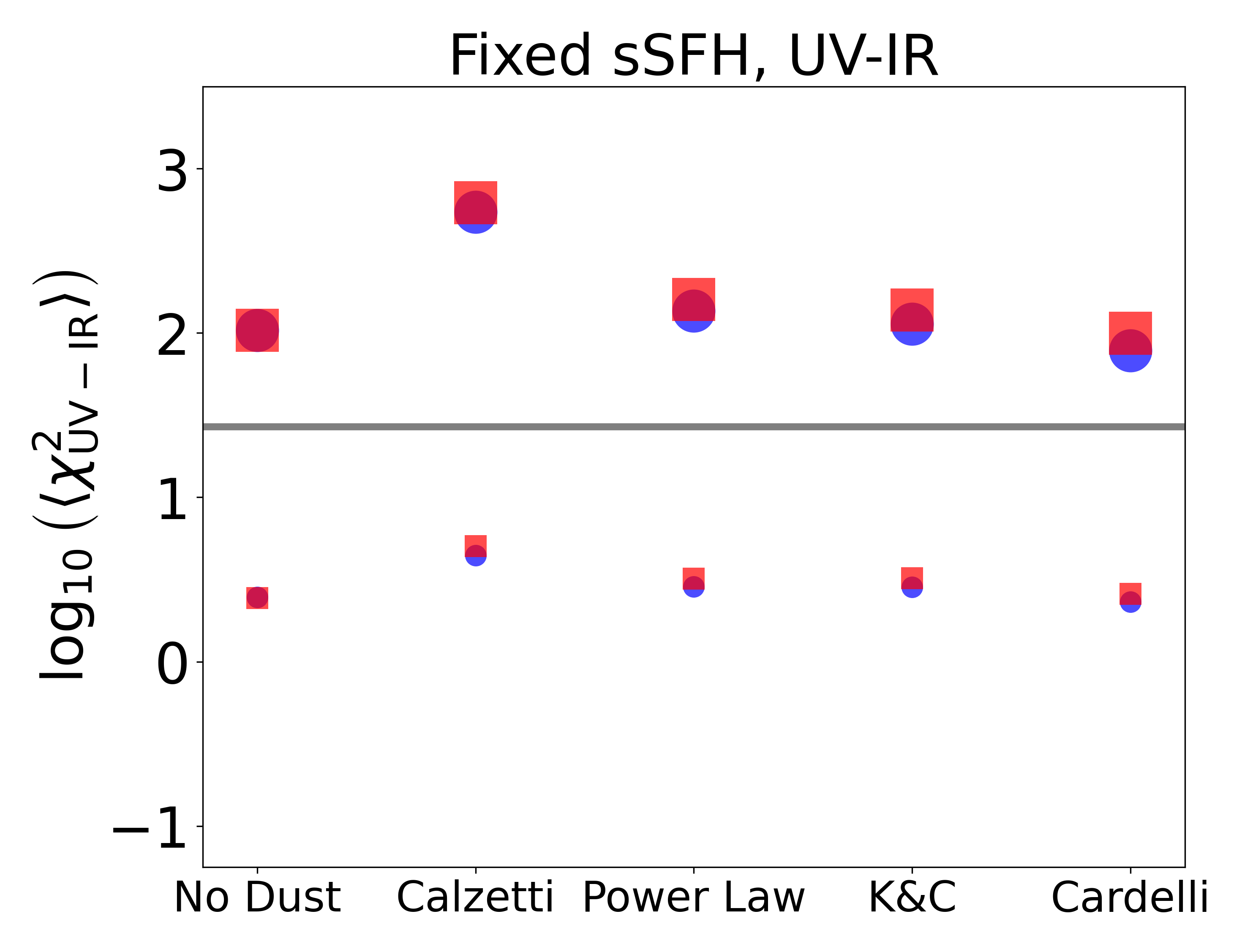}{0.25\textwidth}{}
\fig{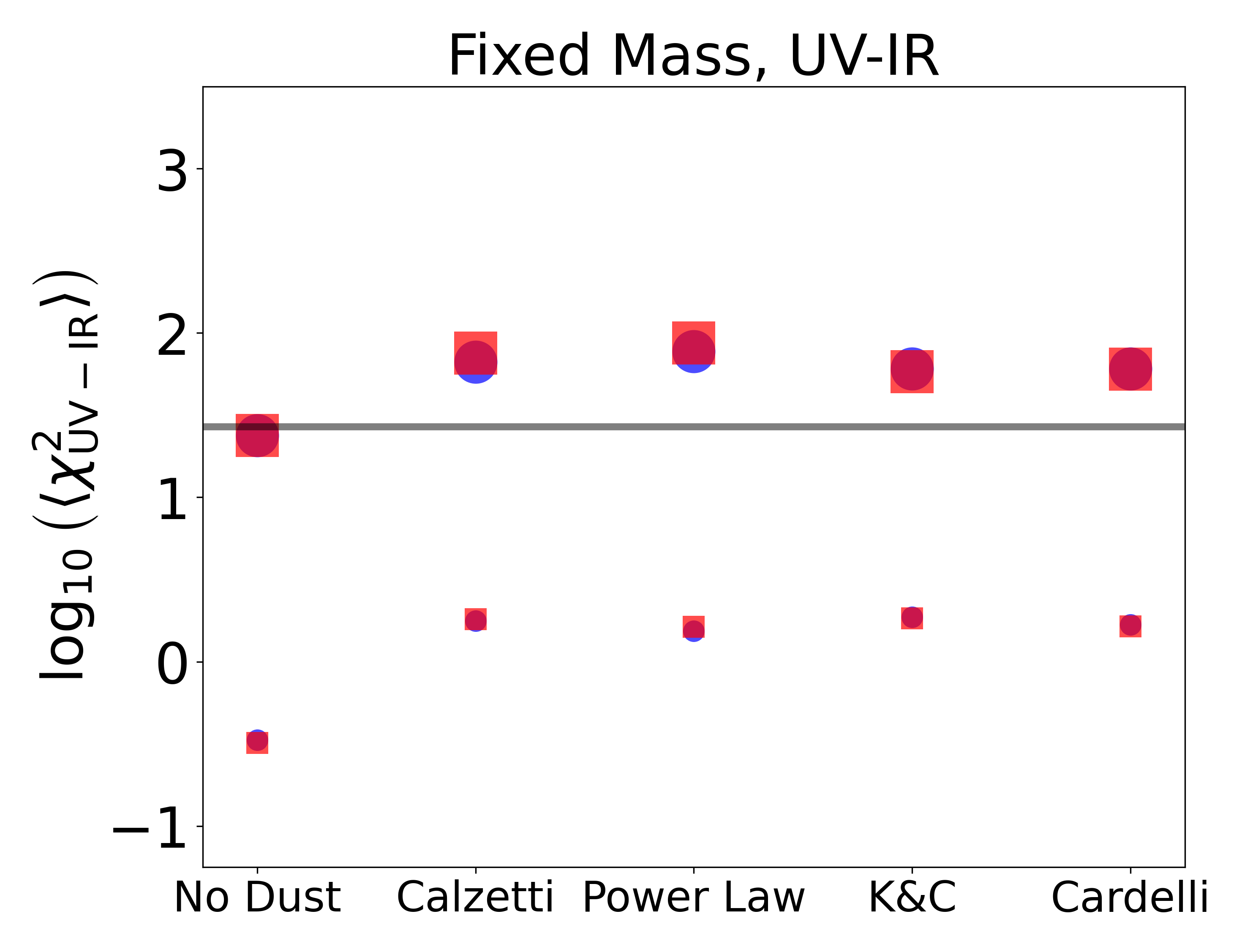}{0.25\textwidth}{}
\fig{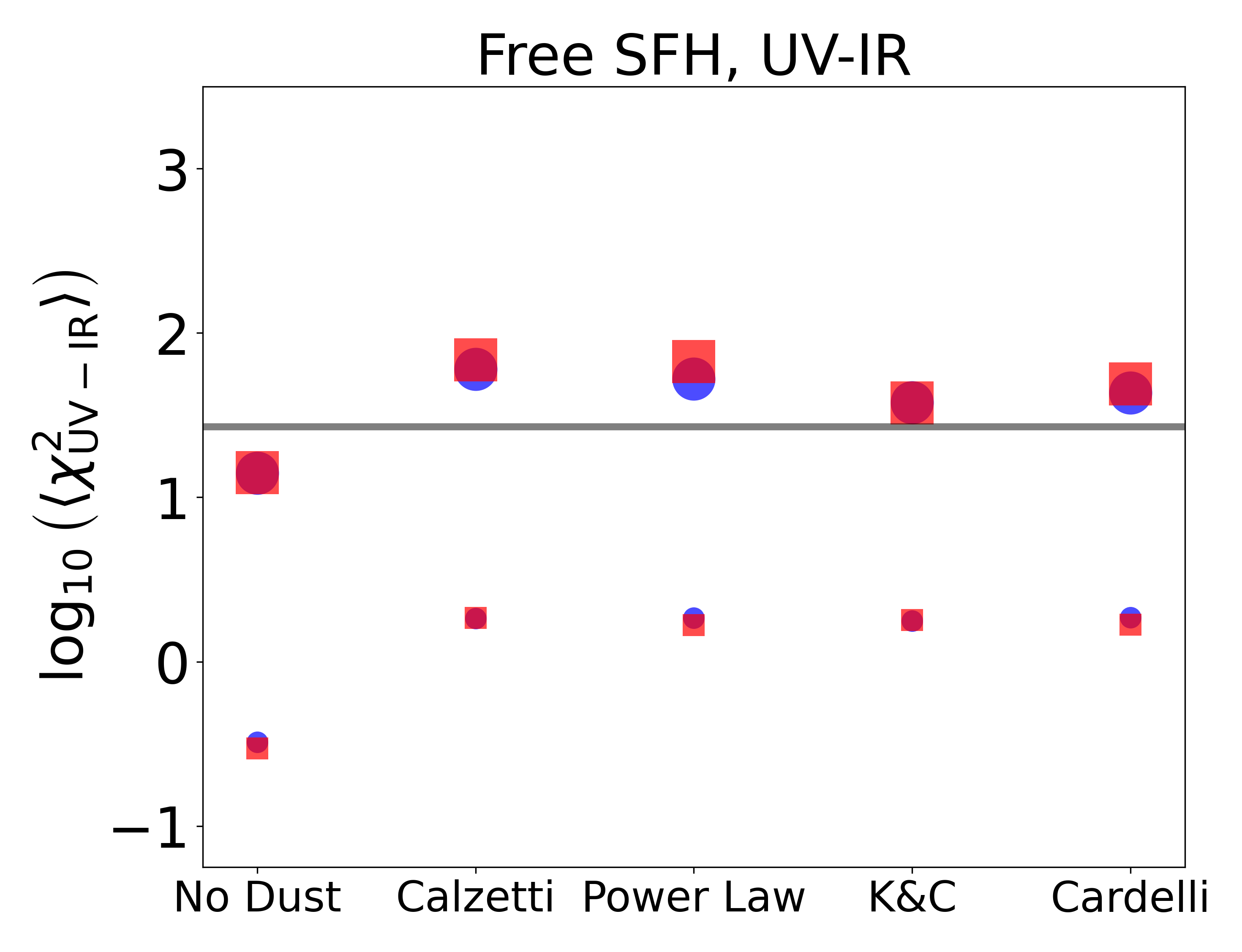}{0.25\textwidth}{}
}
\gridline{\fig{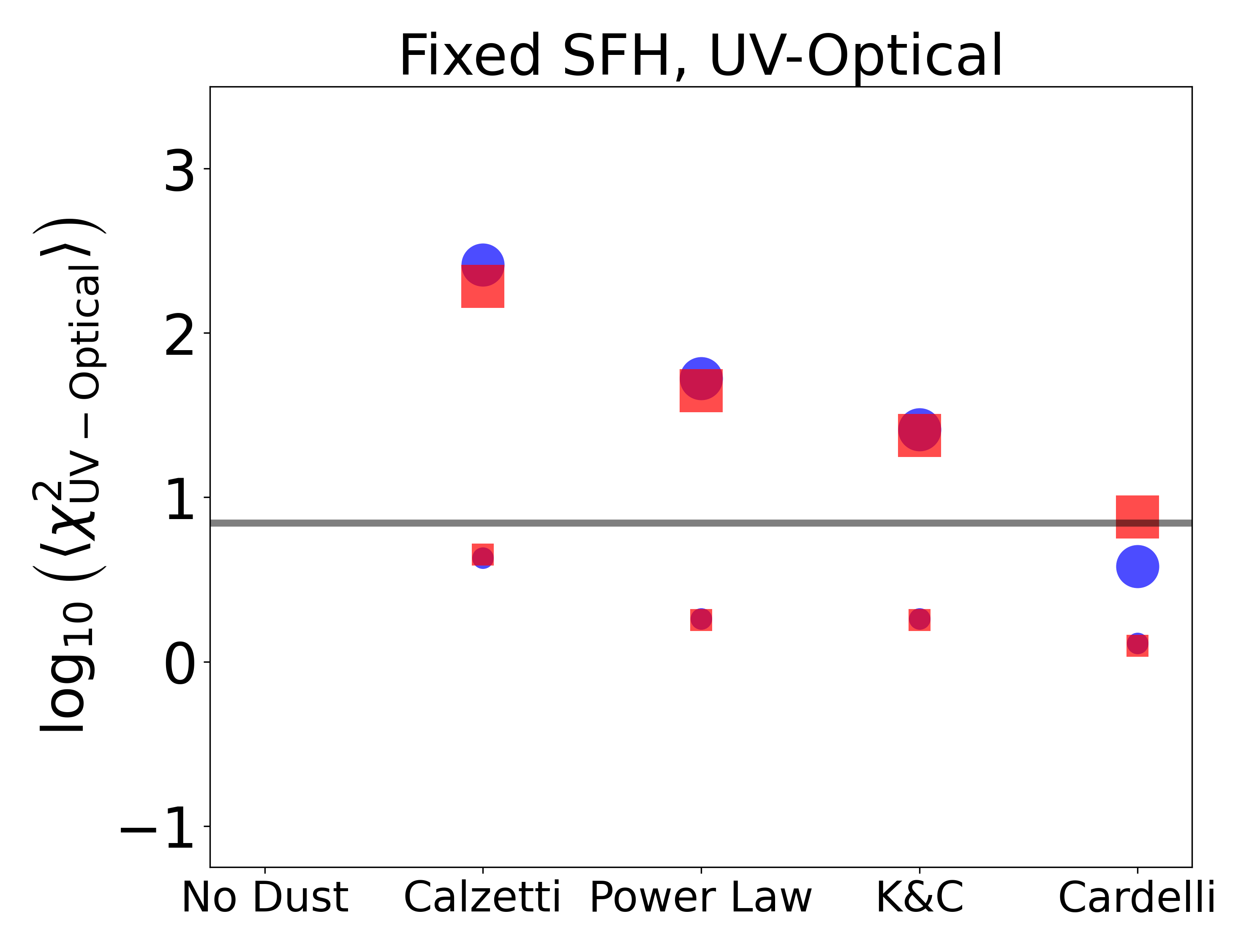}{0.25\textwidth}{}    \fig{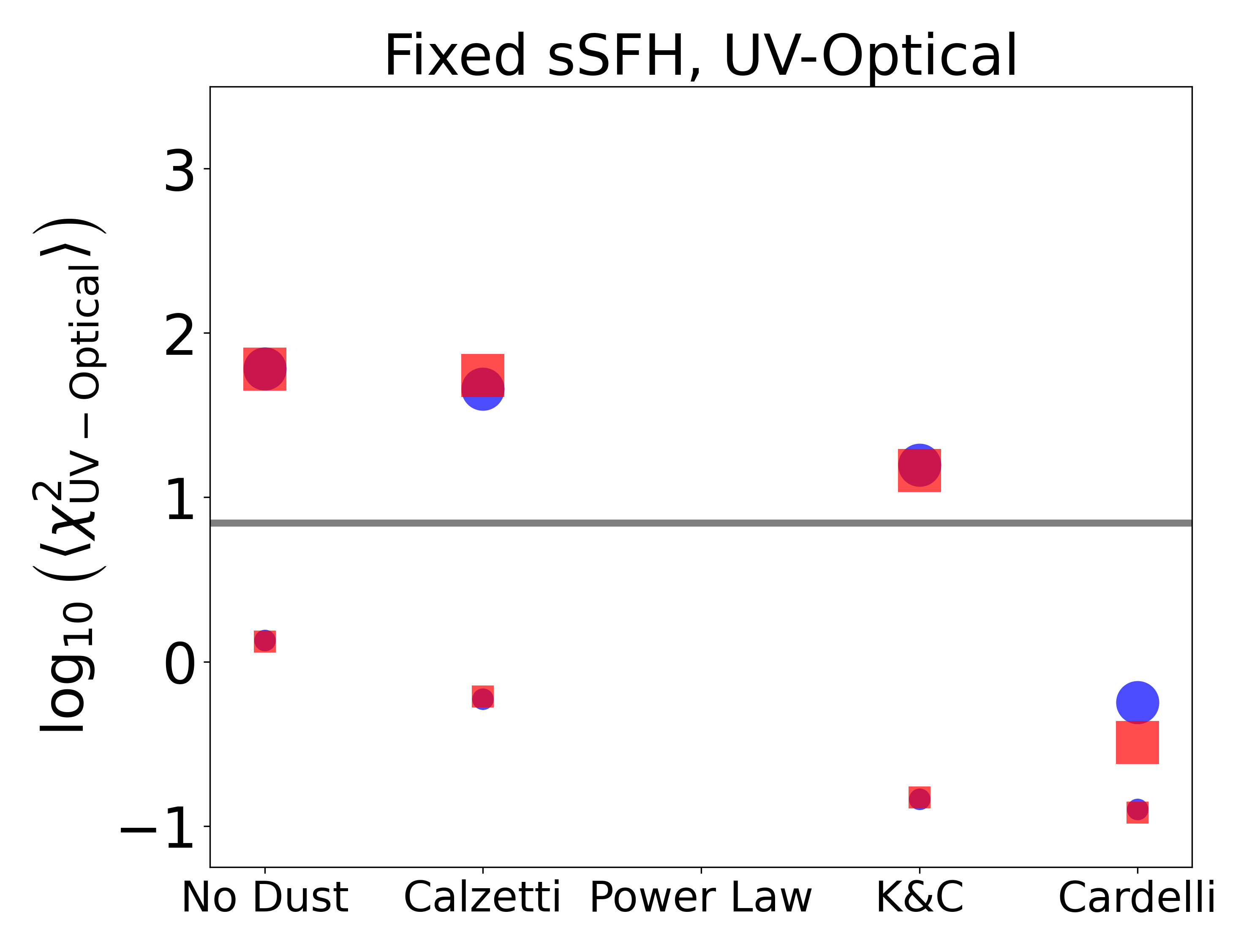}{0.25\textwidth}{}
\fig{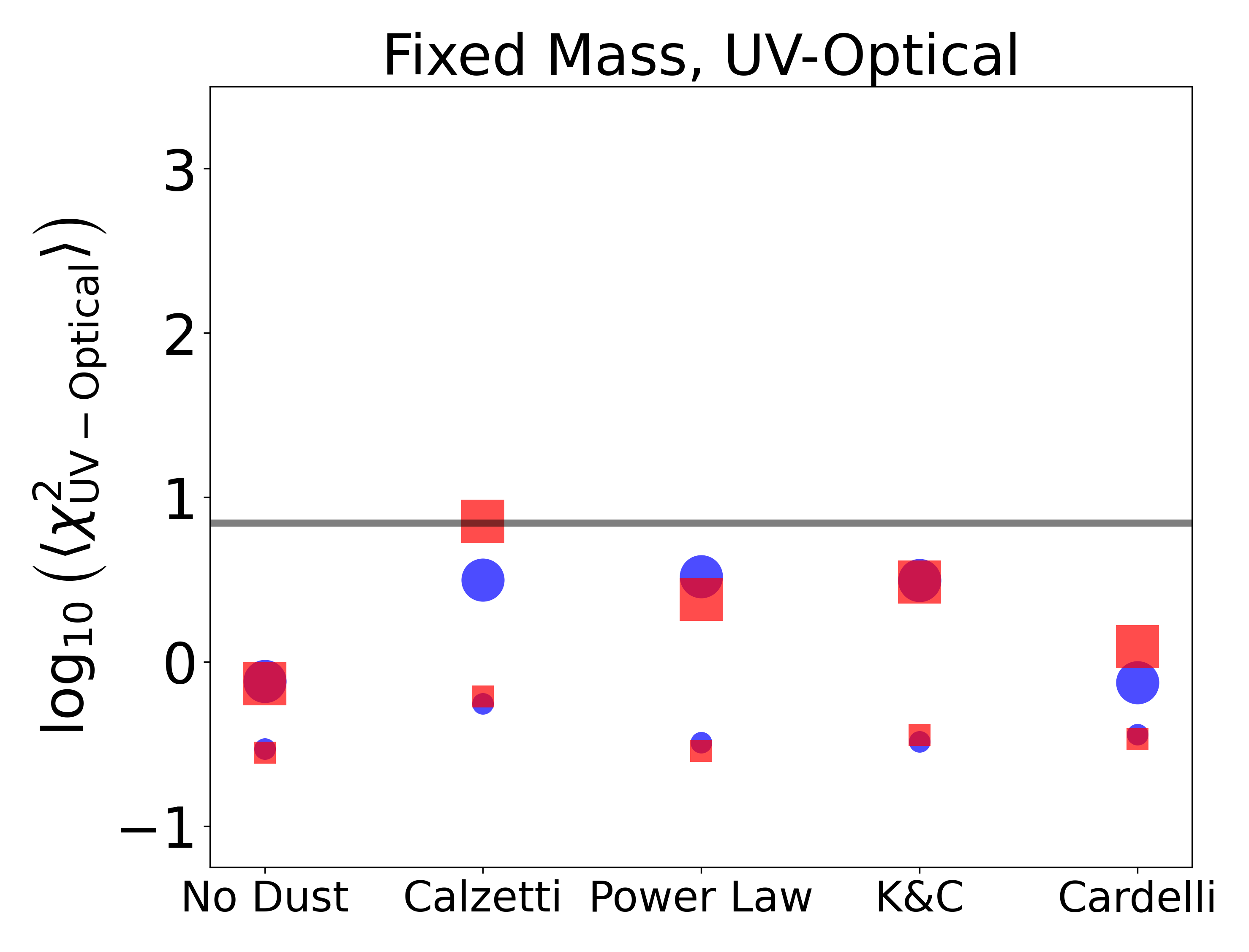}{0.25\textwidth}{}
\fig{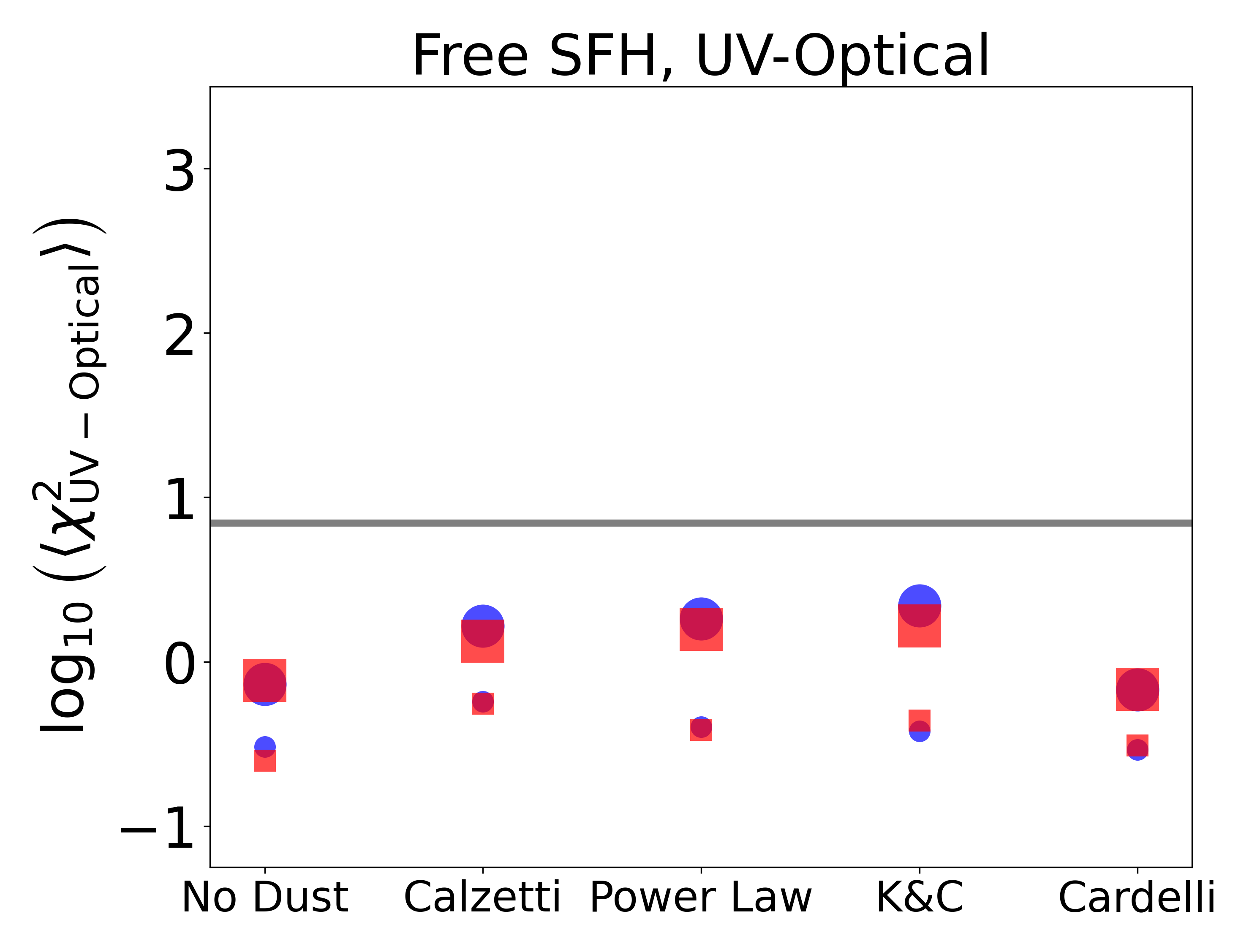}{0.25\textwidth}{}
}
\gridline{\fig{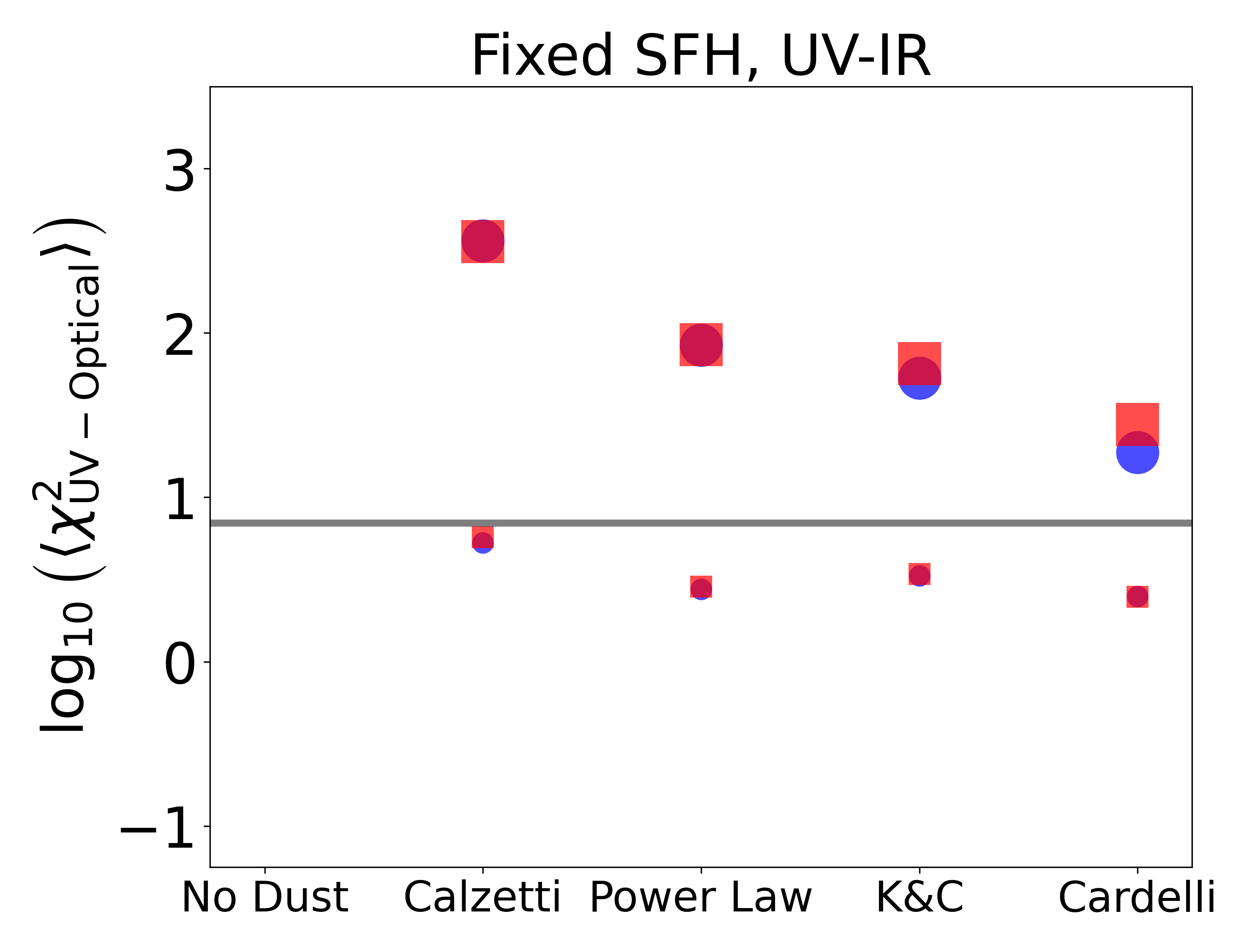}{0.25\textwidth}{}    \fig{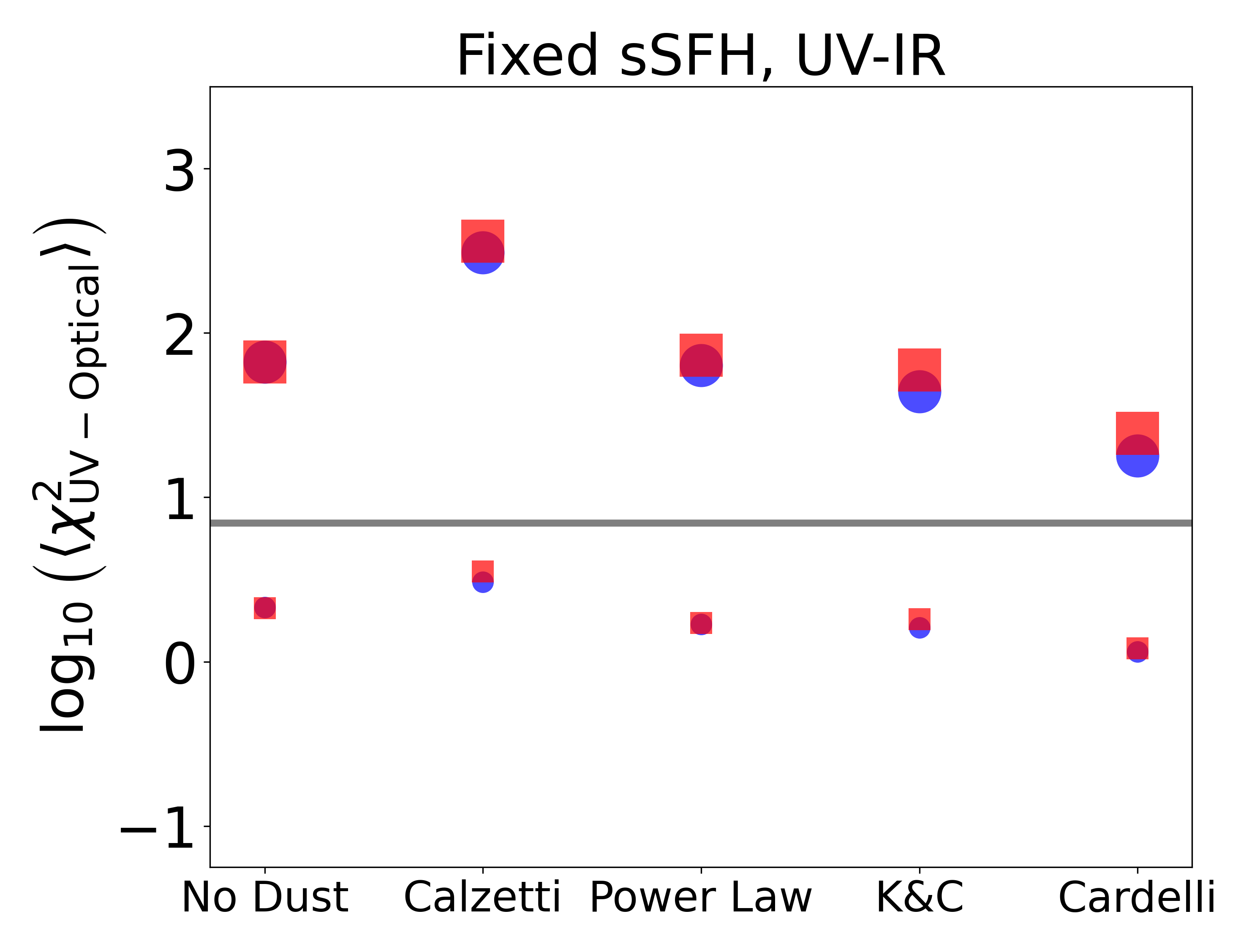}{0.25\textwidth}{}
\fig{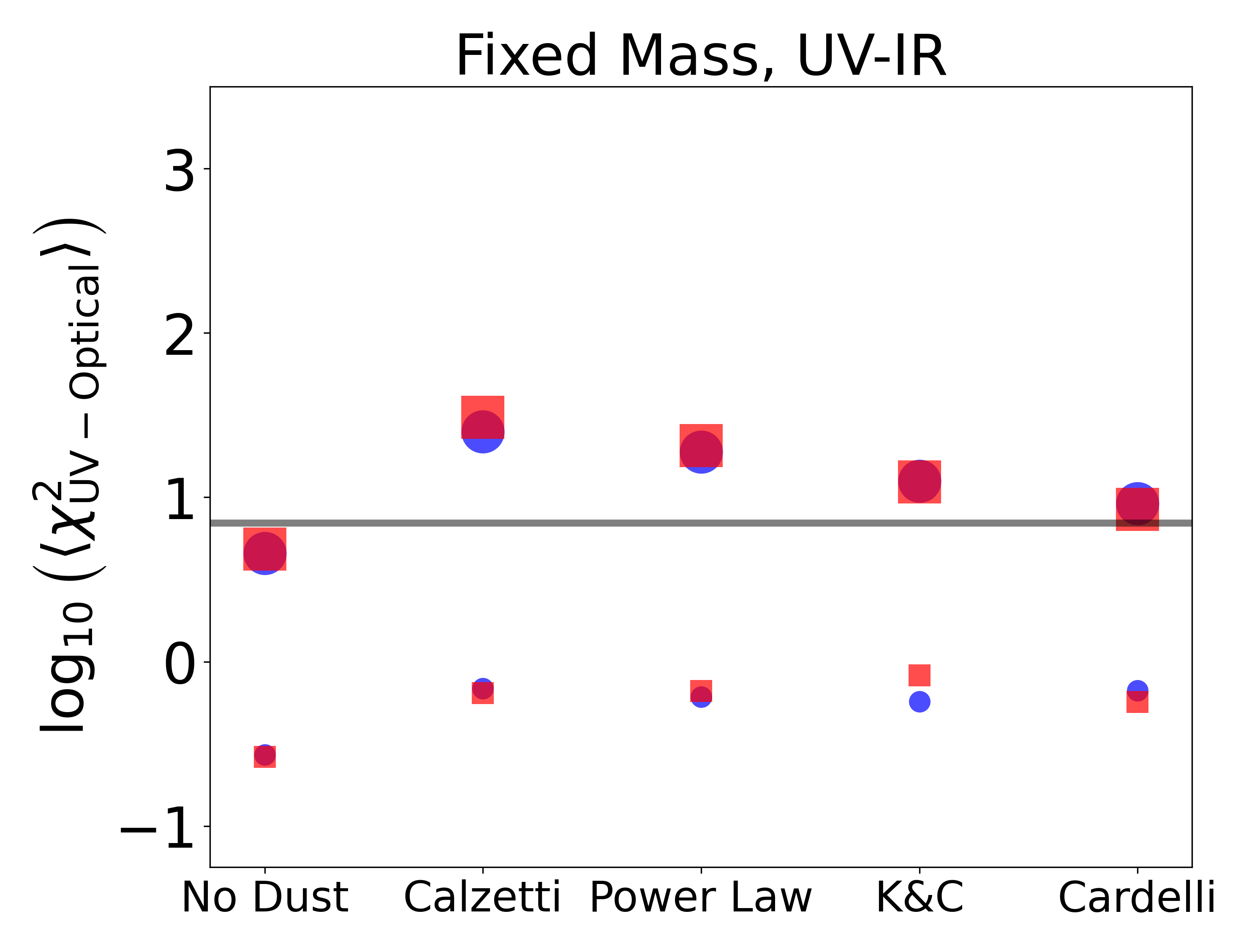}{0.25\textwidth}{}
\fig{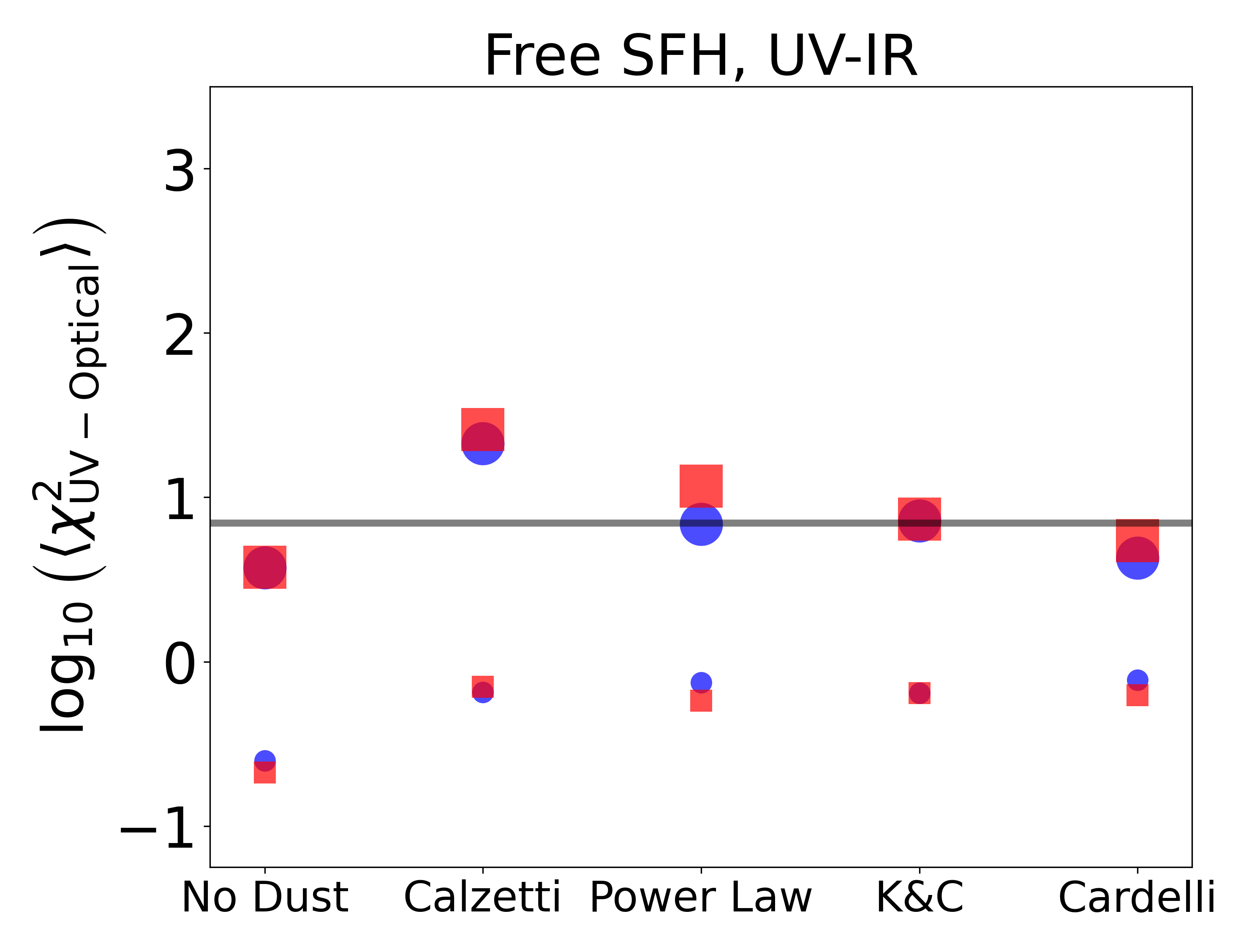}{0.25\textwidth}{}
}
\caption{\label{fig:chi2} $\chi^2$ values for UV-IR fits summed over UV-IR fluxes (top), UV-Optical fits summed over UV-Optical fluxes (middle), and UV-IR fits summed over UV-Optical fluxes (bottom), averaged over all high mass ($> 10^{9.5} M_{\odot}$) galaxies (large points) and low mass ($< 10^{9.5} M_{\odot}$) galaxies (small points) for each dust model and modeling method. From left to right we show fits that have fixed SFHs, fixed sSFHs, fixed masses, and free SFHs. The horizontal lines are located at the number of bands being summed over in each $\chi^2$ calculation and act only as a visual guide for assessing the goodness of fit.}
\end{figure}

\begin{figure}
\gridline{\fig{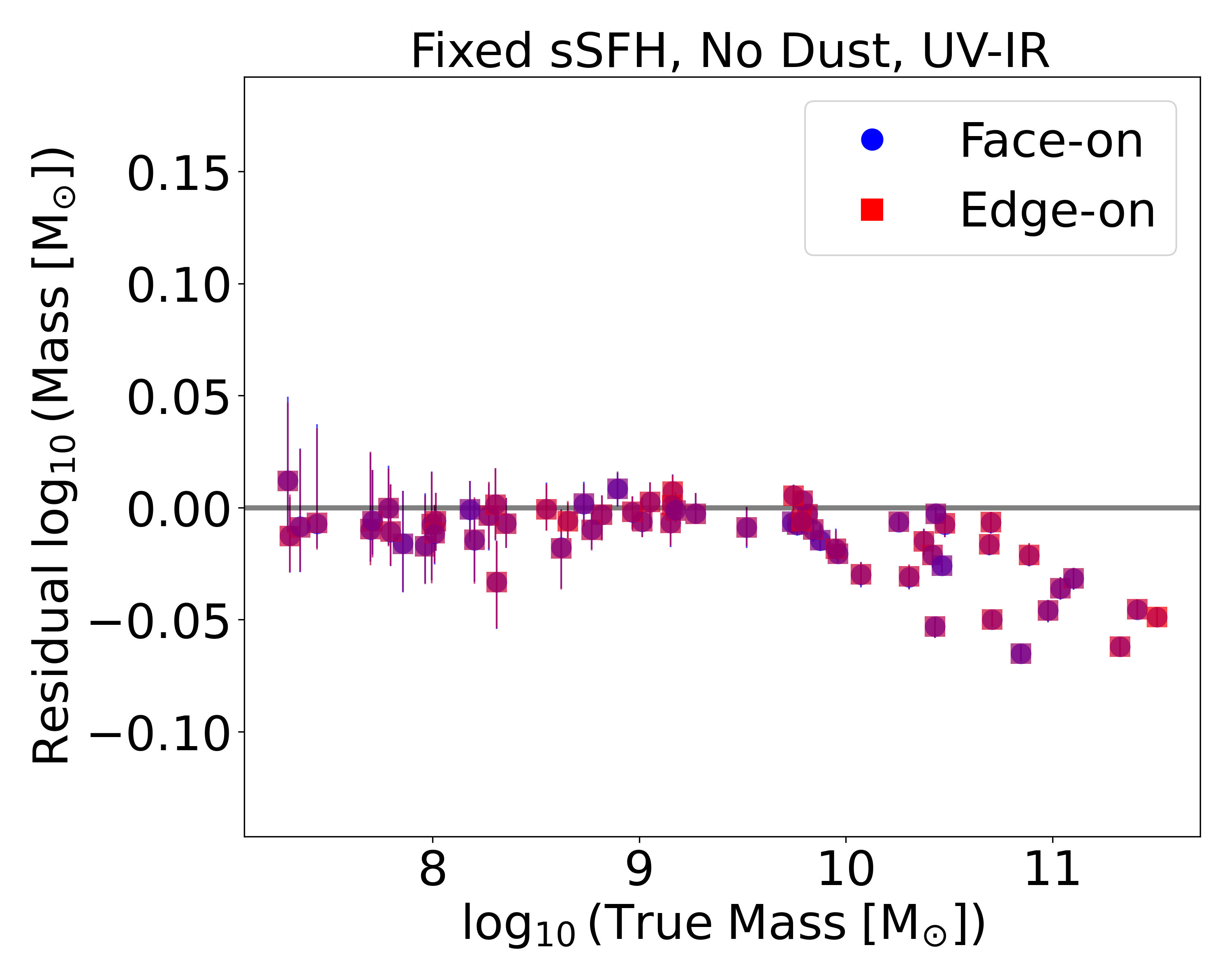}{0.33\textwidth}{}
\fig{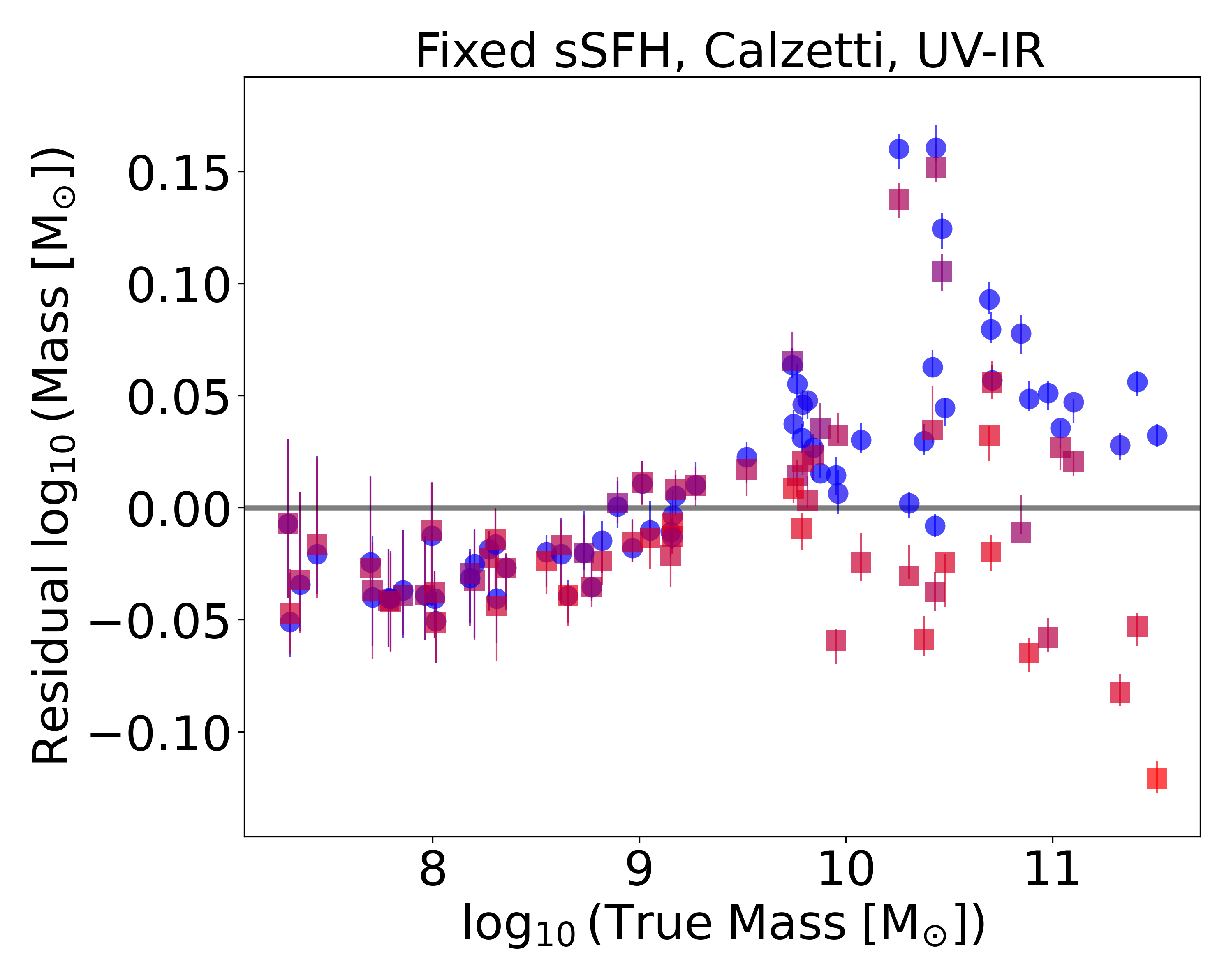}{0.33\textwidth}{}
\fig{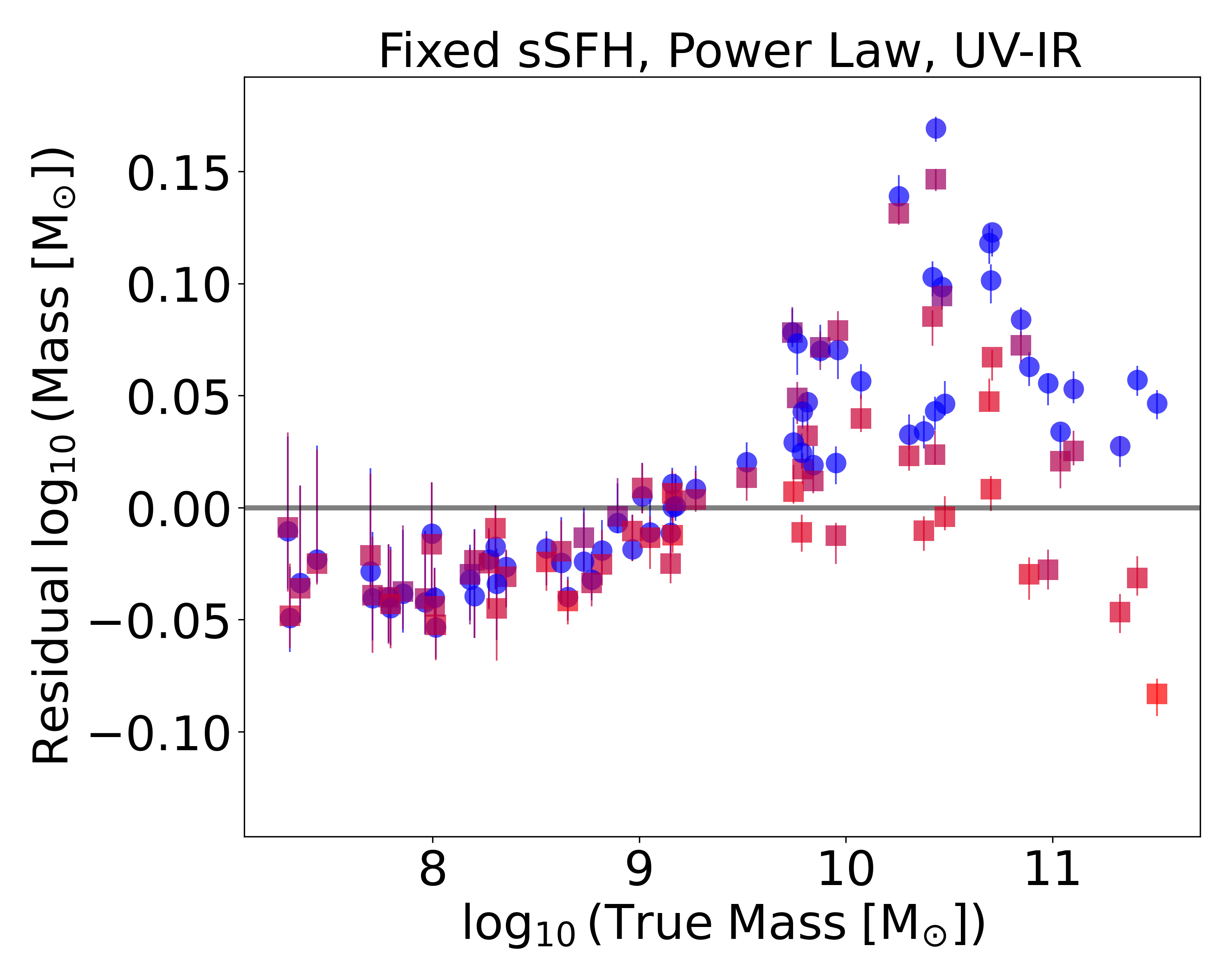}{0.33\textwidth}{}}
\gridline{\fig{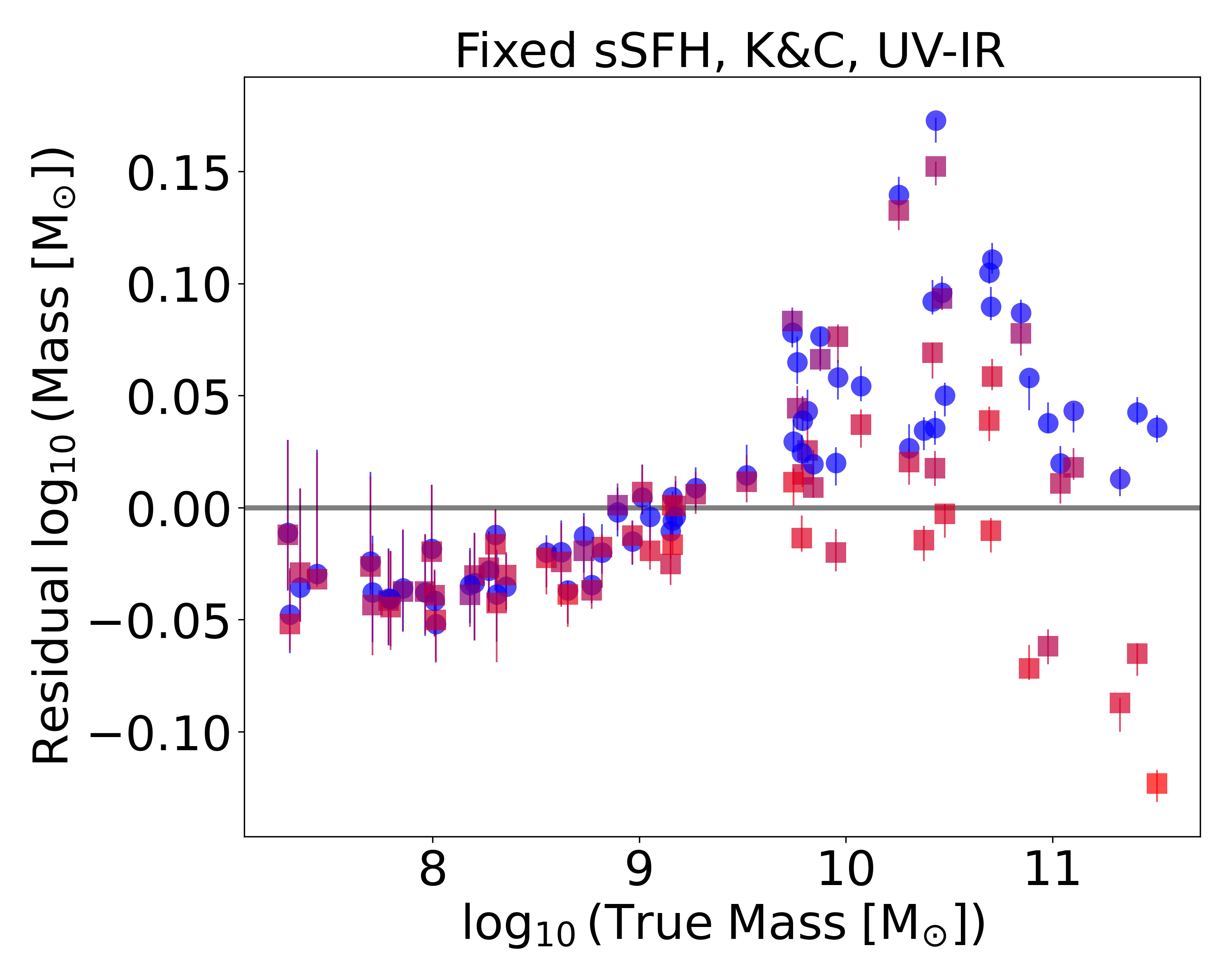}{0.33\textwidth}{}
\fig{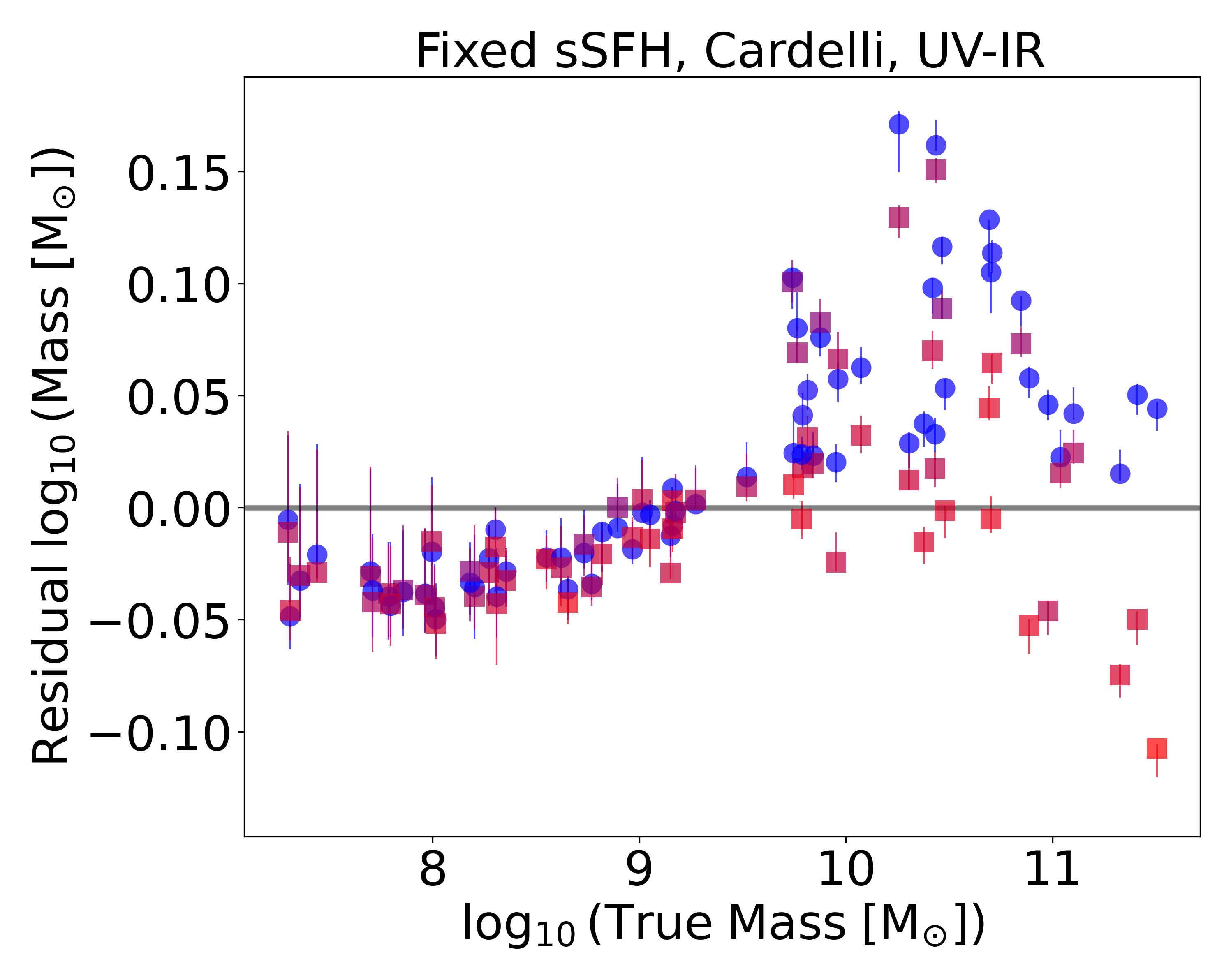}{0.33\textwidth}{}}
\caption{\label{fig:fixed_sSFH_UV-IR_mass_residuals} Distributions of inferred mass residuals as a function of true masses for each dust model using UV-IR filters with sSFHs fixed to their true values. Face-on orientations are denoted by circles and edge-on orientations by squares, with the colors indicating their axis ratios as shown in Figure \ref{fig:axisRatioColors}.}
\end{figure}

\begin{figure}
\gridline{\fig{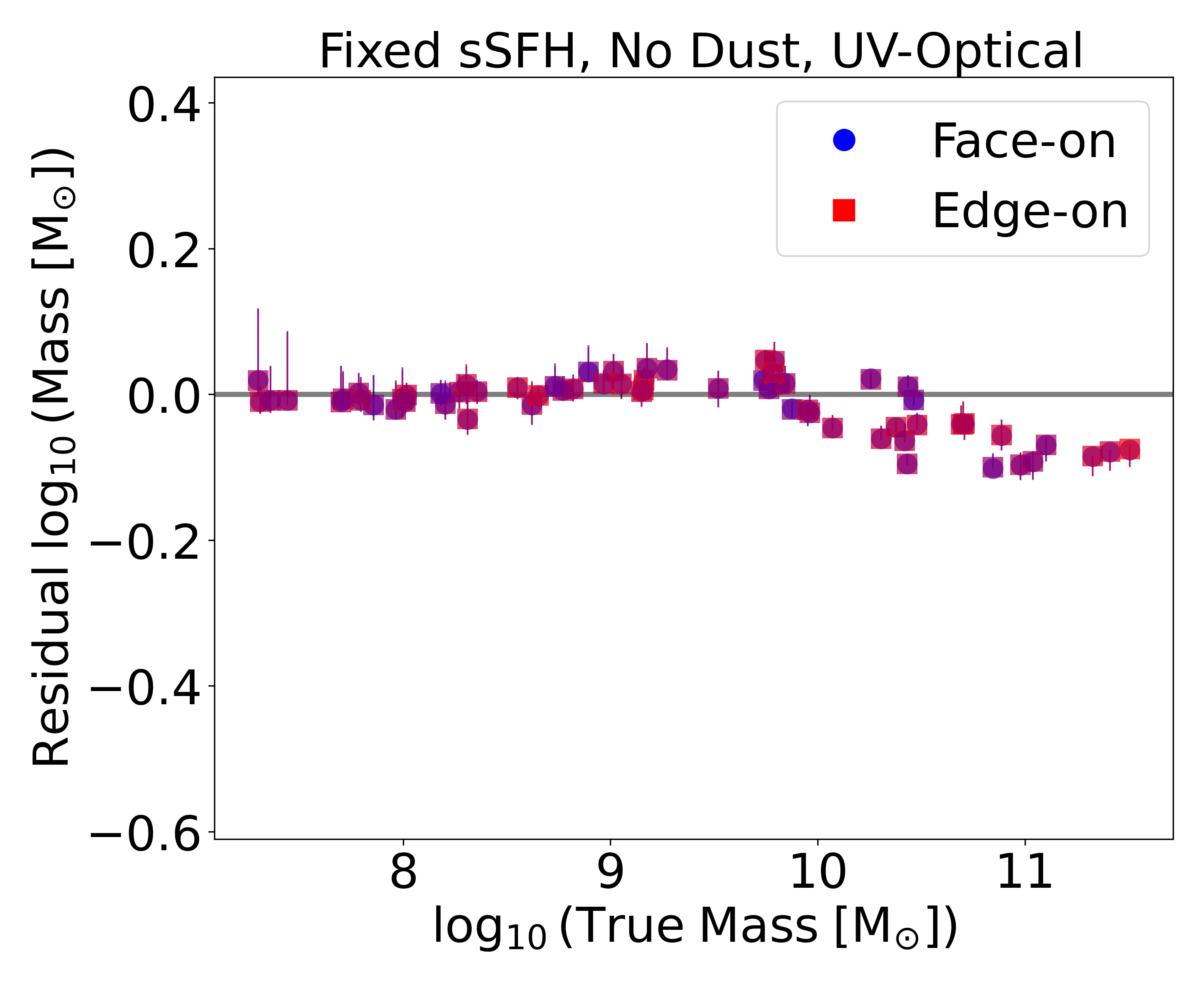}{0.33\textwidth}{}
\fig{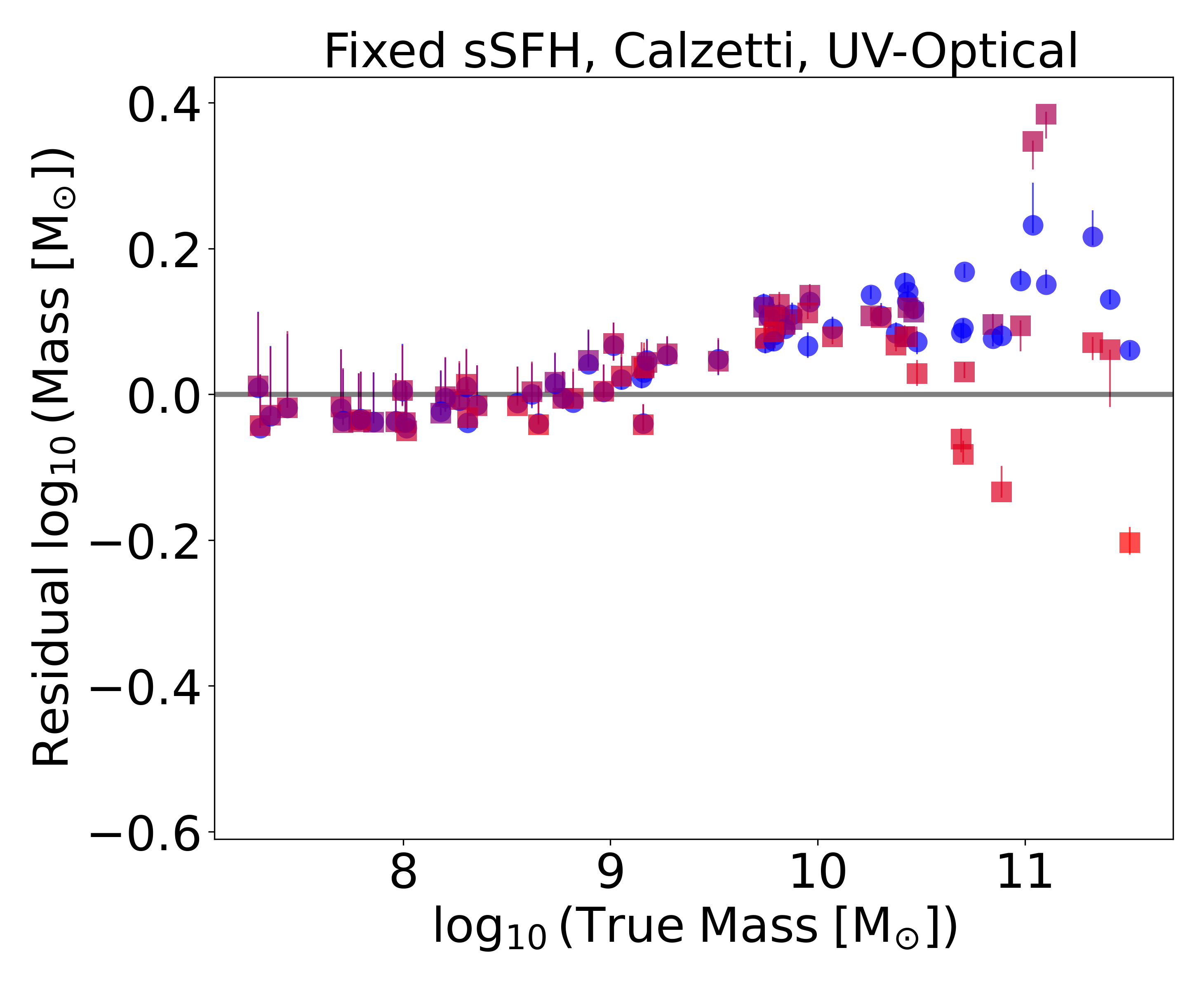}{0.33\textwidth}{}
}
\gridline{\fig{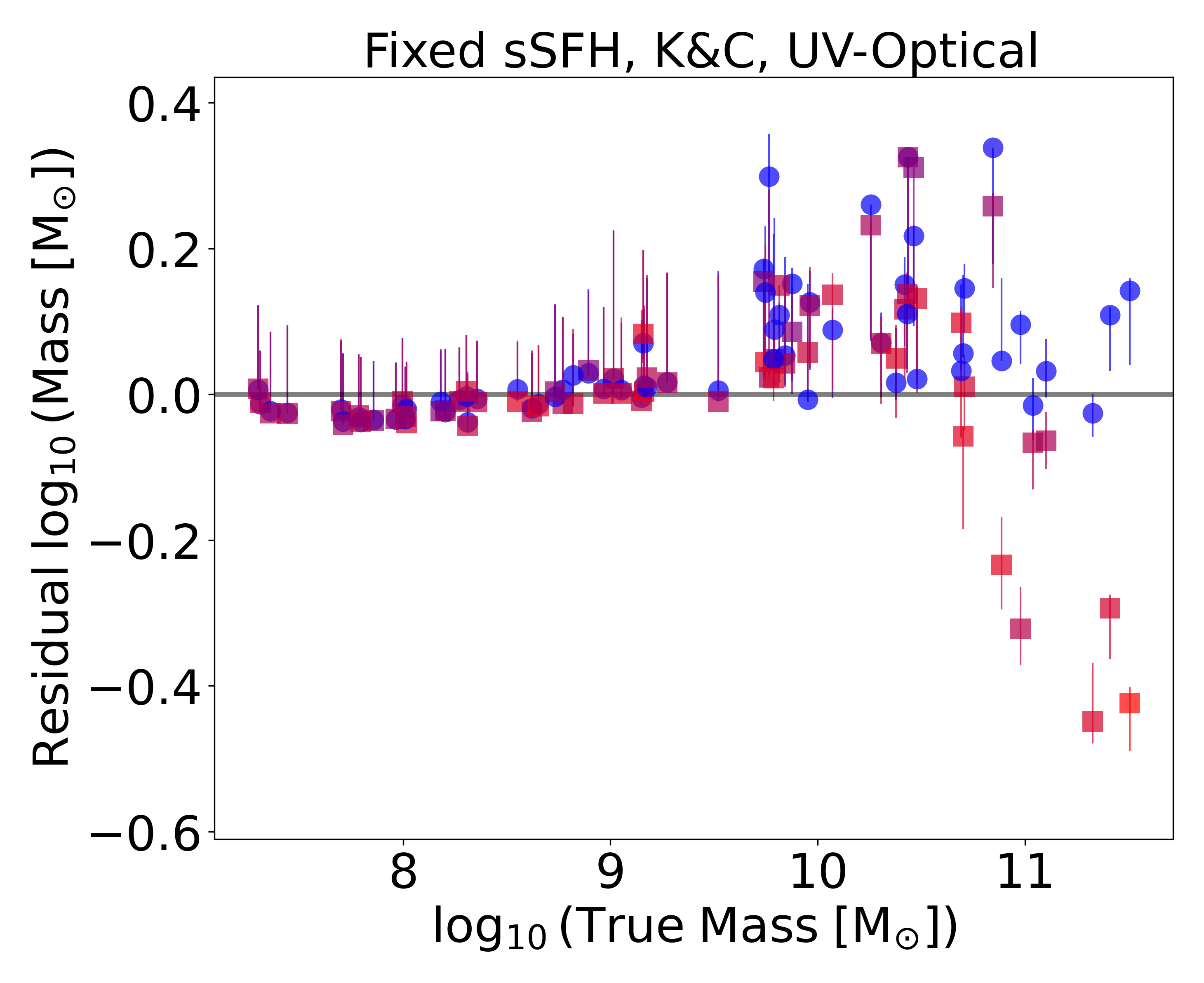}{0.33\textwidth}{}
\fig{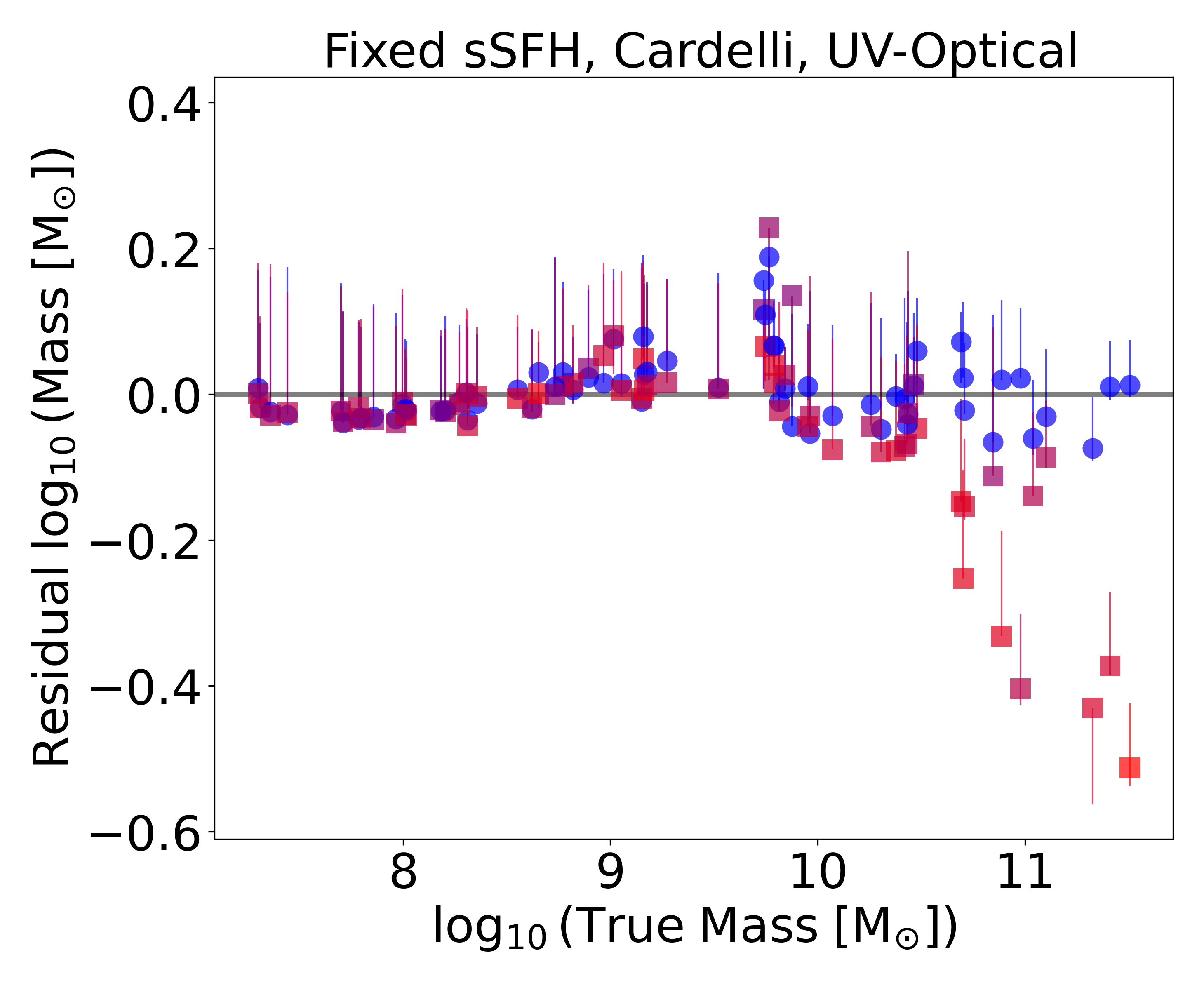}{0.33\textwidth}{}}
\caption{\label{fig:fixed_sSFH_UV-Optical_mass_residuals} Similar to Figure \ref{fig:fixed_sSFH_UV-IR_mass_residuals} but fit to UV-Optical filters. Under these conditions, the power law dust attenuation model was unable to converge in a reasonable amount of time.}
\end{figure}

\noindent To show how various modeling assumptions affect our ability to infer stellar masses and SFRs, we present plots of inferred parameter residuals vs. their true values separately for each dust model and data selection (UV-Optical and UV-IR filters). We further separate this analysis into sections based on the information we give to Prospector, with Fixed SFH in Section \ref{Fixed SFH} having all SFH related parameters fixed to their true values, Fixed sSFH in Section \ref{Fixed sSFH} having the SFH shape (not the total mass) fixed to their true values, Fixed mass in Section \ref{Fixed Mass} having total masses fixed to their true values, and Free SFH in Section \ref{Free SFH} having all SFH related parameters free. 

\subsection{Fixed SFH} \label{Fixed SFH}
\noindent In this section we explore fits in which we fix all SFH parameters including total stellar masses to their true values. 
Because Prospector requires at least two free parameters to perform fits, we do not include dust free fits using this method, since the only free parameter would be the stellar metallicity. This section is qualitatively different than the remaining sections in the sense that all the parameters that we are interested in recovering (masses and SFRs) are fixed to their true values. Therefore, we limit our discussion in this section to the results shown in Figure \ref{fig:chi2}, which show the $\chi^2$ values of each fitting method, summed over the bands denoted in the subscript and averaged over all high mass ($> 10^{9.5} M_{\odot}$) galaxies and low mass ($< 10^{9.5} M_{\odot}$) galaxies separately using the equation

\begin{equation}
\langle \chi_{\rm bands}^{2} \rangle = \frac{1}{N_g} \sum_{g=1}^{N_g} \left( \sum_{b=1}^{N_b}  \frac{\delta_{b,g}^{2}}{\sigma_{b,g}^{2}} \right)
\label{eq:chi2}
\end{equation}

\noindent where $N_g$ is the number of galaxies being averaged over, $N_b$ is the number of bands being summed, $\delta_{b,g}$ is $b^{th}$ flux residual of the $g^{th}$ galaxy, and $\sigma_{b,g}$ is the standard deviation of the flux uncertainty. We can see from this equation that $\chi^2$ values that are summed over more bands will tend to be larger. Although we don't expect these values to be $\chi^2$-distributed, since our residuals are not independent, our model is not linear, and flux residuals are not drawn from a standard normal distribution, we nevertheless plot a horizontal line located at the mean value of a $\chi^2$-squared distribution with $N_b$ degrees of freedom, which is just $N_b$, as a visual guide for assessing the goodness of fit. We note that since low mass galaxies have lower $S/N$, their $\chi^2$ values will be systematically smaller than high mass galaxies. 

The general trend we see in Figure \ref{fig:chi2} is that as we move from fixing parameters to their true values to allowing all parameters to be free (going left to right), we get better fits to the data. This is perhaps not too surprising given that we are essentially adding flexibility to the model by allowing parameters to be free, but it is also a clear demonstration of the effect that model mismatches have on our ability to reproduce the data. 

Comparing the middle and bottom rows of Figure \ref{fig:chi2}, we can see that when fitting to only UV-Optical bands (middle row), the $\chi^2$ values in the UV-Optical bands are consistently lower than the $\chi^2$ values in the UV-Optical bands when fitting to UV-IR photometry (bottom row). That is, the UV-IR fits sacrifice UV-Optical accuracy in order to better match the photometry in the mid- and far-infrared. This gives us a sense of the degree to which model mismatches affect our ability to match the data between these two data selection regimes when fixing the SFH to the true values.

We can also see that going from left to right within each Fixed SFH plot, there is a trend at high masses for more flexible dust models to more accurately reproduce the UV-Optical photometry, while the accuracy of UV-IR photometry stays relatively constant. Comparing the top and bottom rows, we see that UV-Optical bands from UV-IR fits also show increasing accuracy with increasing dust model flexibility, although the trend is weaker than in the UV-Optical fits. This tells us that the inclusion of longer wavelength photometry limits the amount that dust attenuation curve flexibility is able to compensate for model mismatches in the UV-Optical bands. Since only the total energy emitted by dust (not its shape) is constrained by the total energy attenuated by dust, we can attribute this limited ability to compensate for model mismatches to the assumption of energy balance.  

\subsection{Fixed sSFH} \label{Fixed sSFH}
\noindent In this section, we explore how well Prospector is able to recover the physical parameters of our simulated galaxies when fixing the sSFHs (SFH shape), not including the total stellar masses, to their true values. We note that when fixing these parameters and using UV-Optical bands, our fitting procedure was unable to converge in a reasonable amount of time ($\sim 10$ days of CPU time per fit) under the Power Law dust attenuation model. The cause of this will be discussed further in Section \ref{Free SFH}.

Figure \ref{fig:fixed_sSFH_UV-IR_mass_residuals} shows inferred mass residuals as a function of the 
true masses using UV-IR bands with sSFHs fixed to their true values for all dust models. At top left, we show the results of applying
the dust-free Prospector fits to the dust-free simulated SEDs, which have face-on and edge-on 
spectra that are identical. The corresponding agreement between the face-on and edge-on 
inferences indicates that the dust free fits with fixed sSFHs are fully converged in this 
highly constrained parameter space. There is a still a residual in stellar mass, but a 
relatively small one ($< 0.05$ dex).

When dust is included in the modeling applied to the simulated SEDs with dust, 
the mass residuals for high mass galaxies have significantly more scatter than 
the dust free case though the agreement remains within $\pm 0.15$ dex. For low 
mass galaxies, the mass residual scatter is similar to  the dust free case but 
with a slightly more negative bias. 
Comparing face-on and edge-on orientations, we can see that for the high mass 
gaalxies, edge-on orientations tend to have underestimated masses and face-on galaxies 
have overestimated masses. We can understand this by recalling that in the simulated 
galaxies, edge-on orientations tend to have more energy attenuated than emitted by 
dust, while in the fits strict energy balance is imposed. Therefore, in order for 
the fits to match the IR bands, the energy attenuated by dust will tend to be 
underestimated for edge-on orientations, since IR light can only come from the dust. 
With the attenuated energy being underestimated, the fits compensate for this excess 
in UV-Optical light by lowering the total stellar mass. For the face-on orientations,
the opposite of these effects is taking place.

Figure \ref{fig:fixed_sSFH_UV-Optical_mass_residuals} shows inferred mass 
residuals as a function of the true masses using UV-Optical bands with sSFHs fixed to their true values, omitting the 
IR bands included in Figure \ref{fig:fixed_sSFH_UV-IR_mass_residuals}.
In this case, the assumption of energy balance no longer has any effect on our 
inferences. In general the mass inferences become substantially worse (note the 
change in the $y$-axis scale between  Figure \ref{fig:fixed_sSFH_UV-IR_mass_residuals} and 
Figure \ref{fig:fixed_sSFH_UV-Optical_mass_residuals}).

Edge-on orientations at the high mass end ($> 10^{10.5} M_{\odot}$) tend to have 
large negative mass residuals for dust attenuation models that contain a UV-bump 
(K\&C and Cardelli). If we look back to Figure \ref{fig:chi2}, we can see that 
these two dust models, and especially Cardelli, show the best agreement with the 
catalog fluxes. We draw the reader's attention to Figure 12 of \cite{faucher24a}, 
which shows a strong tendency for more heavily attenuated galaxies to have grayer 
attenuation curves. The grayer attenuation allows Prospector to adopt a fit
with lower stellar mass and a lower total  attenuated energy, allowing the dust 
attenuation  model additional flexibility to compensate for the discretization of 
the SFH. 

Even for the Calzetti mass inferences in this range, we still see negative mass biases 
in the edge-on orientations with the lowest axis ratios. In these heavily attenuated cases, 
it is likely that the true attenuation curves are grayer than the Calzetti attenuation 
curve. If we think about lowering the stellar mass as being equivalent to increasing 
the attenuation using a completely gray attenuation curve, we can see how the fits 
might prefer simultaneously lowering the attenuated energy and the stellar mass to 
effectively  give a grayer version of the Calzetti curve. 
Conversely, we might also expect  steeper true attenuation curves to be compensated by an 
increase in the attenuated  energy and stellar mass. Indeed, we do see a trend for 
the Calzetti curve to overestimate the stellar masses at higher axis ratios. 

Note  that when including IR bands in the fit, this trade-off becomes disfavored due to 
the energy balance assumption, since all fluxes are linearly proportional to the 
total stellar mass. That is, changing the attenuated energy cannot be compensated by
a change in stellar mass, because the IR supplies an alternative measure of attenuated
energy.

\subsection{Fixed Mass} \label{Fixed Mass}

\noindent In this section, we explore how well Prospector is able to recover the physical parameters of our simulated galaxies when fixing the total stellar masses to their true values,
but allowing the SFH to vary otherwise.

\begin{figure}
\gridline{\fig{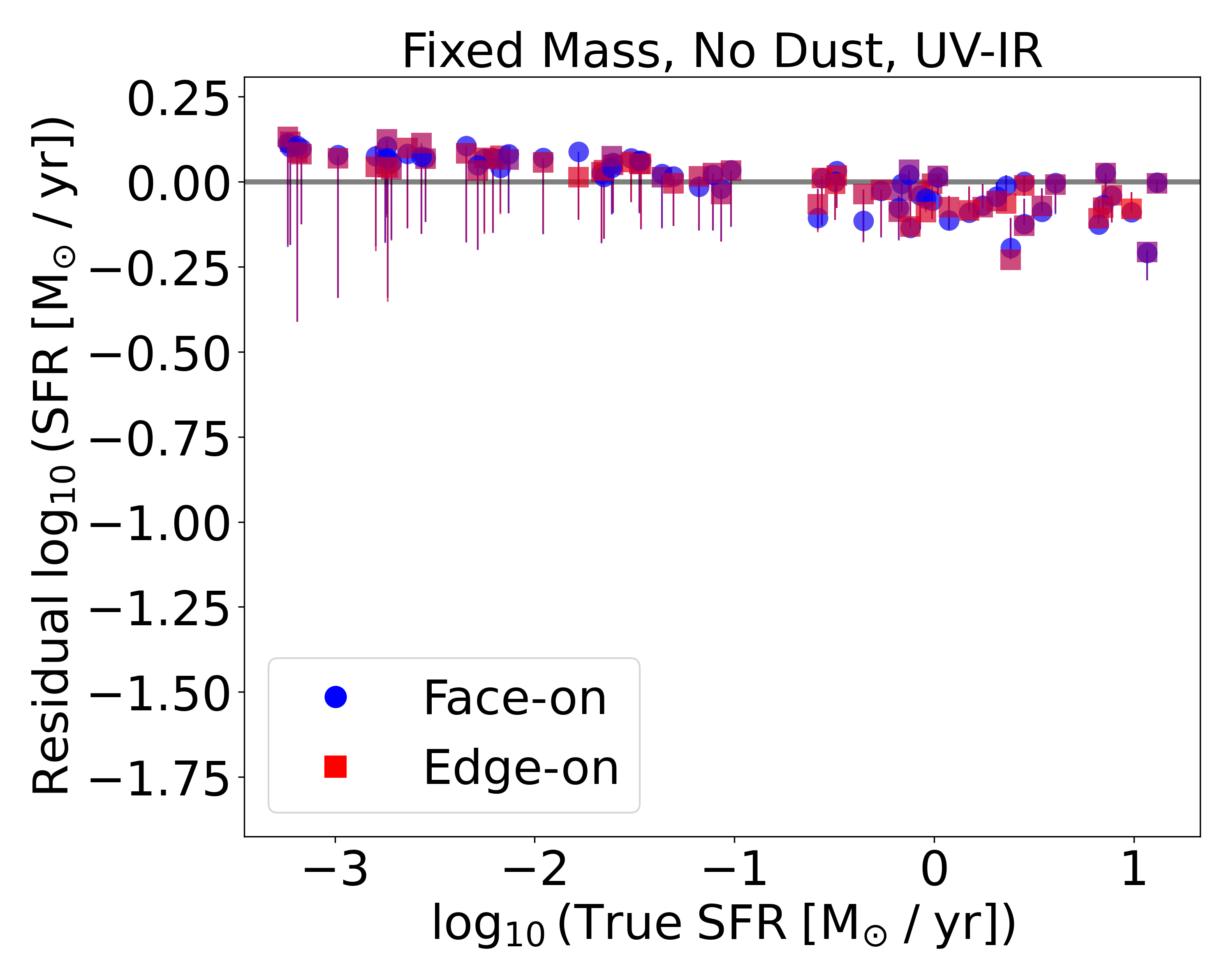}{0.33\textwidth}{}
\fig{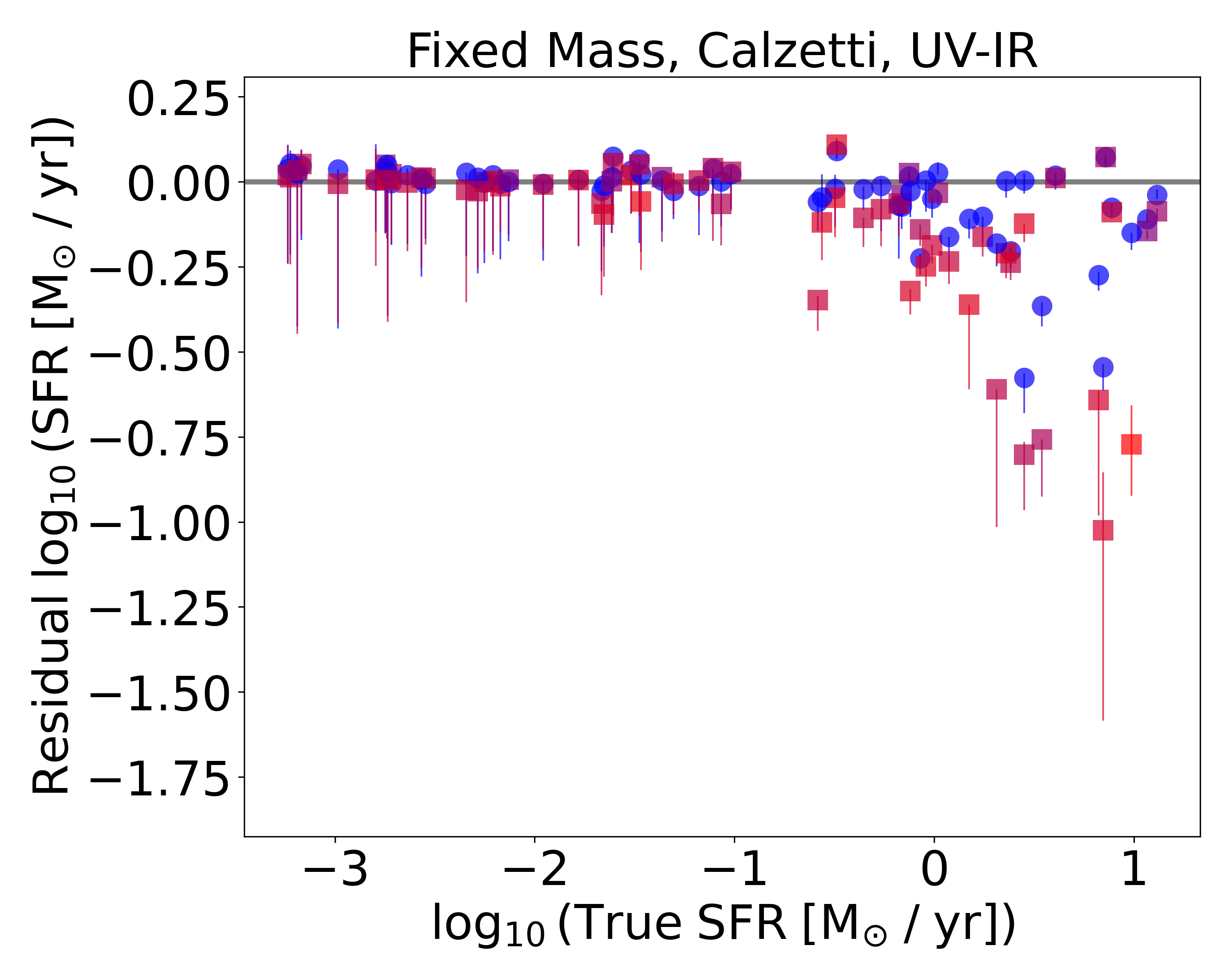}{0.33\textwidth}{}
\fig{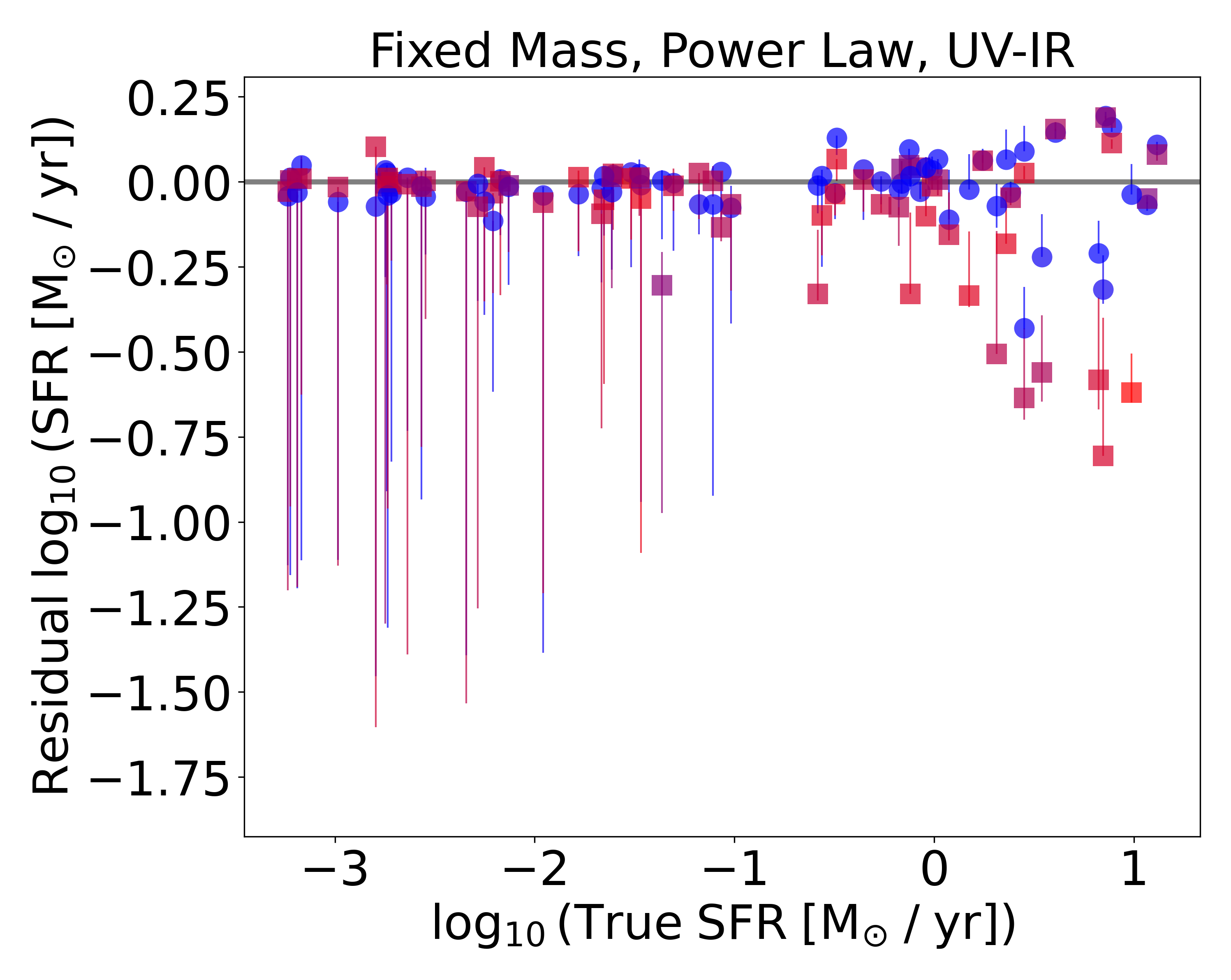}{0.33\textwidth}{}}
\gridline{\fig{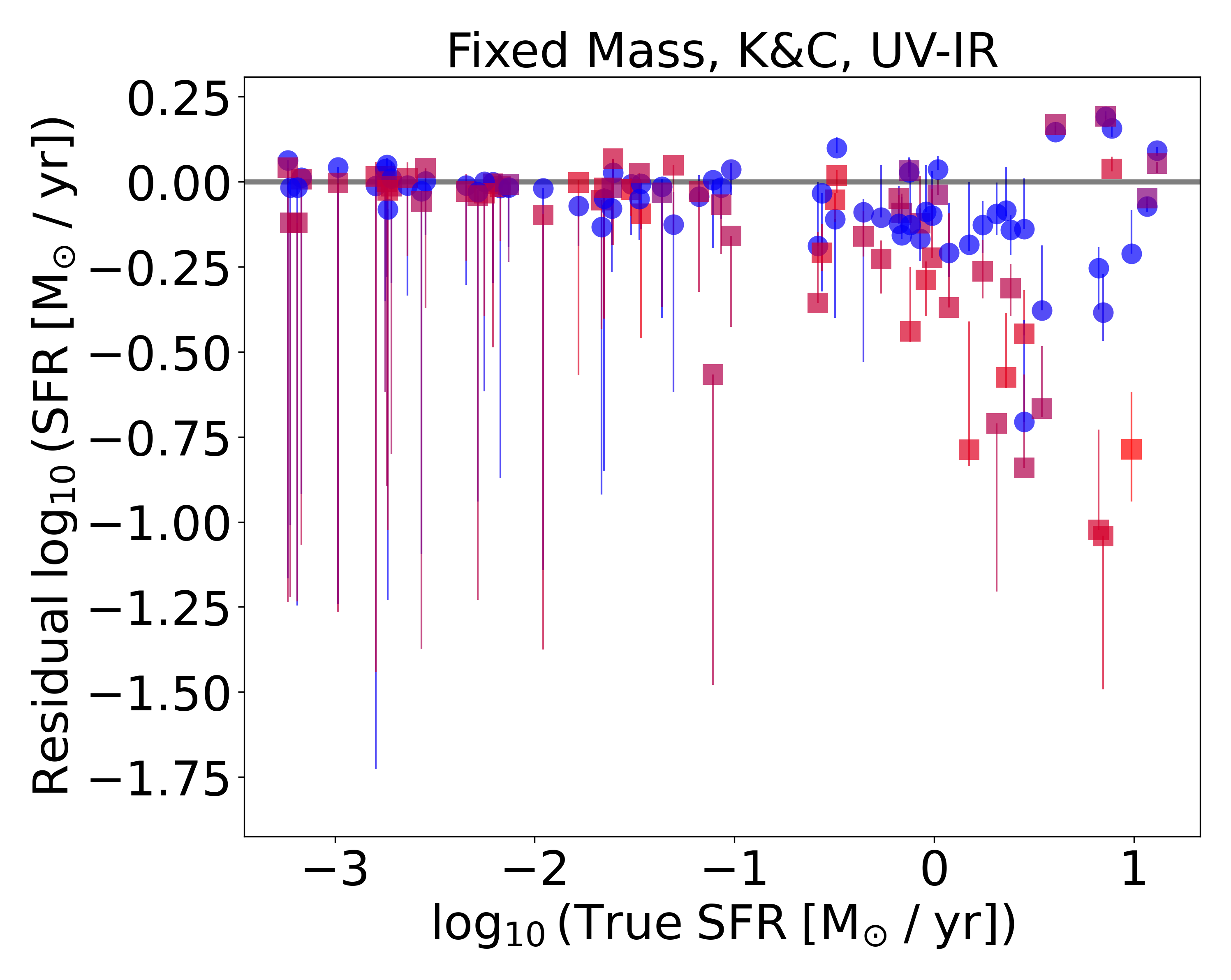}{0.33\textwidth}{}
\fig{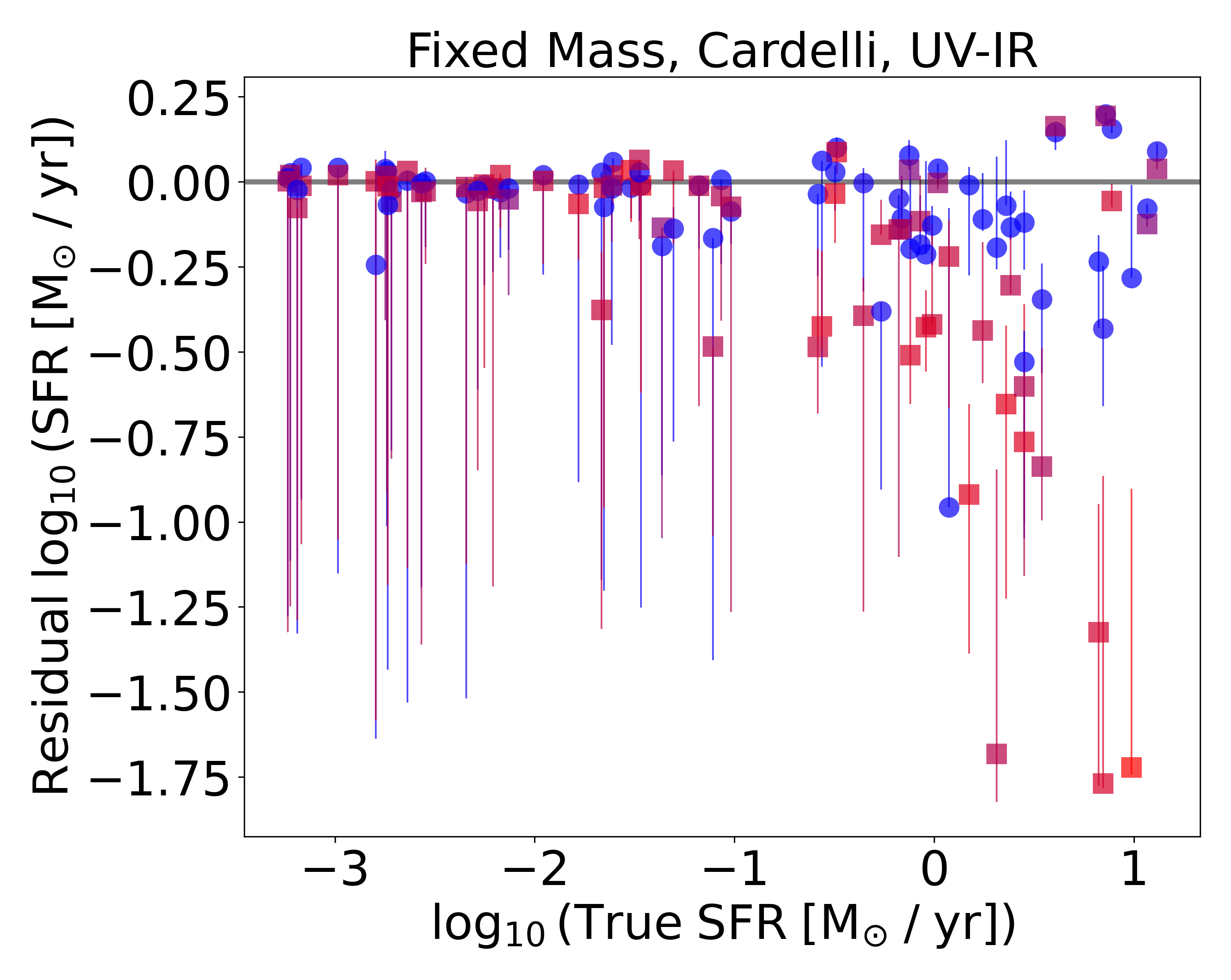}{0.33\textwidth}{}}
\caption{\label{fig:fixed_mass_UV-IR_SFR_residuals} Distributions of inferred SFR residuals as a function of true SFRs using UV-IR filters with total stellar masses fixed to their true values. Face-on orientations are denoted by circles and edge-on orientations by squares, with the colors indicating their axis ratios as shown in Figure \ref{fig:axisRatioColors}.}
\end{figure}

\begin{figure}
\gridline{\fig{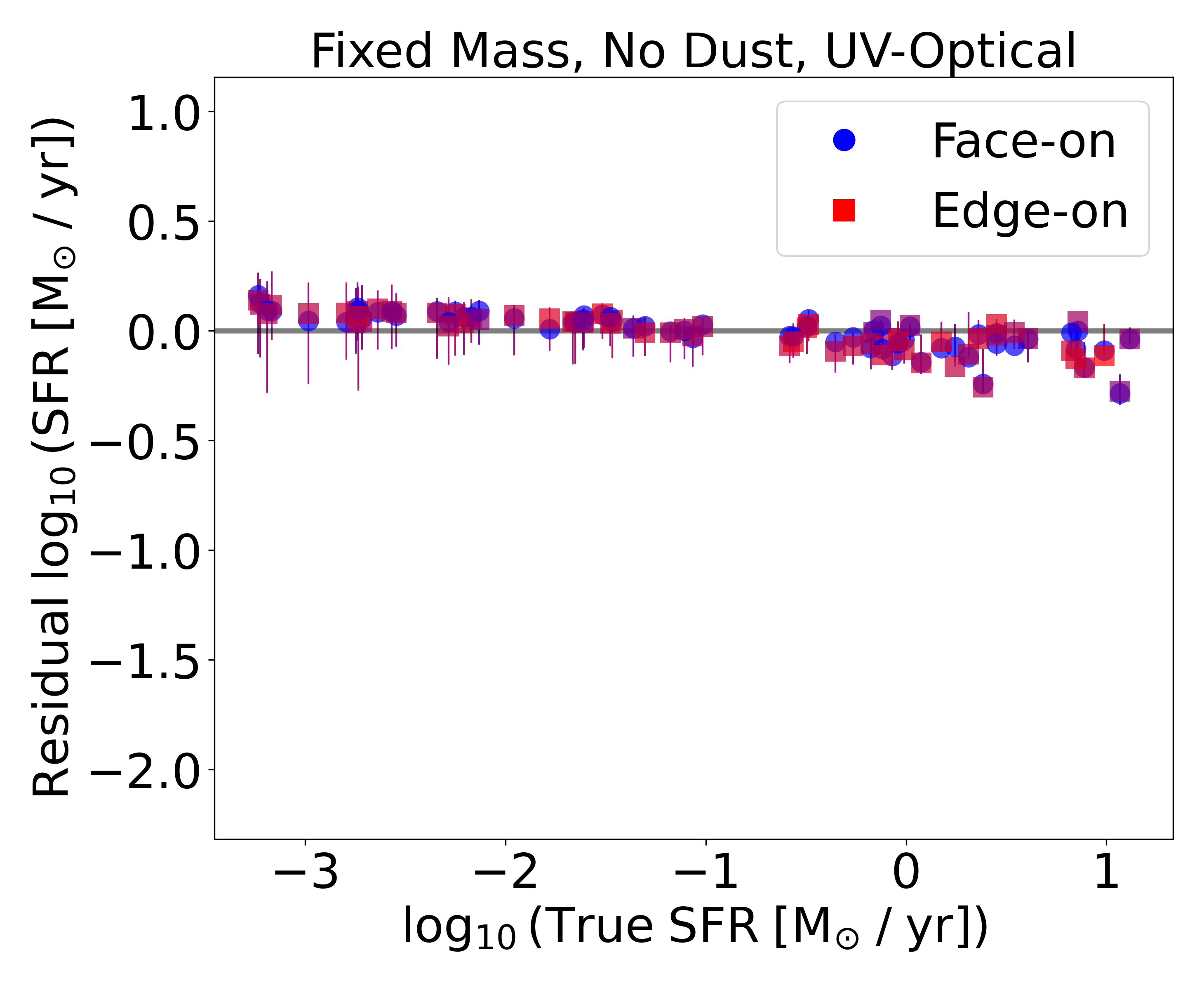}{0.33\textwidth}{}
\fig{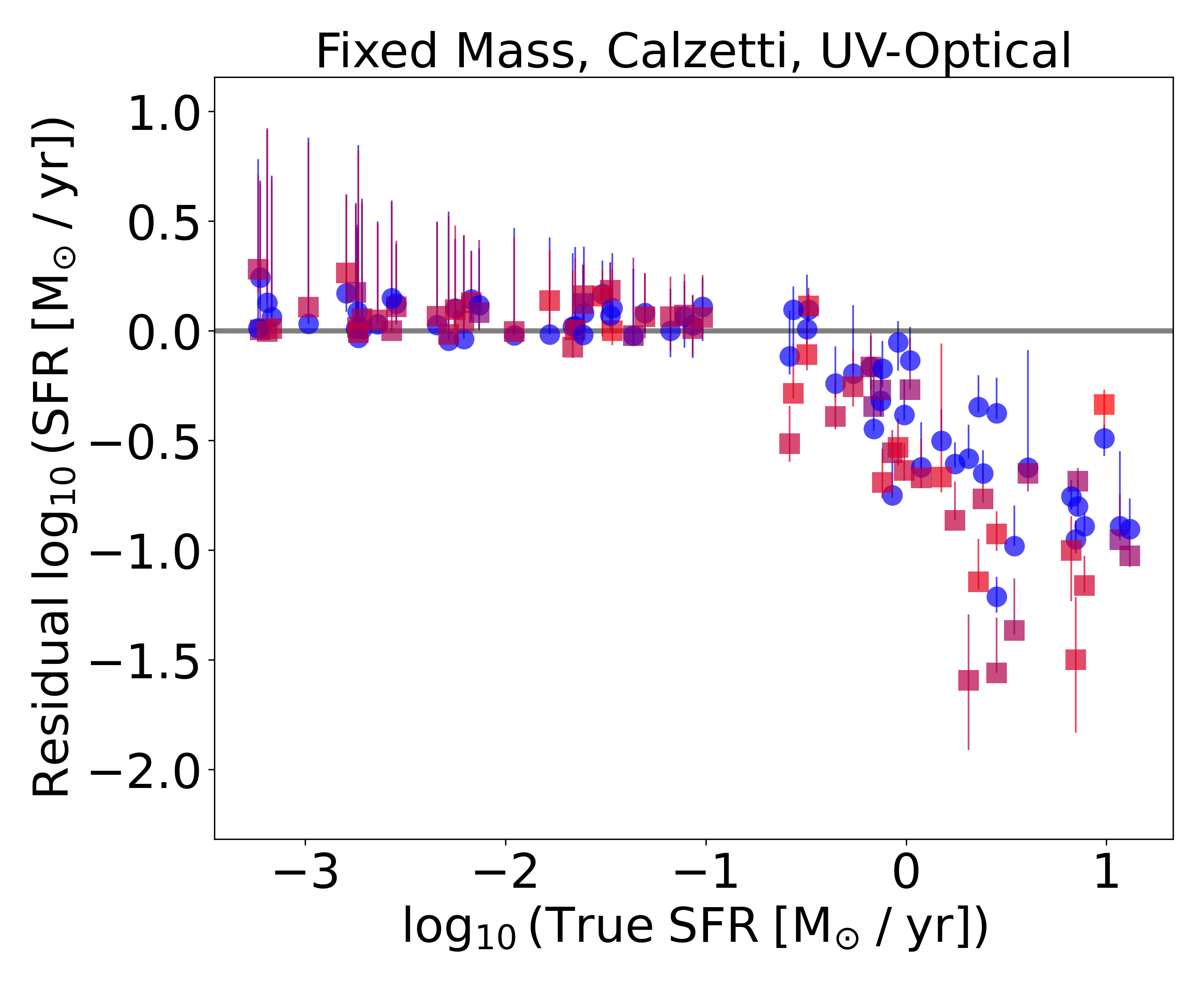}{0.33\textwidth}{}
\fig{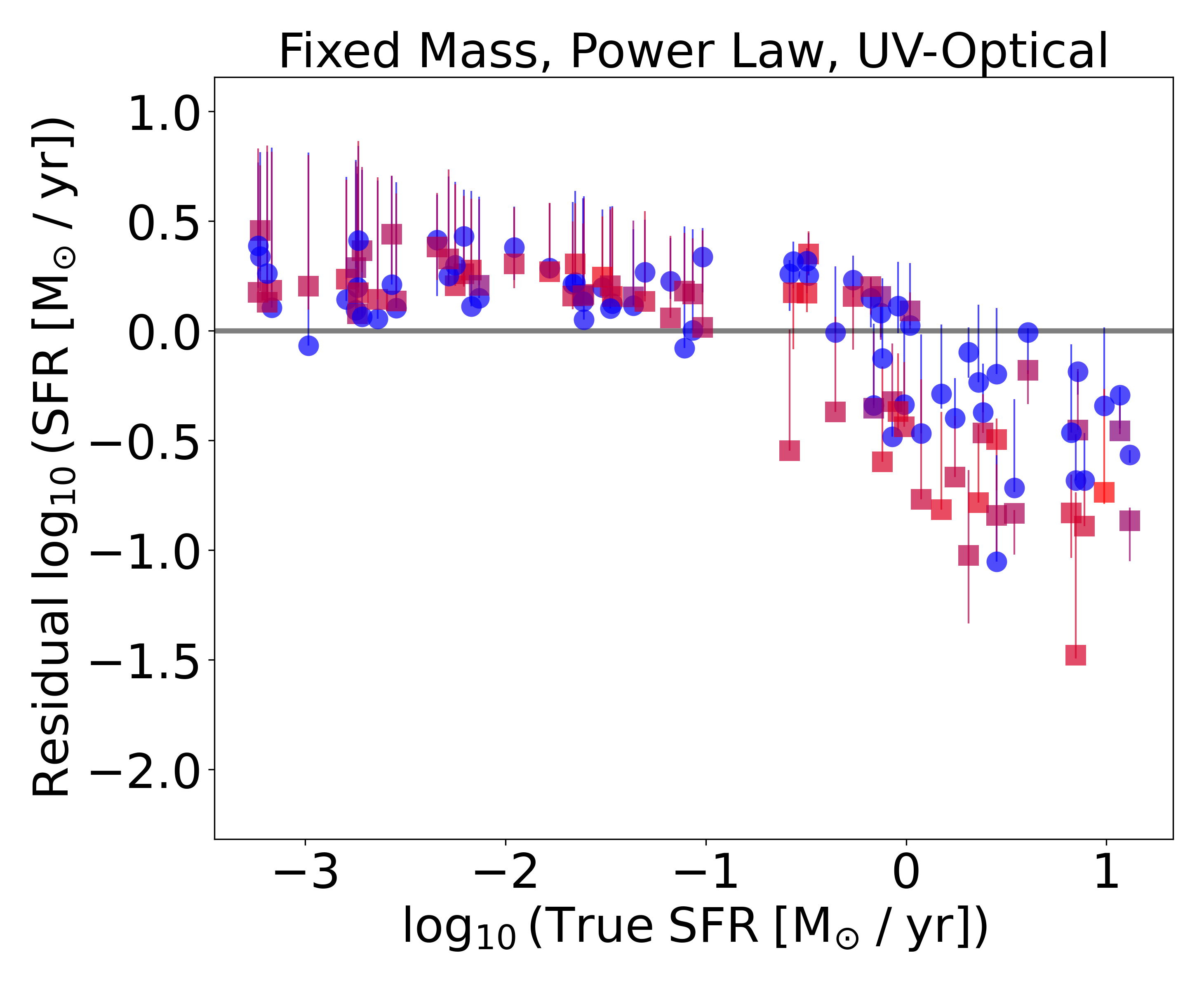}{0.33\textwidth}{}}
\gridline{\fig{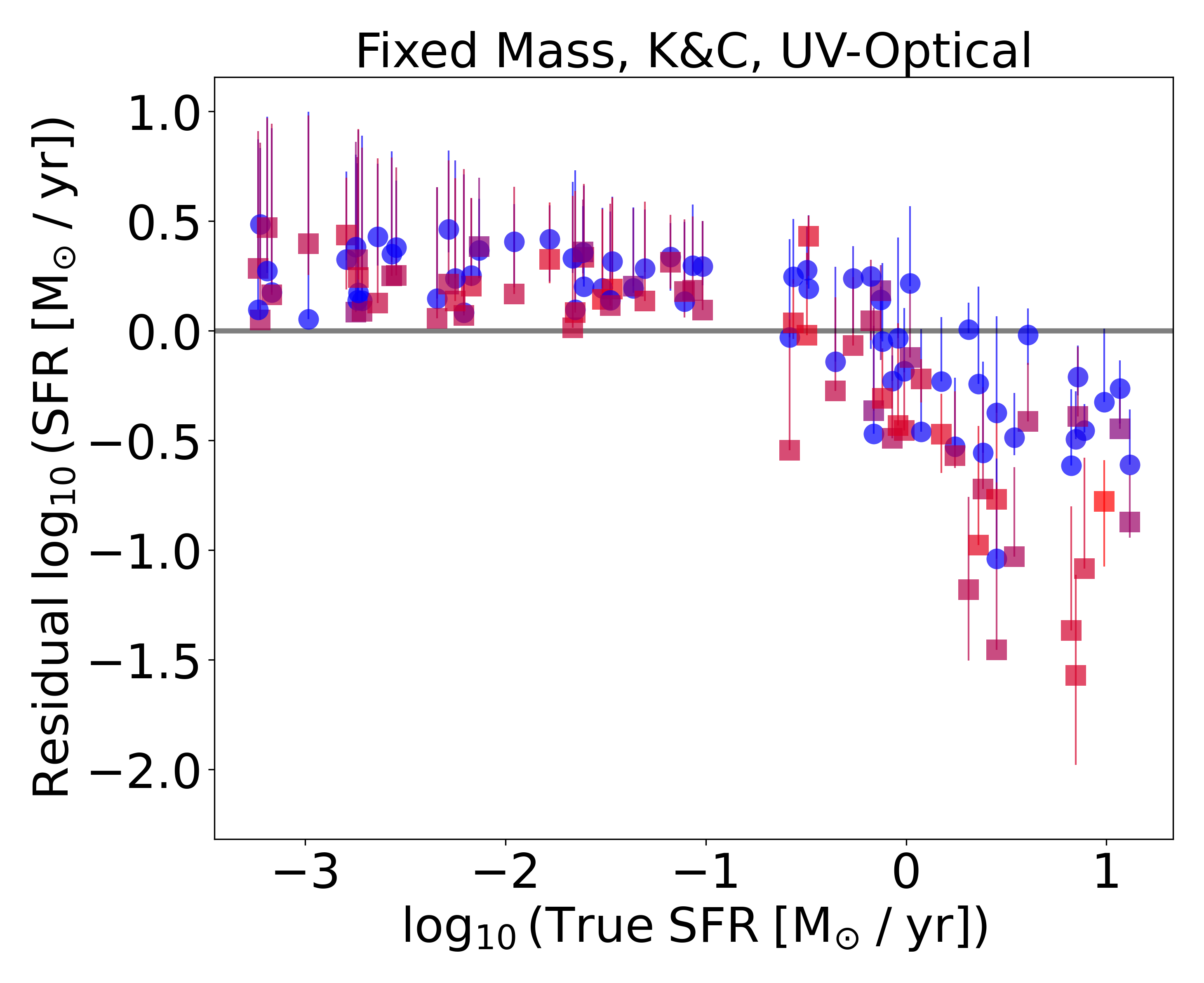}{0.33\textwidth}{}
\fig{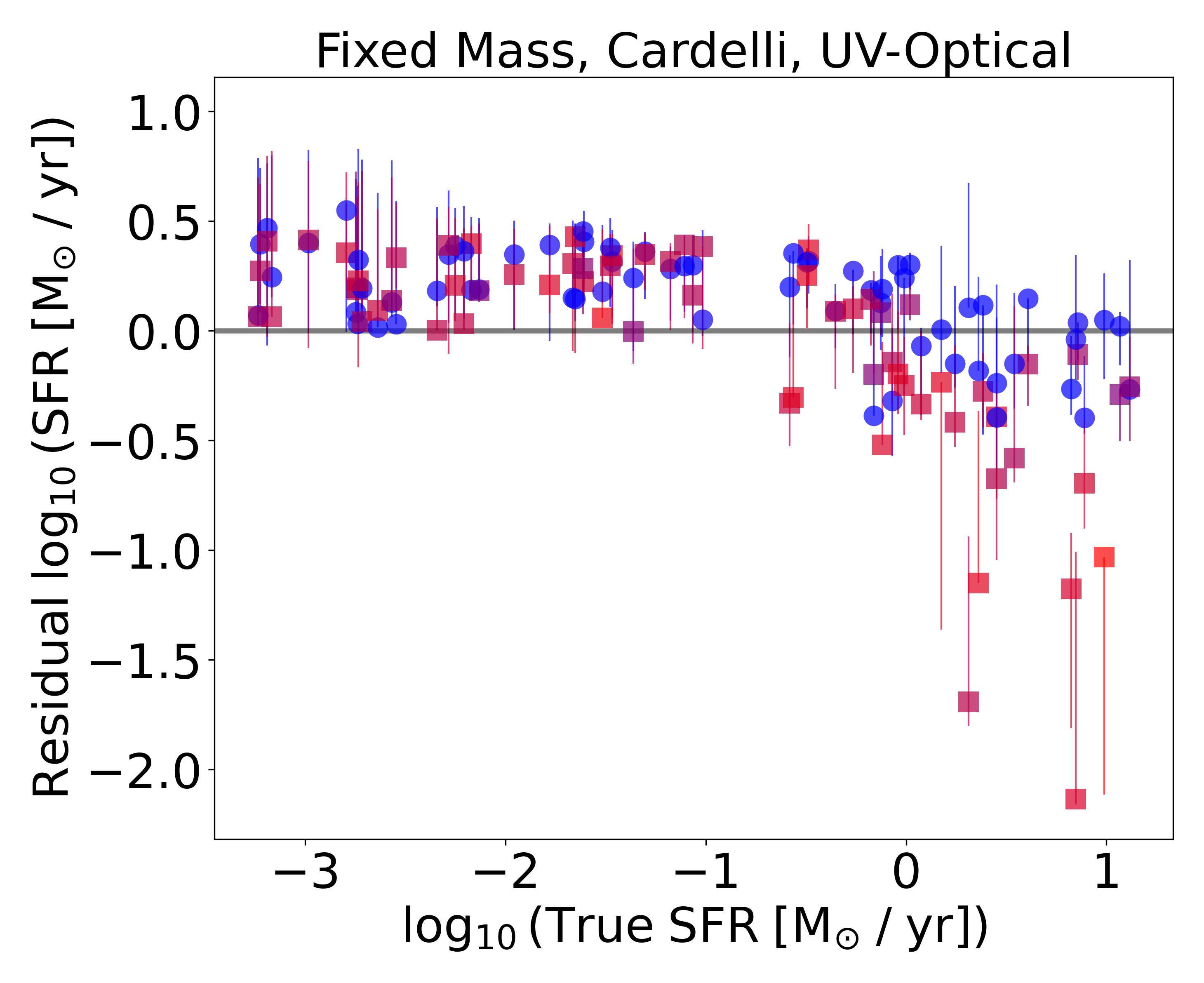}{0.33\textwidth}{}}
\caption{\label{fig:fixed_mass_UV-Optical_SFR_residuals} Similar to Figure \ref{fig:fixed_mass_UV-IR_SFR_residuals} but fit to UV-Optical filters.}
\end{figure}

Figure \ref{fig:fixed_mass_UV-IR_SFR_residuals} shows SFR residuals as a function of true SFRs using UV-IR bands with total stellar masses fixed to their true values for all dust models including the dust free cases. The most noticeable feature of these plots is the strong negative SFR bias for edge-on high mass galaxies. As in the previous section, we can understand this bias by considering the effect of the energy balance assumption between dust attenuation and emission. 
For edge-on dusty galaxies, the IR is emitting less energy than appears attenuated. Matching 
the IR bands then means underestimating the energy  attenuated by dust, which increases UV 
fluxes predicted by SED modeling at a given SFR, leading to underestimated SFRs in order to 
match the UV bands. 

Figure \ref{fig:fixed_mass_UV-Optical_SFR_residuals} shows SFR residuals as a function of 
true SFRs using UV-Optical bands with total stellar masses fixed to their true values for all dust models including the dust free cases. 
This figure shows that taking away the IR data 
causes even stronger negative SFR biases for galaxies at the high SFR end, especially 
for edge-on orientations. Analogously to the explanation of mass residuals in 
Section \ref{Fixed sSFH}, we attribute this effect to the trade-off between SFR and 
total attenuated energy. For more heavily attenuated galaxies (higher SFRs), the SFR 
residuals are more negatively biased. We interpret this as a tendency for the fits to 
prefer using the SFH shape to account for SED changes that are in reality due to the 
dust attenuation. This tendency results in less attenuation and lower SFR estimates, 
since both increasing attenuated energies and lowering SFRs will preferentially decrease 
fluxes at shorter wavelengths. At the low SFR end, we see the opposite trend for the 
more flexible dust models (all except Calzetti), in which the fits overestimate the SFR and the total attenuated energy. This indicates a tendency at low SFRs for the fits 
to prefer using the dust attenuation curve to account for the SFH shape. 

Comparing Figures \ref{fig:fixed_mass_UV-IR_SFR_residuals} and \ref{fig:fixed_mass_UV-Optical_SFR_residuals}, we see that taking away IR data actually reduces the size of our inferred SFR credible intervals. This is perhaps counter-intuitive, since in general  one would expect adding data to tighten constraints. Our interpretation of this is that when only considering UV-Optical bands, there is so much flexibility in the model relative to the data that Prospector is able to find parameters that can compensate for model mismatches and yield accurate fluxes, resulting in a posterior distribution that is sharply peaked near those parameters and narrow credible intervals. However, because of the model mismatches, these parameters may be significantly biased. 
Comparing these two figures, we see that the low mass galaxies fit with UV-IR data yield more accurate SFR inferences despite having larger error bars. This is likely due to the fact that in our analysis we didn't actually add any noise to our mock observations. Relative to UV-IR bands, the model is no longer flexible enough to compensate for model mismatches and it becomes necessary to sacrifice some bands in order to fit others better, leading to a larger set of parameters with similar posterior probability densities. We will discuss this situation further in the next section. 


\subsection{Free SFH} \label{Free SFH}
\noindent In this section, we explore how well Prospector is able to recover the physical parameters of our simulated galaxies when allowing all SFH related parameters to be free. We note that this is the only method of the four explored in this work that can be extrapolated to assess our ability to infer physical parameters of galaxies in the real Universe. However, it is also important to keep in mind that even in this case we still tell Prospector the correct luminosity distance and redshift ($z=0$) of each galaxy.  

\begin{figure}
\gridline{\fig{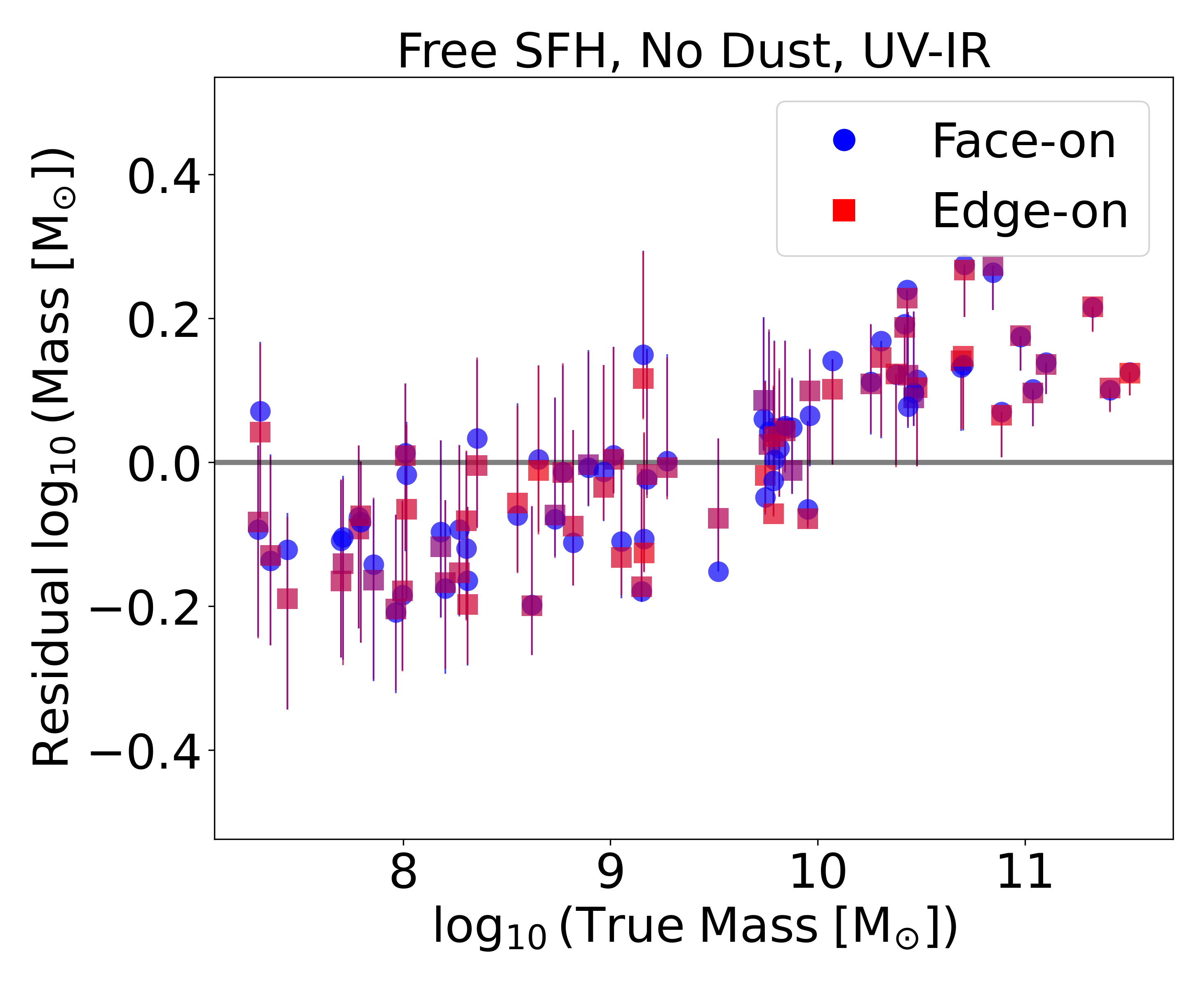}{0.33\textwidth}{}
\fig{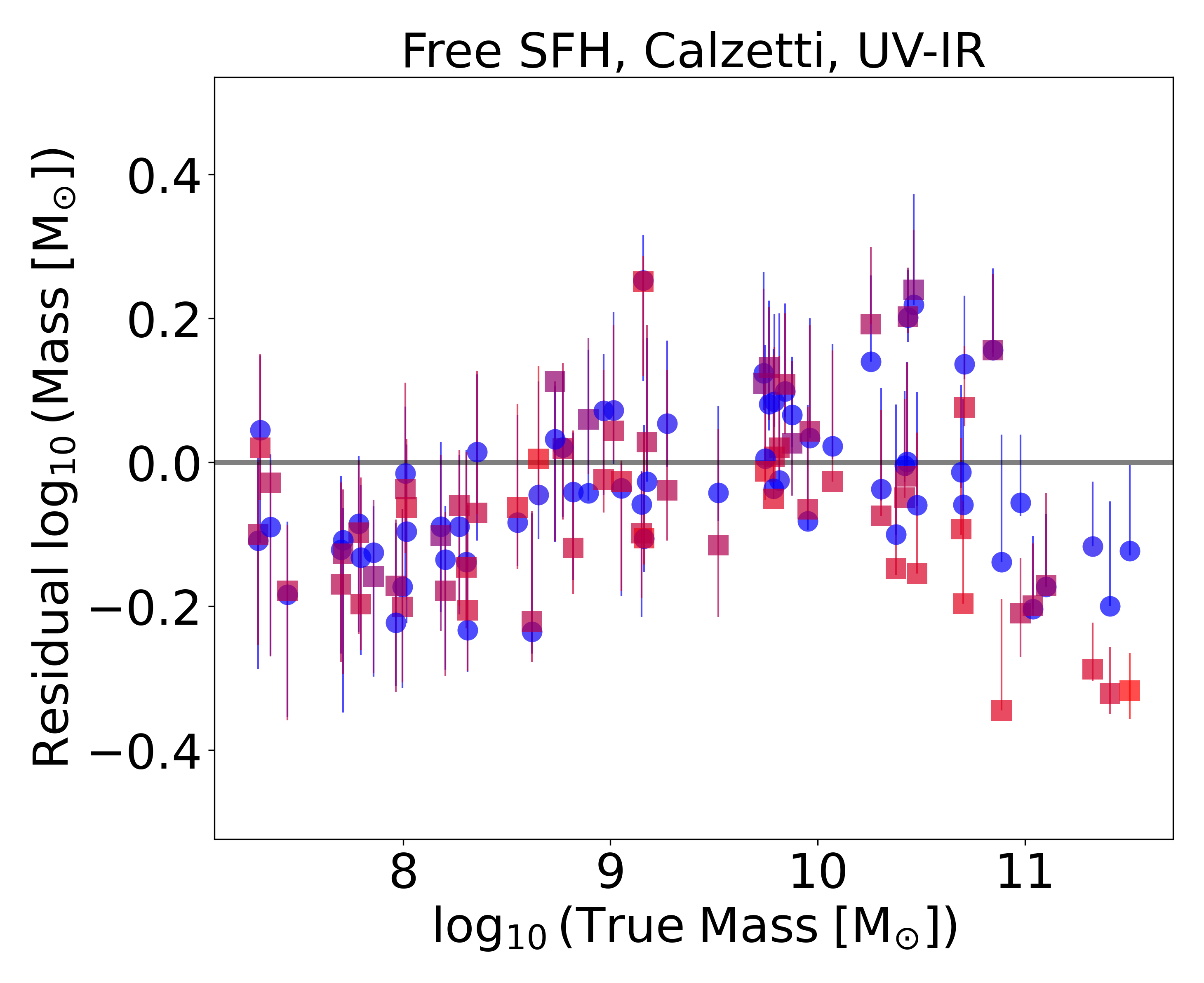}{0.33\textwidth}{}
\fig{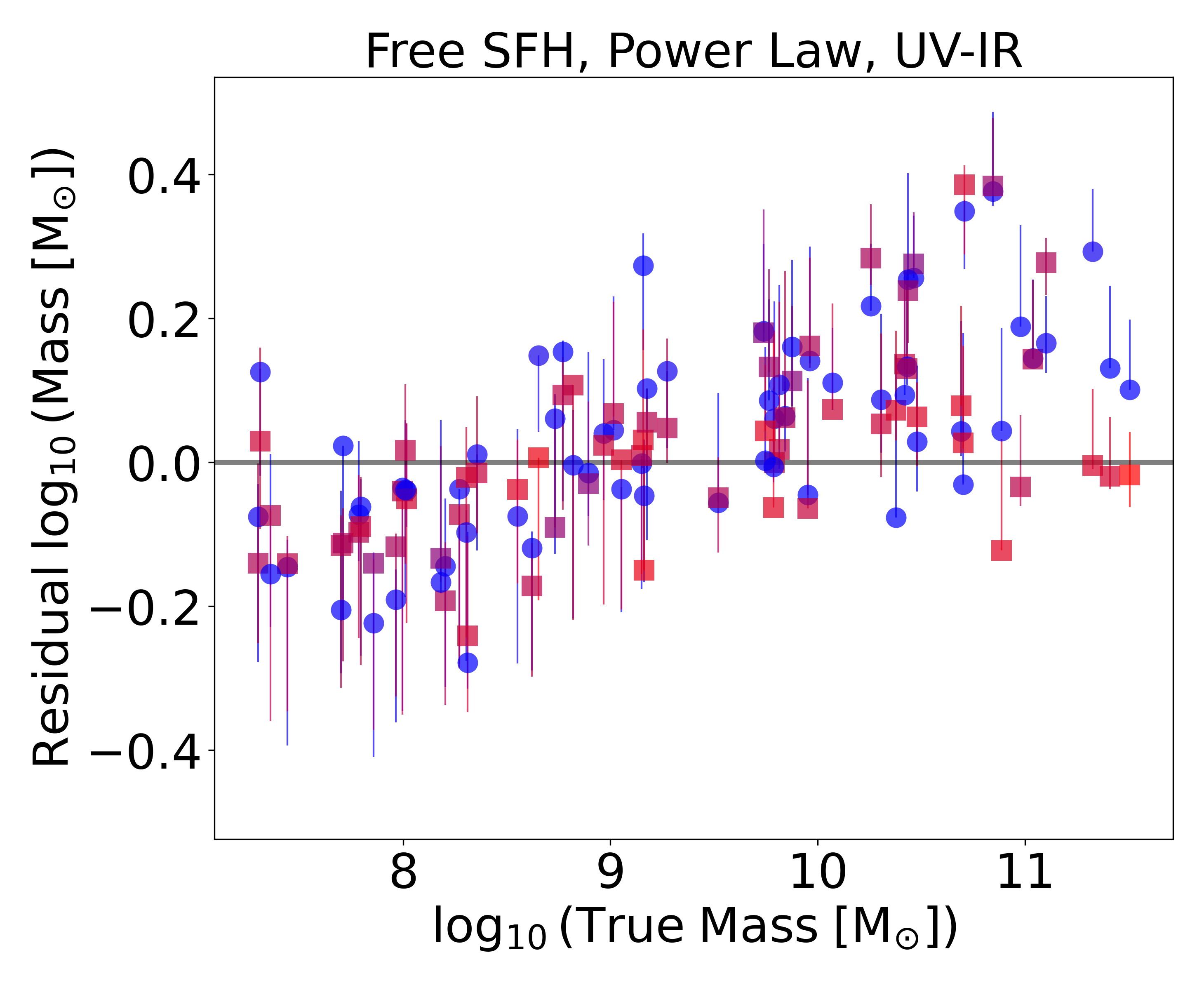}{0.33\textwidth}{}}
\gridline{\fig{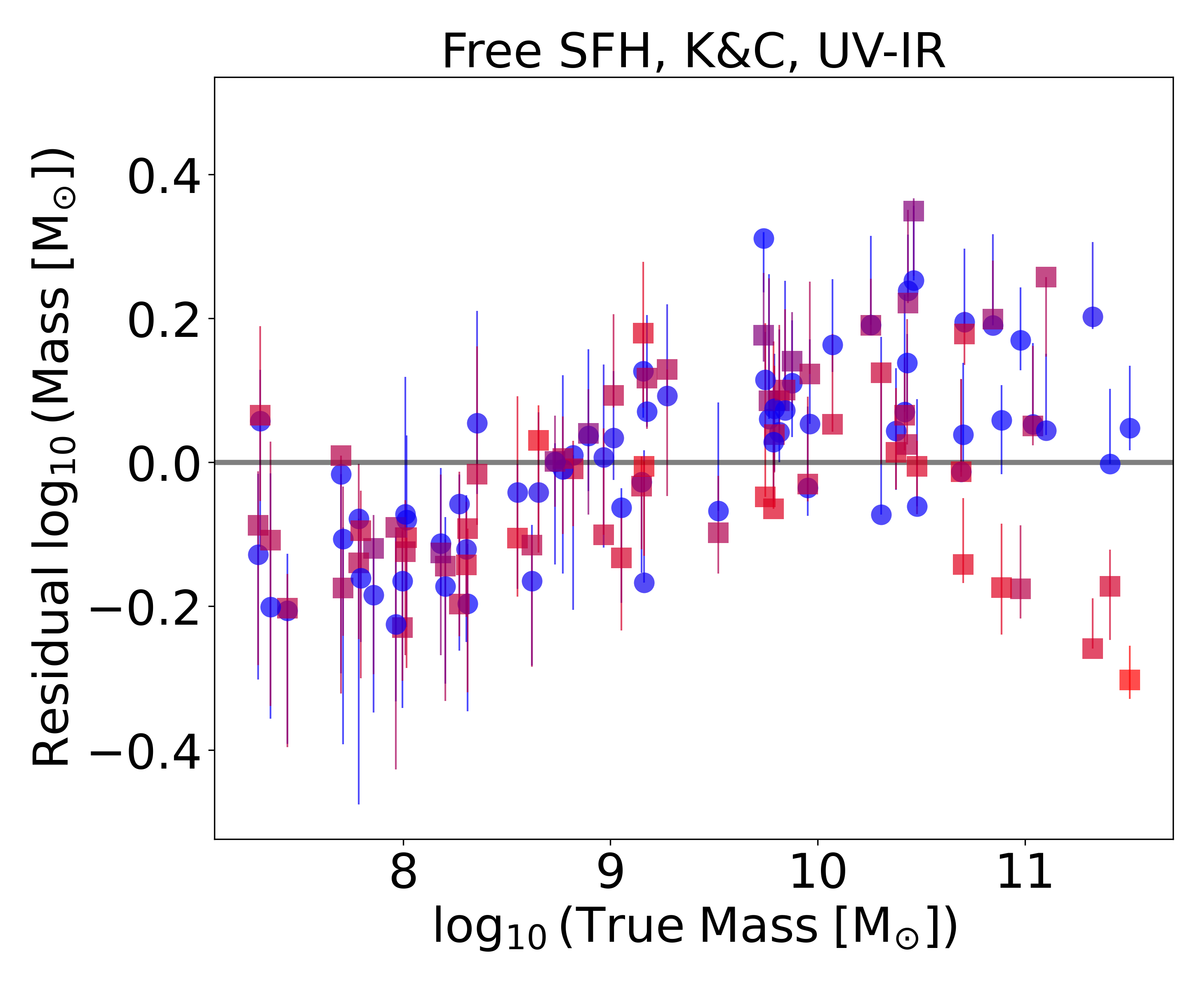}{0.33\textwidth}{}
\fig{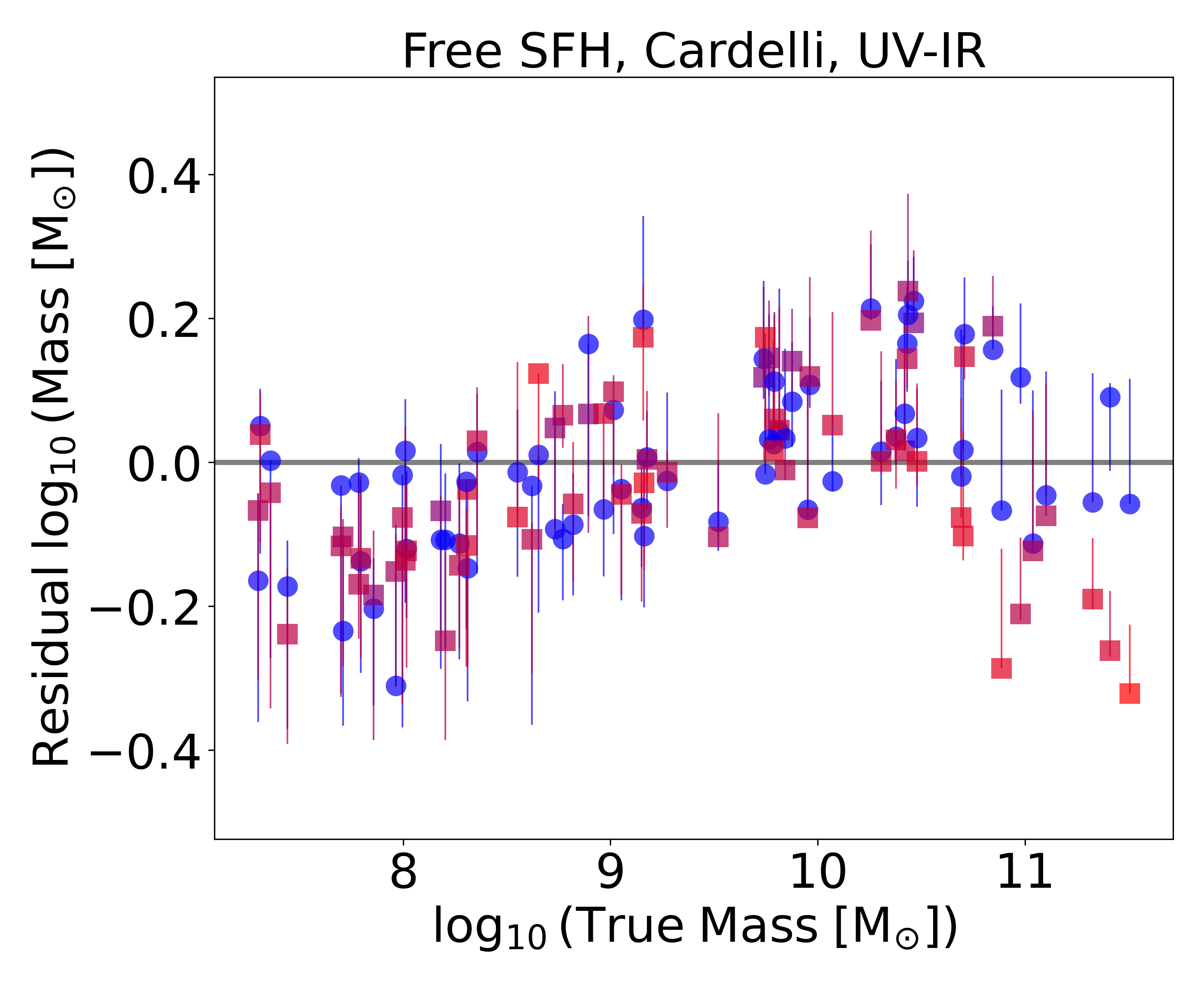}{0.33\textwidth}{}}
\caption{\label{fig:free_SFH_UV-IR_mass_residuals} Distributions of inferred mass residuals as a function of true masses using UV-IR filters with all SFH parameters free. Face-on orientations are denoted by circles and edge-on orientations by squares, with the colors indicating their axis ratios as shown in Figure \ref{fig:axisRatioColors}.}
\end{figure}

\begin{figure}
\gridline{\fig{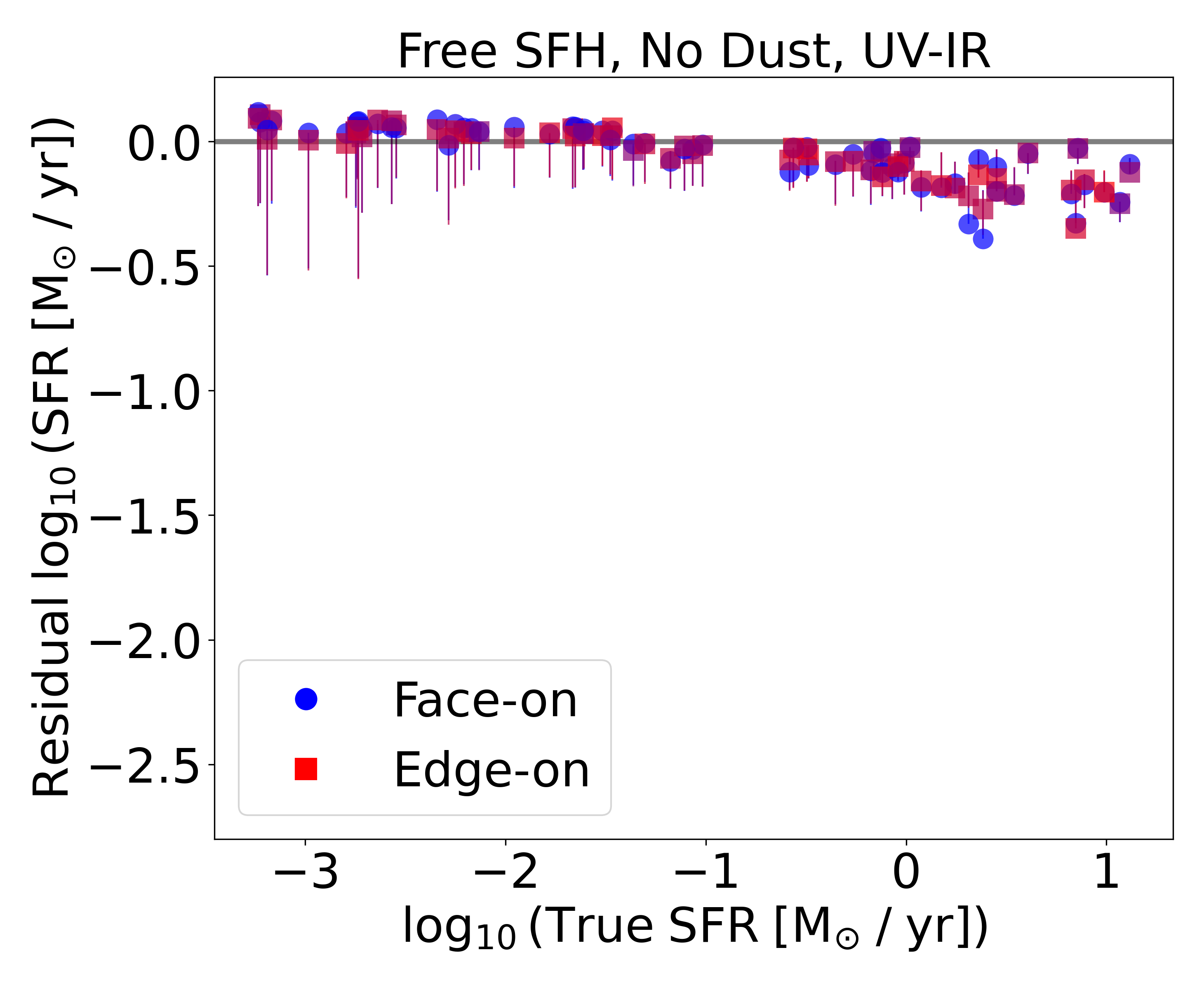}{0.33\textwidth}{}
\fig{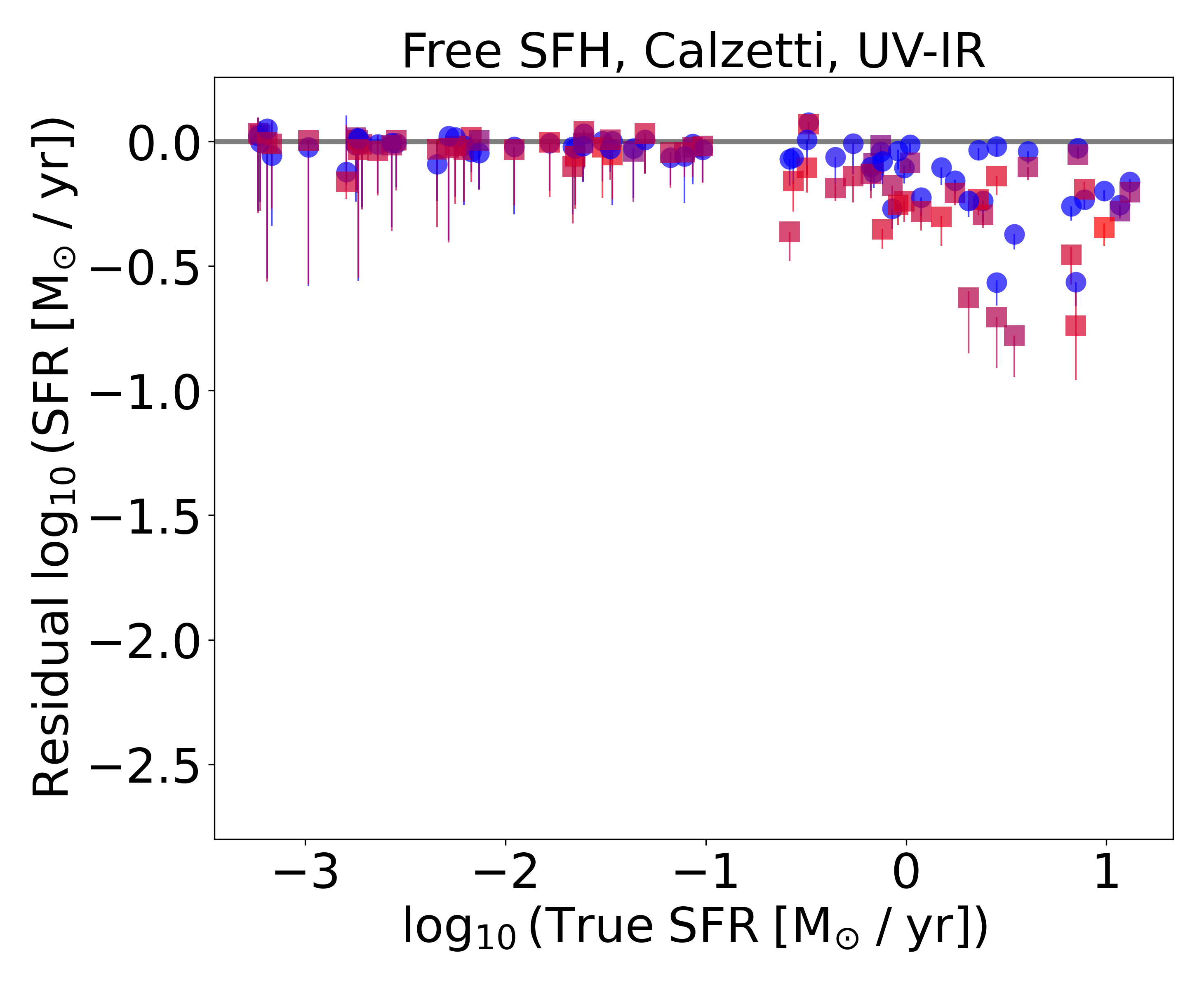}{0.33\textwidth}{}
\fig{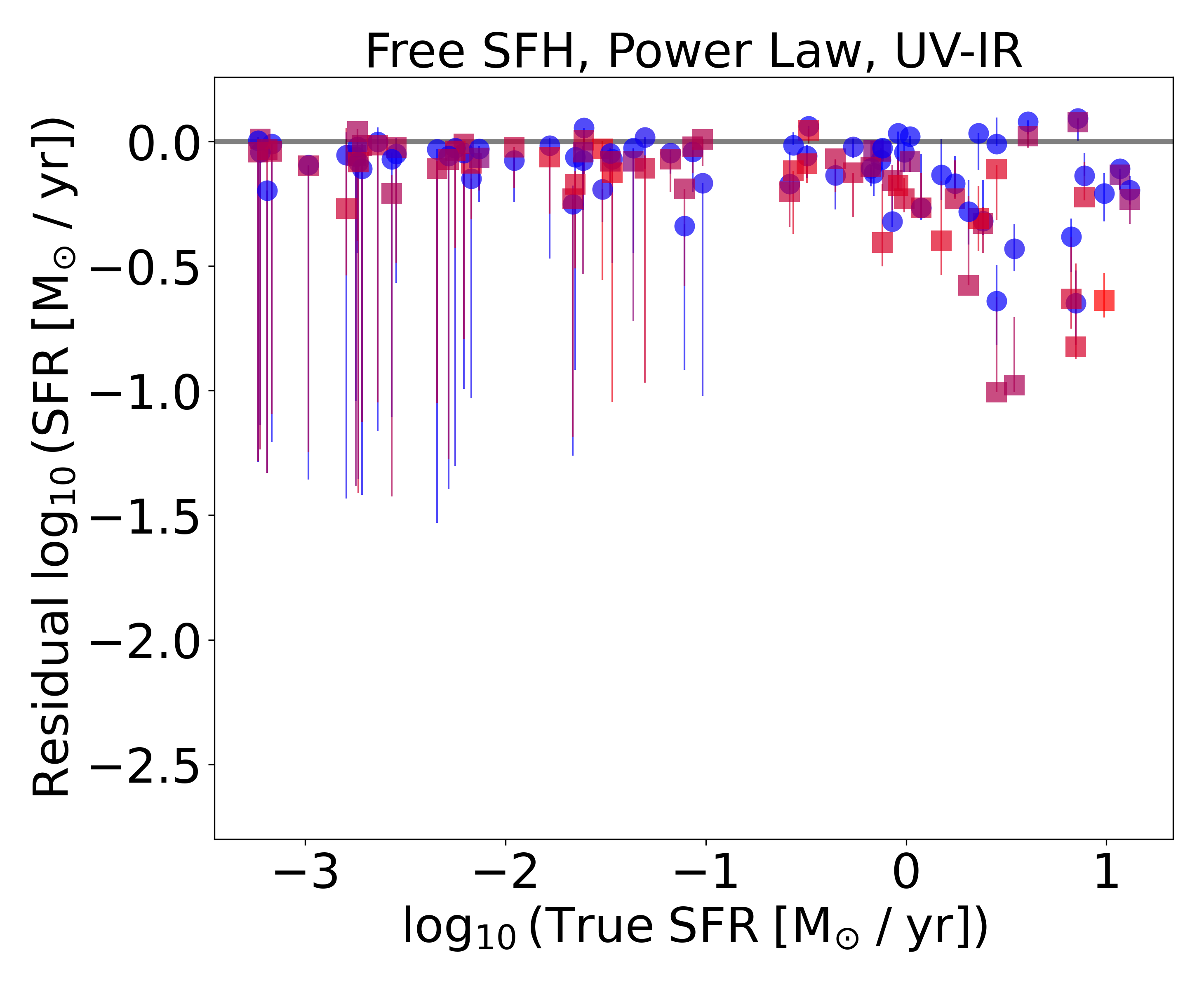}{0.33\textwidth}{}}
\gridline{\fig{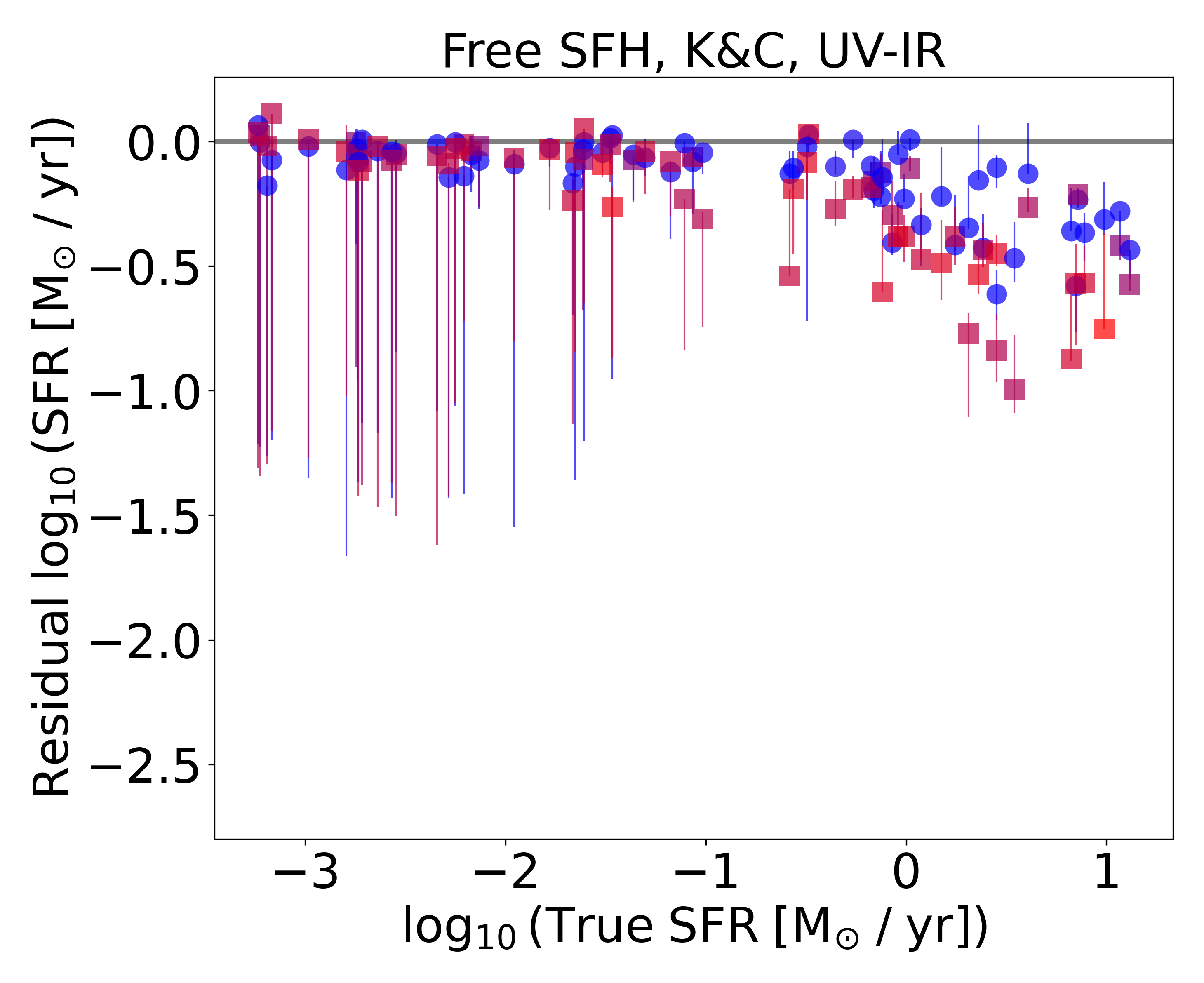}{0.33\textwidth}{}
\fig{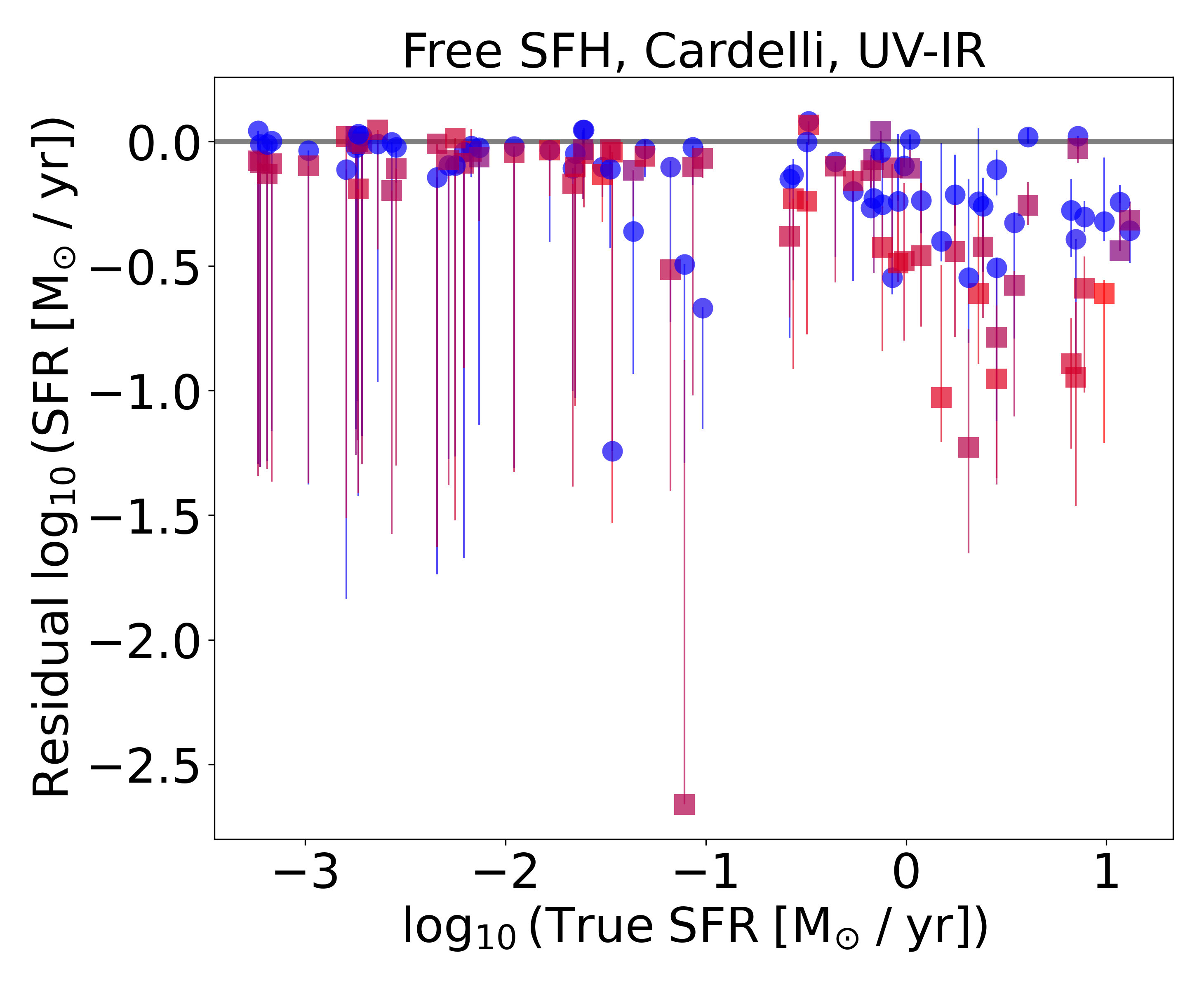}{0.33\textwidth}{}}
\caption{\label{fig:free_SFH_UV-IR_SFR_residuals} Distributions of inferred SFR residuals as a function of true SFRs using UV-IR filters with all SFH parameters free. Face-on orientations are denoted by circles and edge-on orientations by squares, with the colors indicating their axis ratios as shown in Figure \ref{fig:axisRatioColors}.}
\end{figure}

\begin{figure}
\begin{center}
\includegraphics[width=0.5\textwidth]{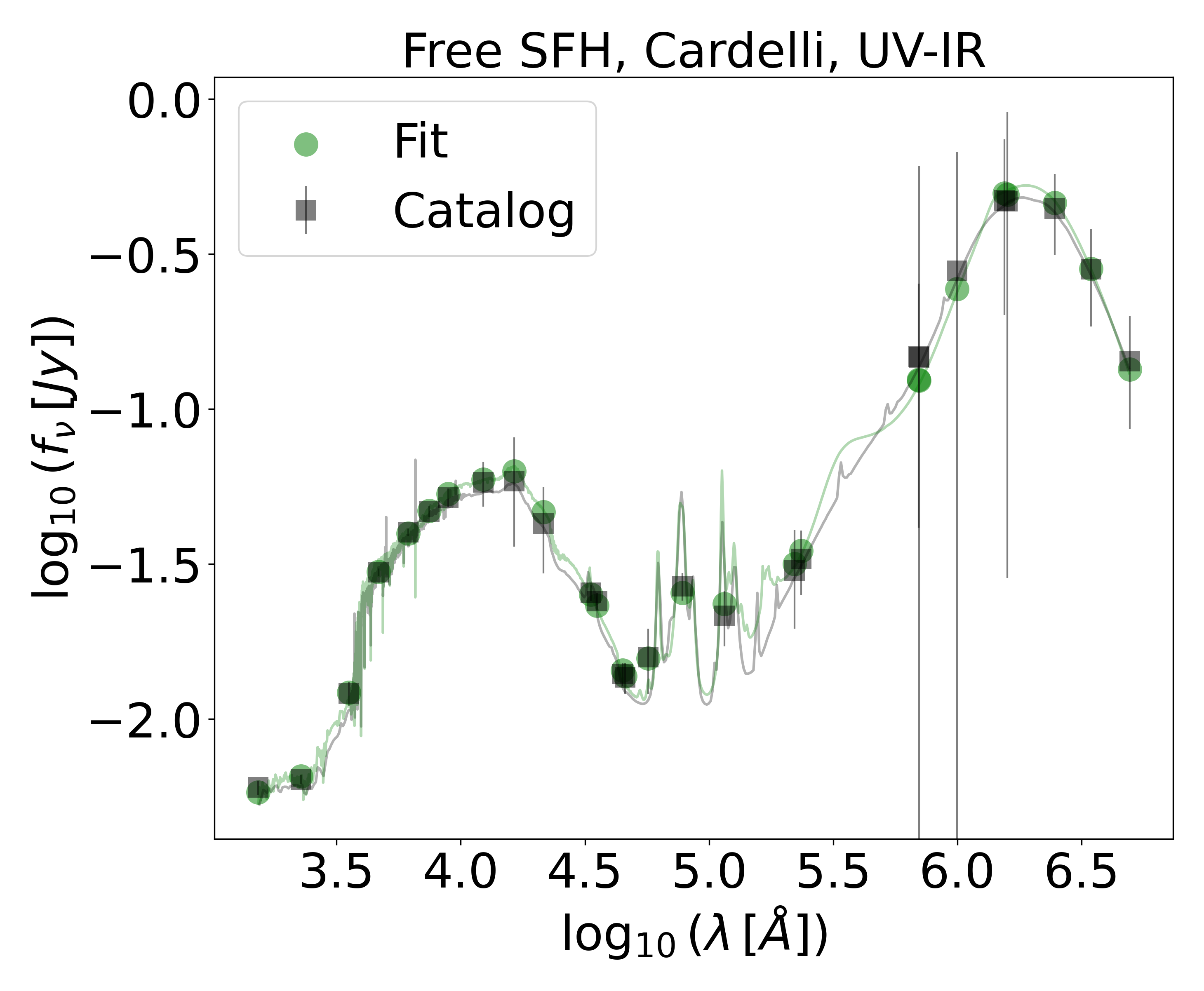}
\end{center}
\caption{\label{fig:wild_Cardelli_DustPedia_g2.19e11_edge_fit} \small 
Fit of the edge-on orientation of the galaxy that has $-2.66$ dex residual in $\log_{10}(\rm SFR \; [M_{\odot} / \rm yr])$ using the Cardelli dust attenuation model, and UV-IR bands (see Figure \ref{fig:free_SFH_UV-IR_SFR_residuals} for reference). Error bars that extend to the bottom of the plot represent fluxes that are consistent with zero up to one sigma.}
\end{figure}

\begin{figure}
\gridline{\fig{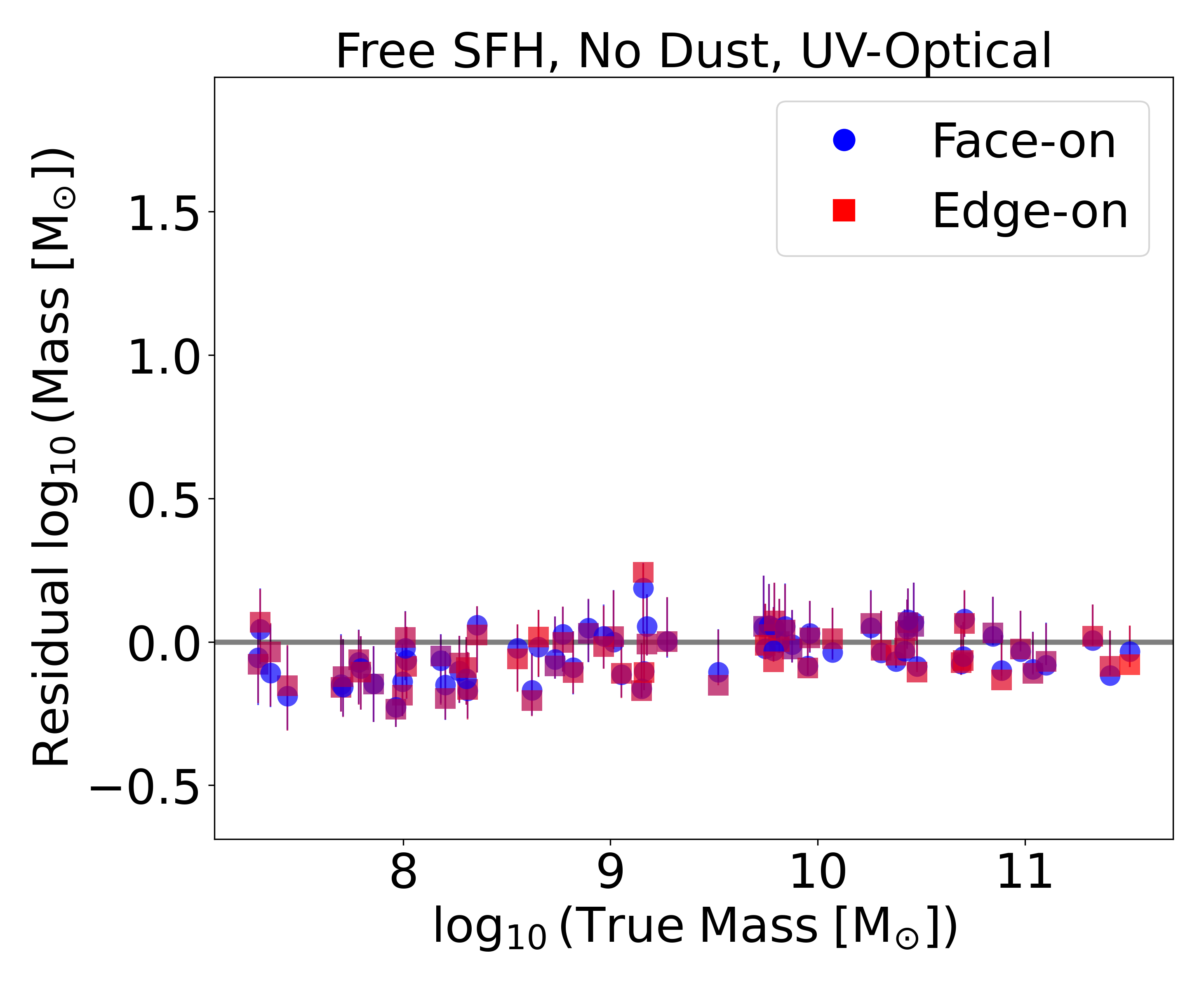}{0.33\textwidth}{}
\fig{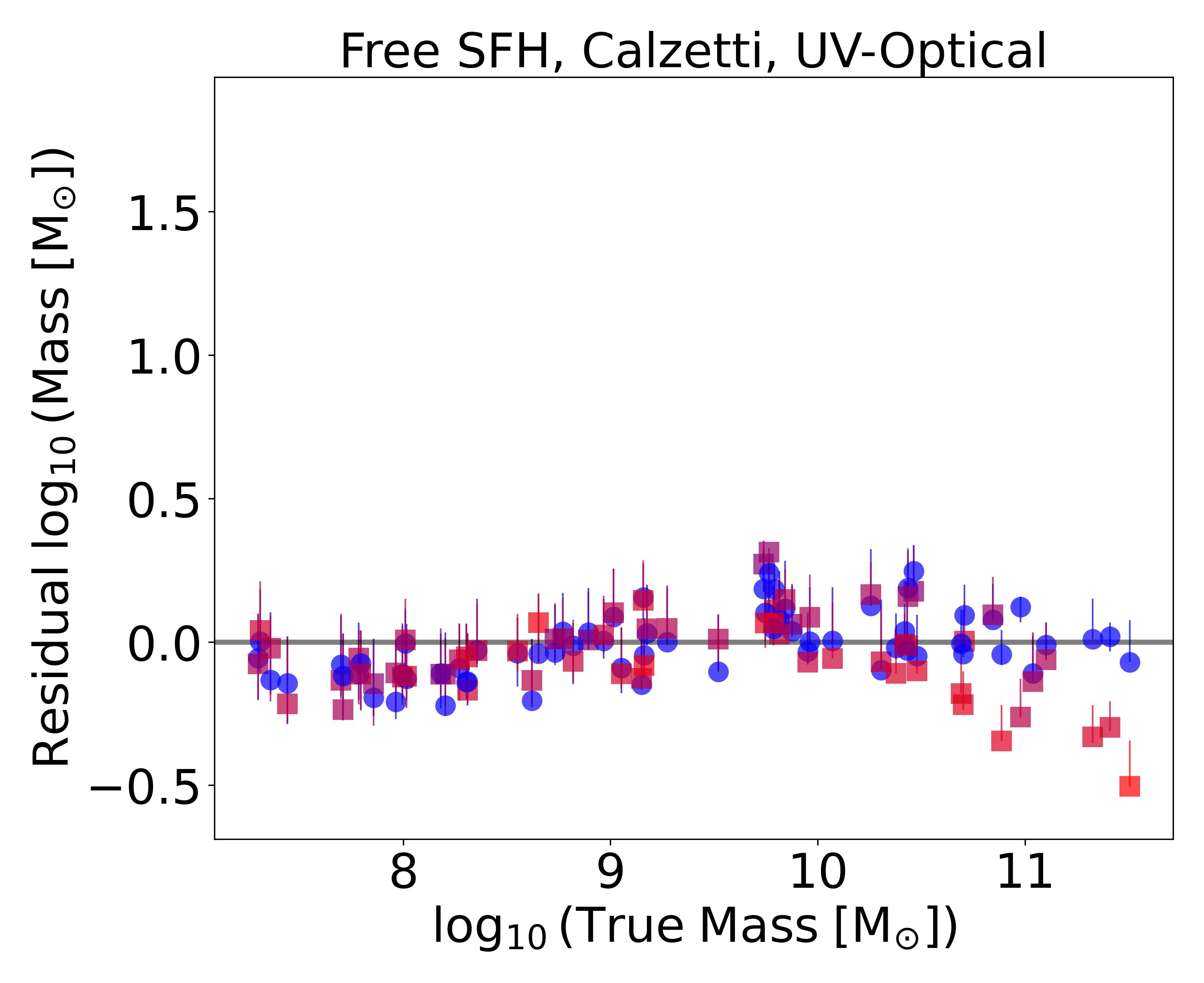}{0.33\textwidth}{}
\fig{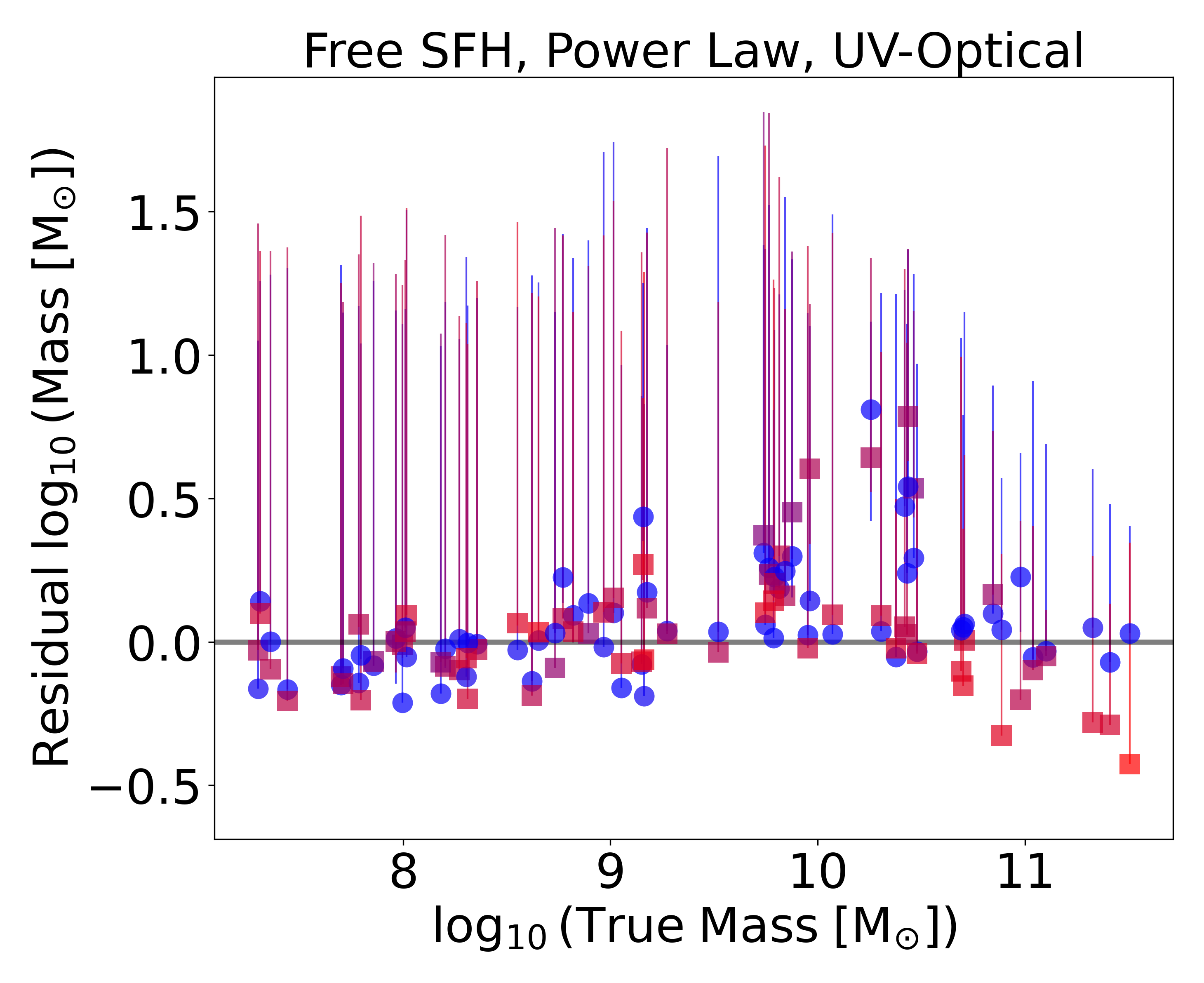}{0.33\textwidth}{}}
\gridline{\fig{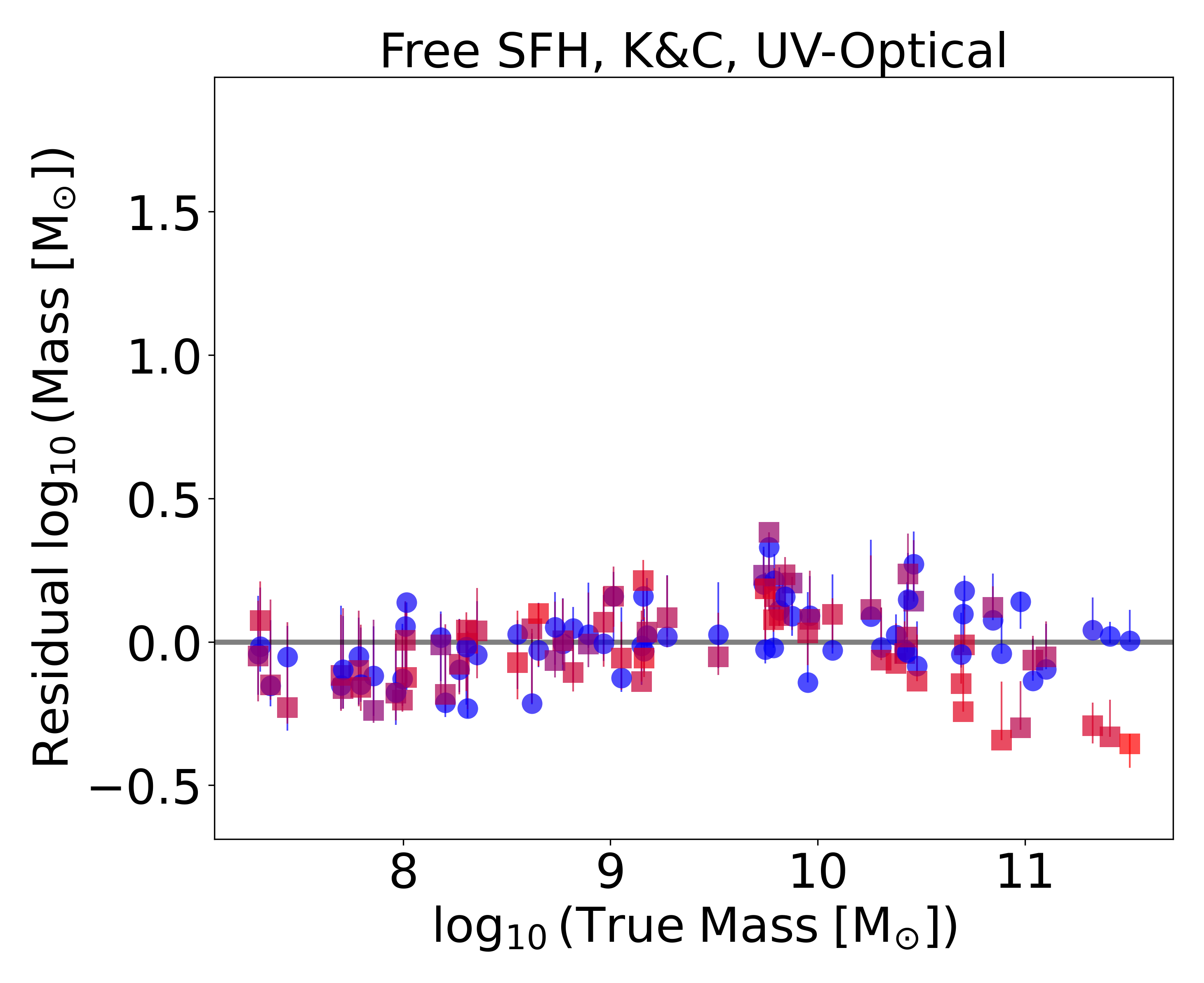}{0.33\textwidth}{}
\fig{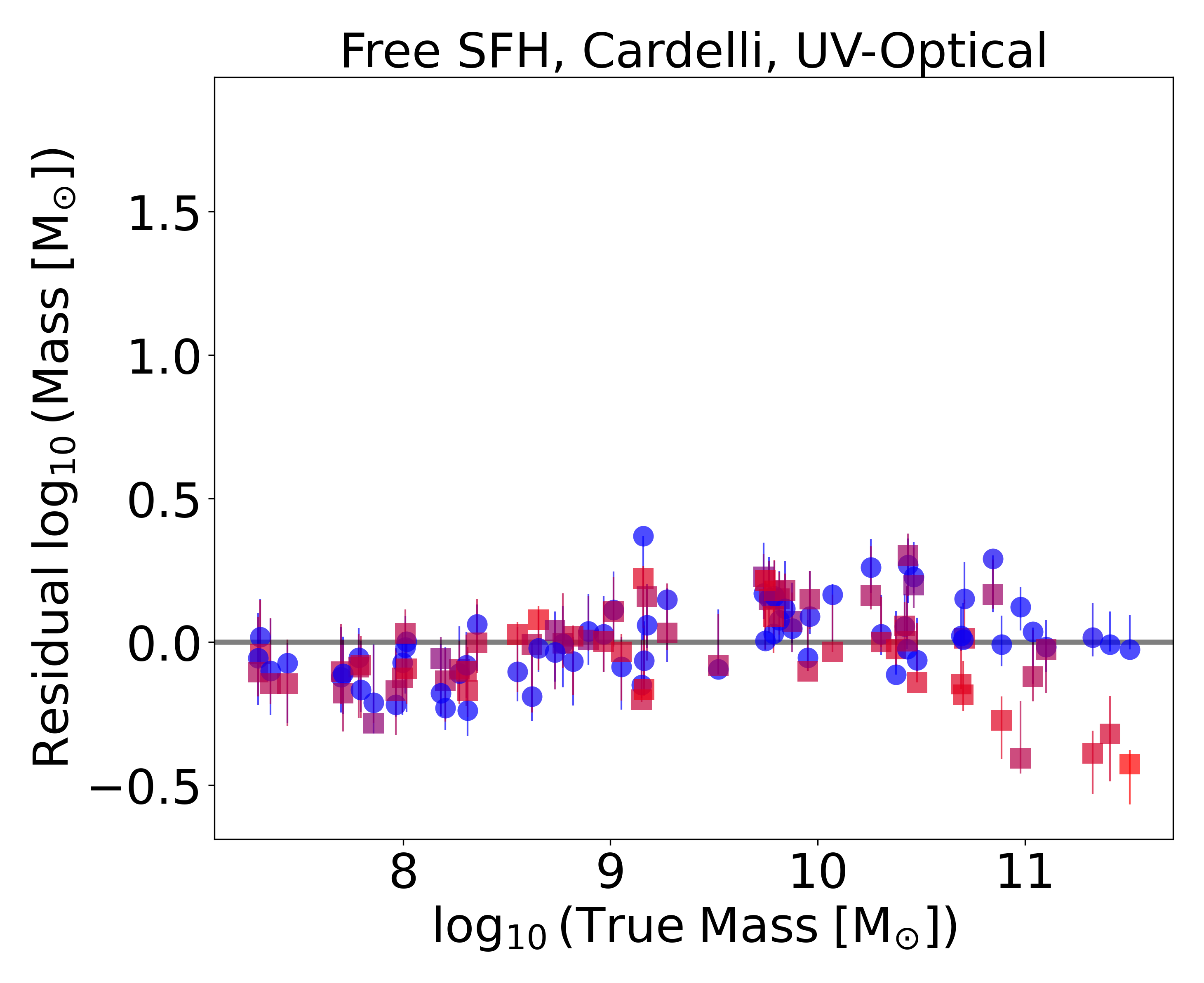}{0.33\textwidth}{}}
\caption{\label{fig:free_SFH_UV-Optical_mass_residuals} Similar to Figure \ref{fig:free_SFH_UV-IR_mass_residuals} but fit to UV-Optical filters.}
\end{figure}

\begin{figure}
\gridline{\fig{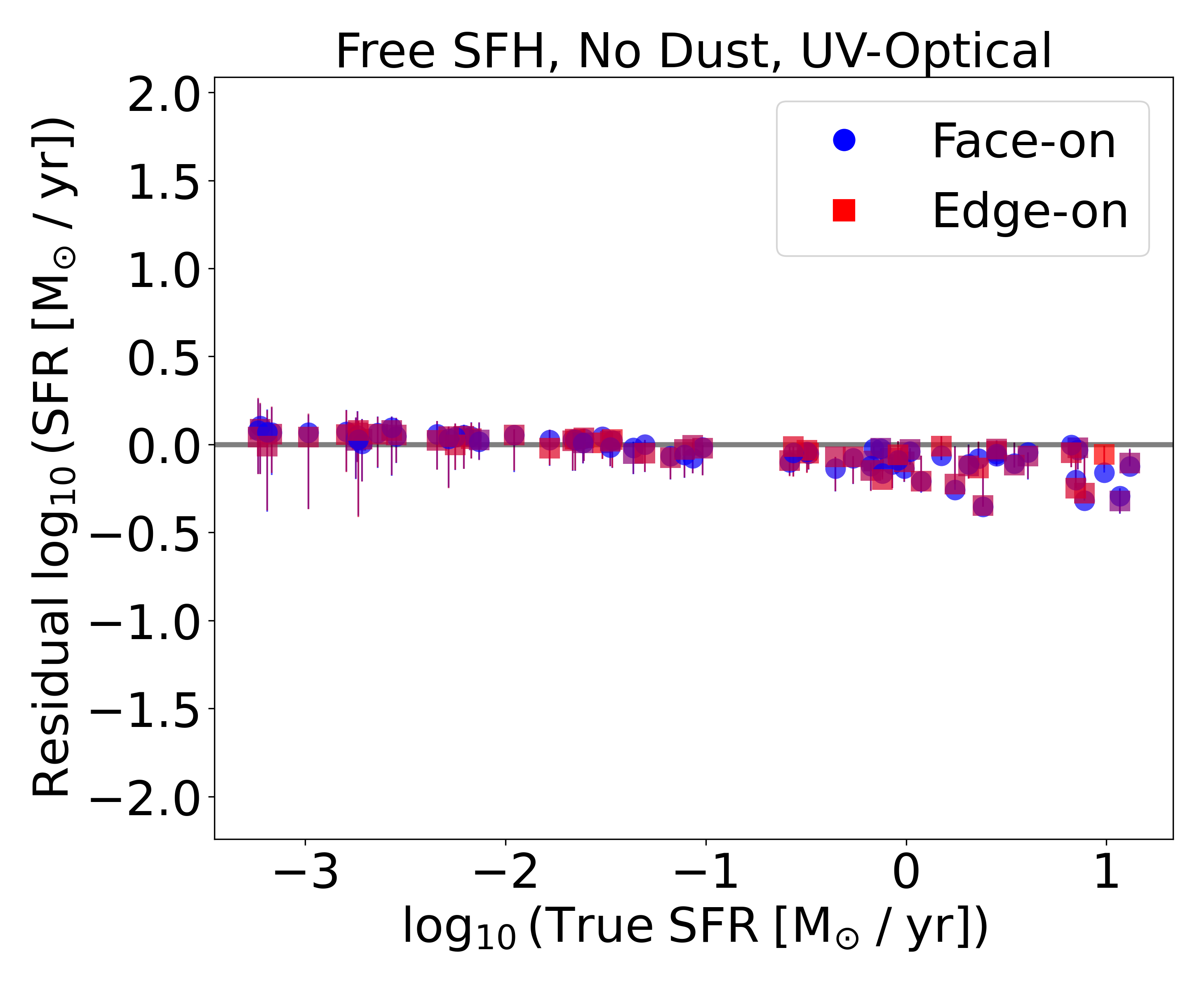}{0.33\textwidth}{}
\fig{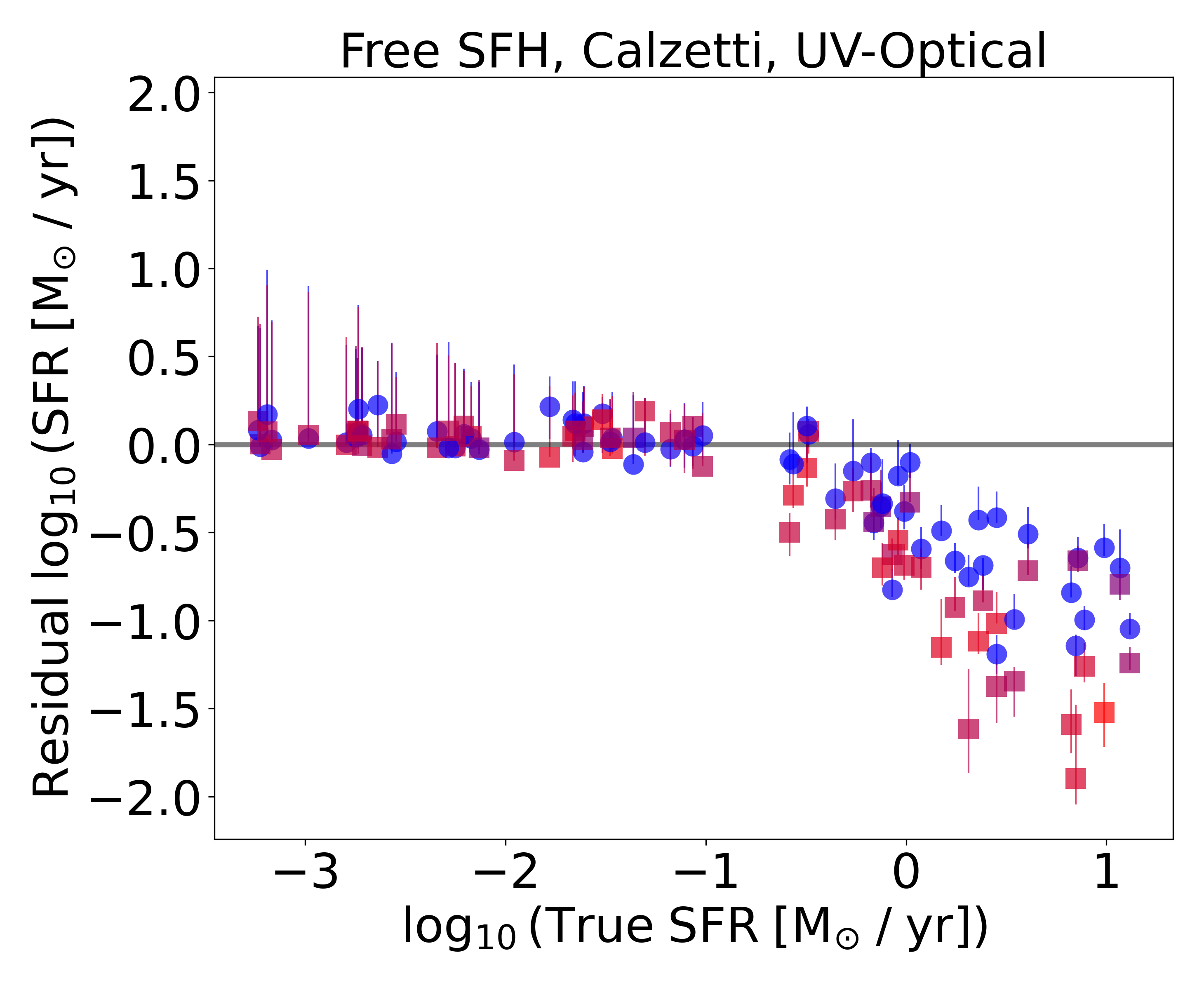}{0.33\textwidth}{}
\fig{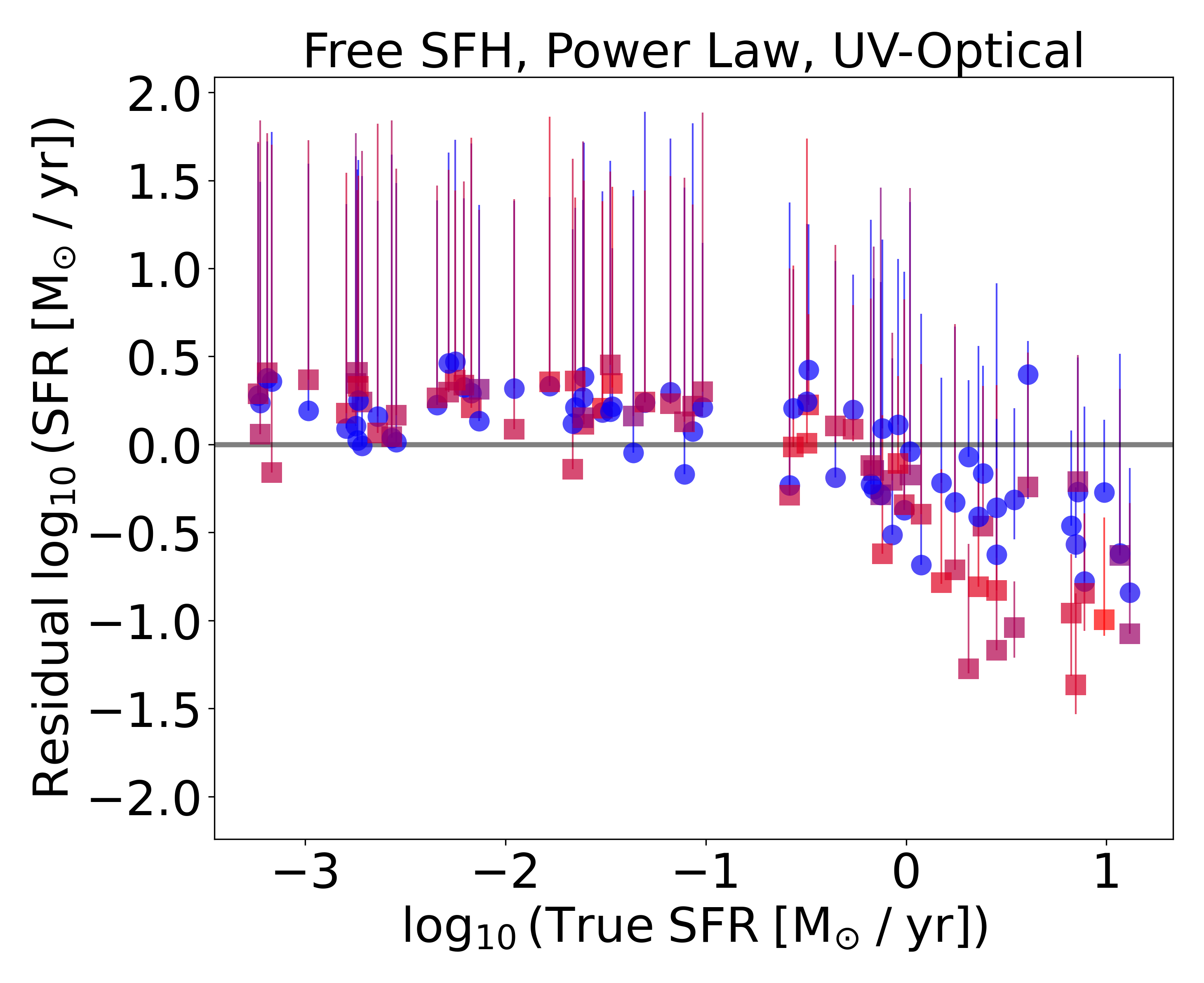}{0.33\textwidth}{}}
\gridline{\fig{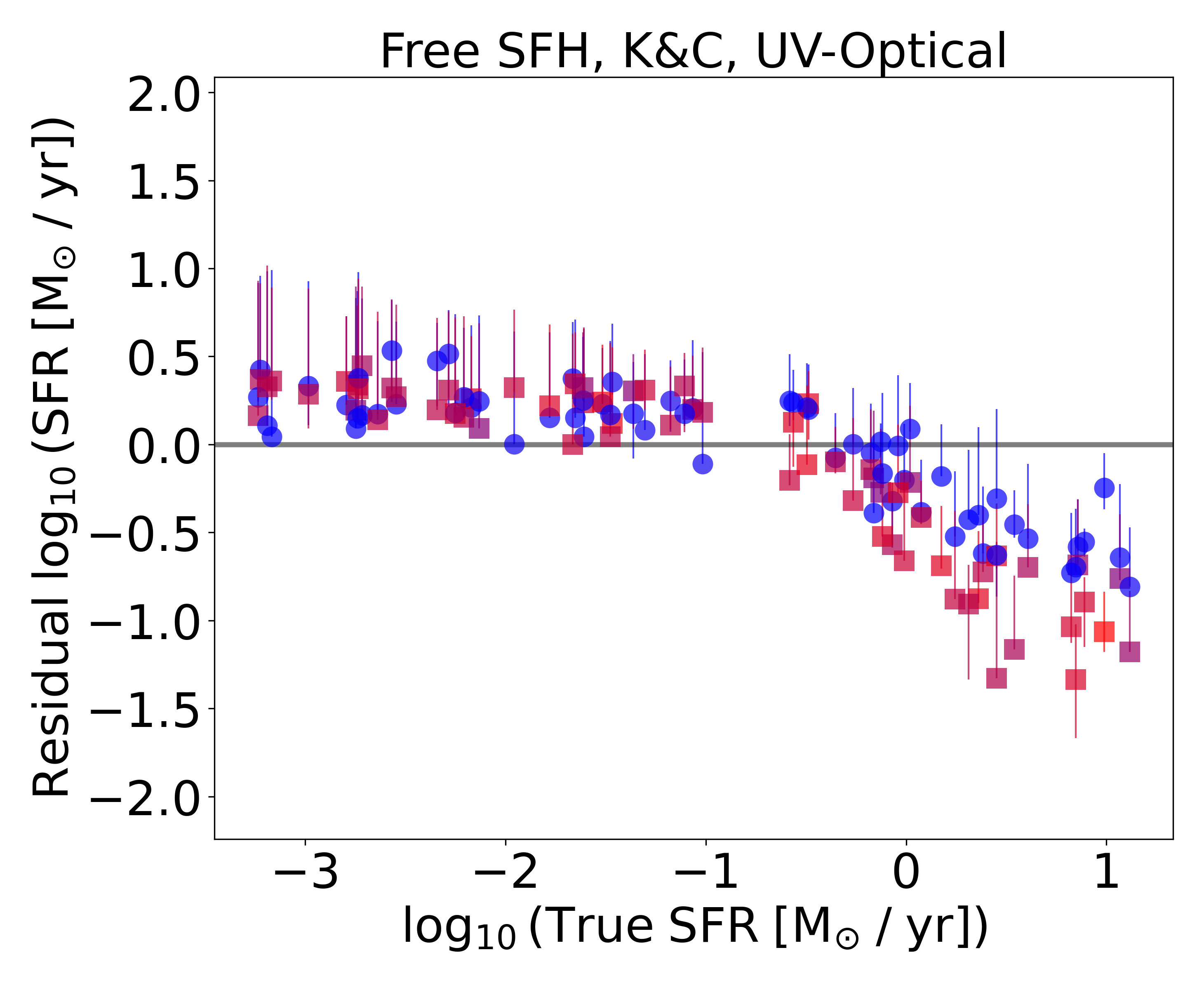}{0.33\textwidth}{}
\fig{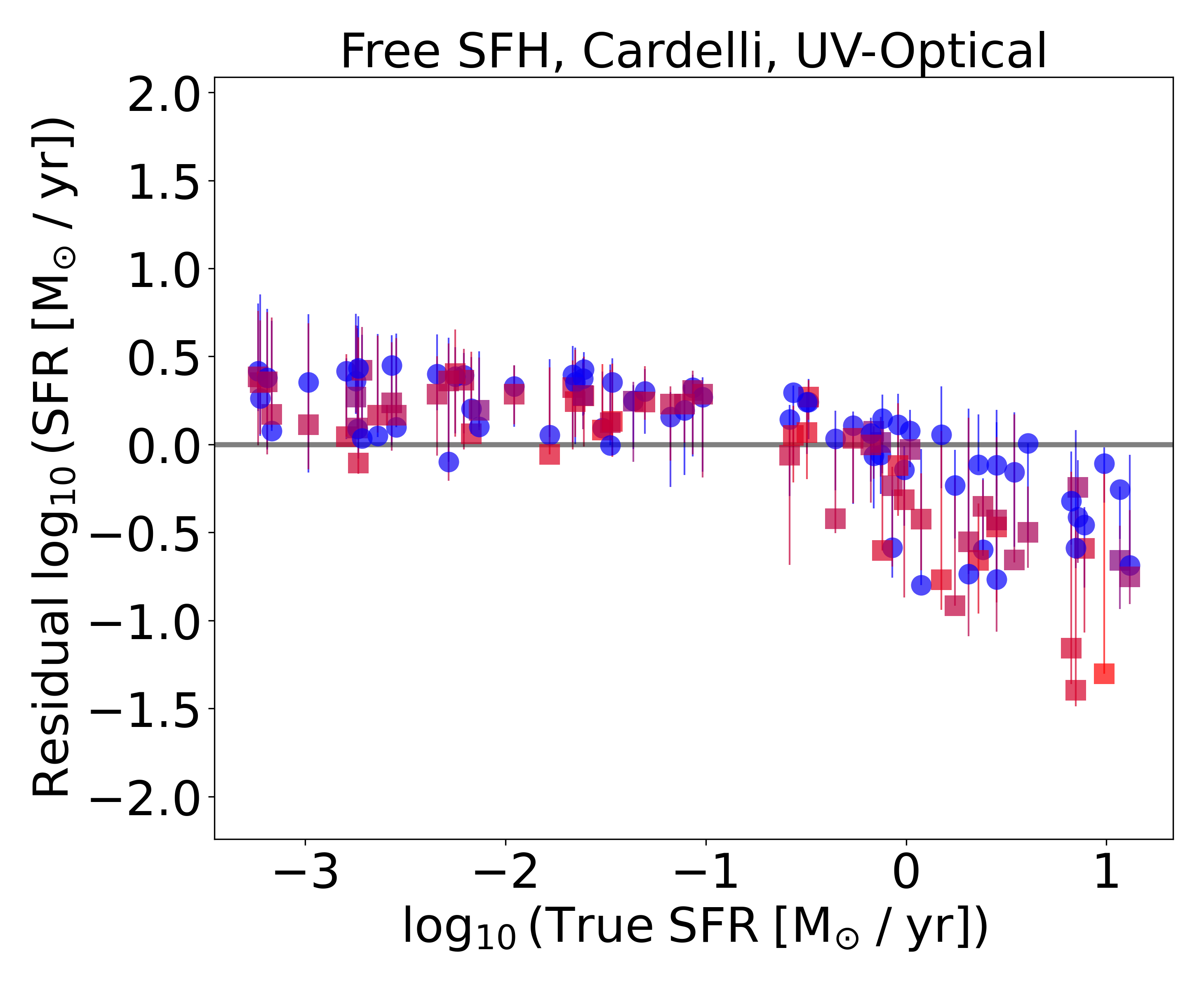}{0.33\textwidth}{}}
\caption{\label{fig:free_SFH_UV-Optical_SFR_residuals} Similar to Figure \ref{fig:free_SFH_UV-IR_SFR_residuals} but fit to UV-Optical filters.}
\end{figure}

Figure \ref{fig:free_SFH_UV-IR_mass_residuals} shows mass residuals as a function of 
the true masses using UV-IR bands with all SFH parameters free in the fits for all dust models including the dust free cases. In the dust free case, we see that the flexibility in the model's SFH and metallicity alone lead to mass residuals that tend have biases of ${\sim}~{-0.1}$ dex at low masses and ${\sim} {0.1}$ dex at high masses. Interestingly, galaxies at the high mass end show credible intervals that are significantly smaller than their residuals, demonstrating that realistic model mismatches can yield posterior distributions that lead to overconfidence in our inferences. 


In the cases that include dust, low mass galaxies show mass residuals that are 
qualitatively similar to the dust free case. This is not too surprising given that 
these low mass galaxies experience only a small amount of dust attenuation. As 
we move to higher masses, we begin to see deviations from the dust free case. 
In particular, we see that edge-on orientations of the most massive galaxies tend 
to have mass residuals of ${\sim}~{-0.3}$ dex. We interpret this as being 
a consequence of the imposed energy balance in the fits, since massive 
edge-on galaxies will have significantly more energy attenuated than emitted 
by dust, leading to both underestimated mass and SFRs (see Figure 
\ref{fig:free_SFH_UV-IR_SFR_residuals} for the corresponding SFR 
residuals). 

Figure \ref{fig:free_SFH_UV-IR_SFR_residuals} shows SFR residuals as a function of 
the true SFRs using the UV-IR bands with all SFH parameters free in the fits.  The SED model fit underestimate SFRs of high 
mass galaxies consistently among all dust models and for face-on and edge-on 
orientations, with edge-on orientations showing even stronger negative biases 
that extend down to ${\sim} {-1.2}$ dex or more. Relative to the results of 
\cite{trcka20a}, which show positive SFR biases of ${\sim} {0.1-0.2}$ dex, our SFR 
inferences show a significant preference for older stellar populations. While 
robustly probing the cause of this discrepancy would require a direct comparison 
using the same SED modeling methods on the same set of simulated mock observations, 
it is likely that the lack of a cold gas phase of the ISM in the EAGLE simulations \citep{schaye15a}, 
which leads to artificially smooth dust density structures and therefore less 
gray attenuation curves, lessens the severity of model mismatches in their inferences. 


In the Cardelli dust model fits one galaxy has an wildly underestimated SFR 
residual of $-2.66$ dex,  corresponding to a factor of $\sim 450$. Given the extreme 
nature of this residual, we include a plot of the corresponding fitted SED in Figure 
\ref{fig:wild_Cardelli_DustPedia_g2.19e11_edge_fit}. Despite 
the highly inaccurate SFR inference, the fitted SED shows surprisingly good agreement 
with the data. We also note that this particular fit has a mass residual of only 
$0.004$ dex, indicating that the large discrepancy in the inferred SFR is primarily 
driven by the energy balance assumption. This is a clear demonstration of 
the potential pitfalls of SFR inferences from broadband photometry alone.

Figure \ref{fig:free_SFH_UV-Optical_mass_residuals} shows mass residuals as a 
function of true masses using UV-Optical bands with all SFH parameters free in the fits for all dust models including the dust 
free cases. The power law dust model fits have large credible intervals that 
extend to extremely high masses. In addition, the SFRs for these fits in 
Figure \ref{fig:free_SFH_UV-Optical_SFR_residuals} show credible intervals that extend 
to high SFRs. That this effect only happens for the power law fits gives us a 
clue to as to the cause. Of the dust models explored in this work, only the 
Calzetti and power law models do not contain a UV-bump, whereas 
the catalog attenuation curves all contain relatively strong UV-bumps. The 
Calzetti model only has the flexibility to change its overall strength, 
not its shape. The power law model, while lacking a UV-bump, is a 
two-component dust model with the ability to change its power law index and 
strength independently for the young and old stellar  populations, giving it 
a fair amount of flexibility in its shape. That these large credible 
intervals extend to large values in both mass and SFR for galaxies of all 
masses and for only the power law dust model leads us to interpret these extended 
posterior samples as being heavily over-attenuated. Without IR bands to 
constrain the energy attenuated by dust, the flexibility of the power law 
dust model is being used to compensate for model mismatches, in particular 
its inability to correctly model the UV-bump. 
This issue is likely the reason for our fitting procedure's inability to converge 
in the Fixed sSFH, UV-Optical, Power Law fits from Section \ref{Fixed sSFH}.  

In the end our fitting procedure does converge on inferences with much higher 
accuracy and precision than the credible intervals would lead one to expect. 
Recall that in our analysis we did not actually add any noise to our 
mock observations. The error bars only have the effect of scaling the 
relative contribution of different bands to the likelihood. It is likely the 
case that repeating this analysis many times with each mock observation being 
drawn from the corresponding Gaussian distribution, as is the case for real 
observations, would yield inferences that show qualitative agreement that 
vary around the true value in a way more consistent with 
these credible intervals. 

Figure \ref{fig:free_SFH_UV-Optical_SFR_residuals} shows the SFR residuals as
a function of the true SFRs using UV-Optical bands with all SFH parameters free in the fits. At higher true SFR, the SFR 
inferences become steadily more negatively biased for all dust models, with 
edge-on orientations showing even stronger negative biases that extend well 
beyond factor of ten discrepancies. Similarly to Section \ref{Fixed Mass}, 
we attribute this effect to the trade-off between increasing total attenuated energy 
and decreasing SFR, since both will preferentially reduce fluxes 
at shorter wavelengths. We 
interpret this as a tendency for the fits to prefer using the SFH shape 
to account for the dust attenuation, which explains the particularly strong 
correlation between true SFR and negative SFR residuals for the inflexible 
Calzetti dust attenuation model. For the more flexible dust models, we 
also see a tendency at low true SFRs to have positive SFR residuals. We 
interpret this as a preference for the fits to use the dust attenuation 
curves to account for the SFH shapes, which also explains why we do
not see this trend in the Calzetti fits.   

As was the case in Section \ref{Fixed Mass}, we can see by comparing 
Figures \ref{fig:free_SFH_UV-IR_SFR_residuals} and \ref{fig:free_SFH_UV-Optical_SFR_residuals} 
that removing the IR data from the fit reduces the size of our inferred 
SFR credible intervals (with the exception of the power law dust model 
discussed previously). We again attribute this non-intuitive change to the 
additional flexibility allowed to compensate for model mismatches when fitting to 
only UV-Optical bands. This situation points to a subtlety related to data selection 
and credible intervals. We have seen that UV-IR bands give more accurate 
inferences with larger credible intervals, while UV-Optical bands give less 
accurate inferences with smaller credible intervals. Let us now consider 
an extreme case in which we fit to a single band. In this case, there will 
be an extremely large set of parameters that exactly match the data, since we 
could pick an arbitrarily shaped SFH and adjust the total mass to match the 
data point. The resulting parameter inferences would be wildly inaccurate 
and have similarly large credible intervals. Relative to UV-IR and single 
band photometry, UV-Optical fits are an interesting case in which the model 
is flexible enough to compensate for model mismatches, but only for a 
narrow range of parameters, which can lead to overconfidence in the 
accuracy of parameter inferences.

\section{Discussion} \label{Discussion}

\begin{figure}
\gridline{\fig{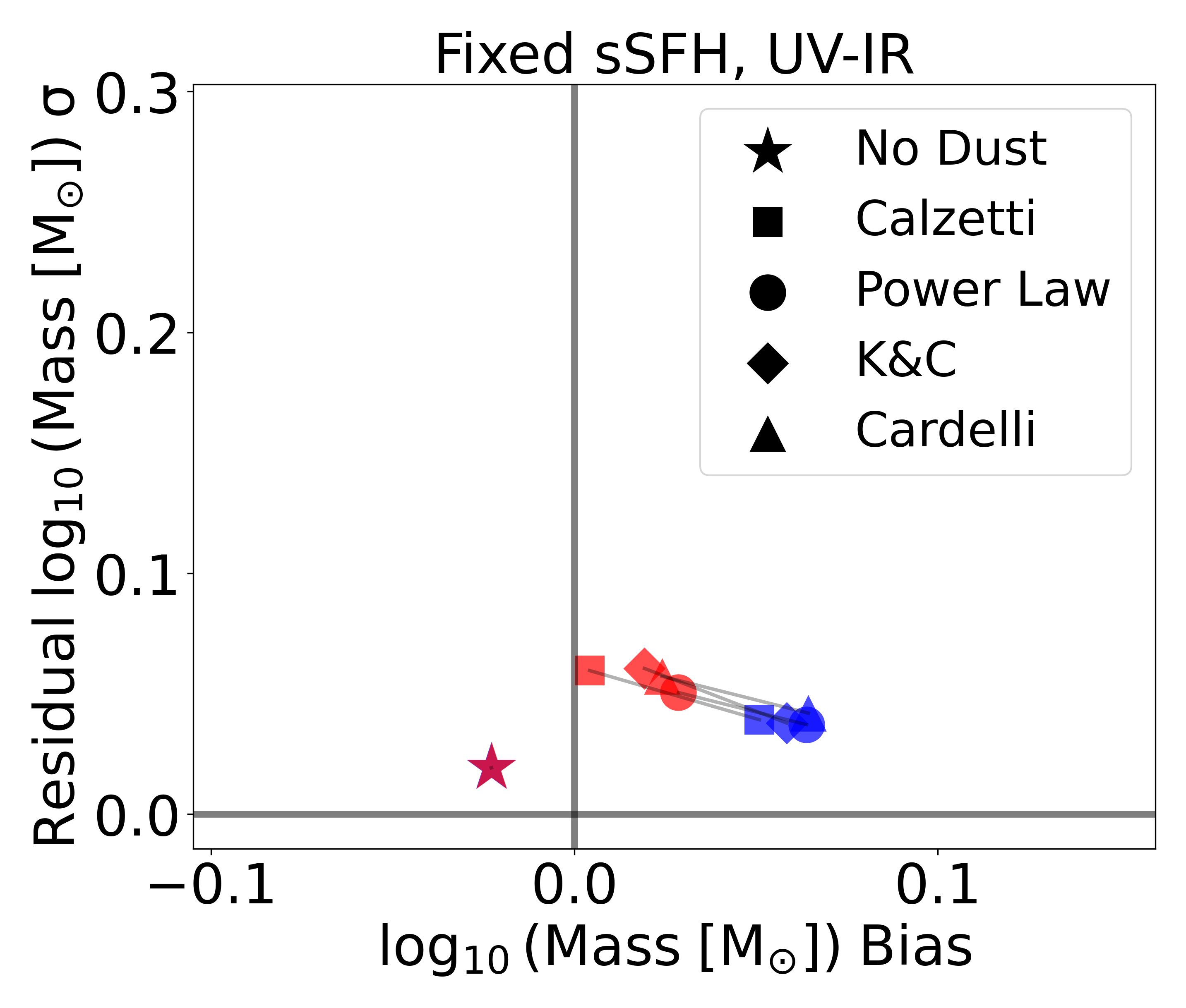}{0.25\textwidth}{}    \fig{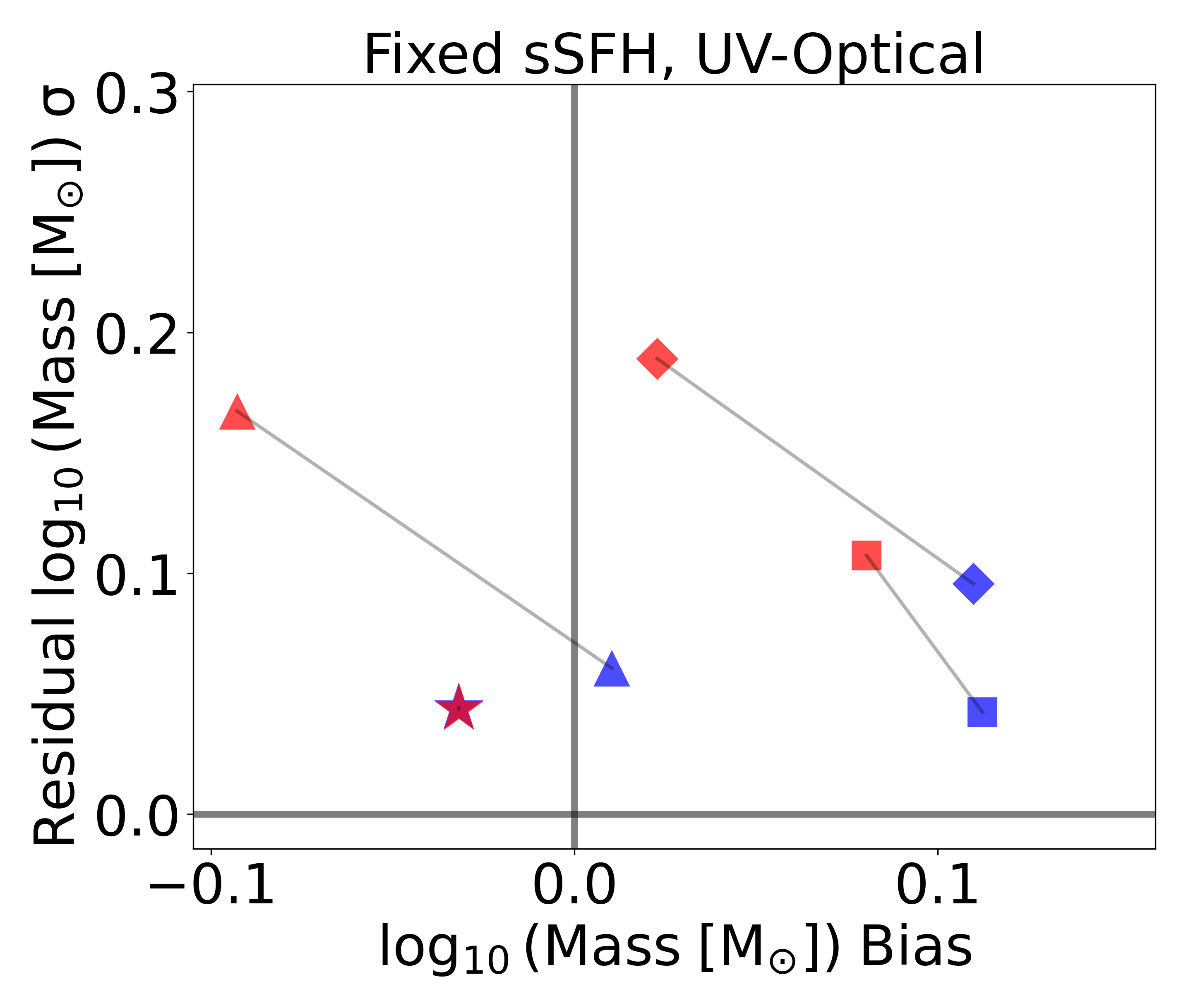}{0.25\textwidth}{}
\fig{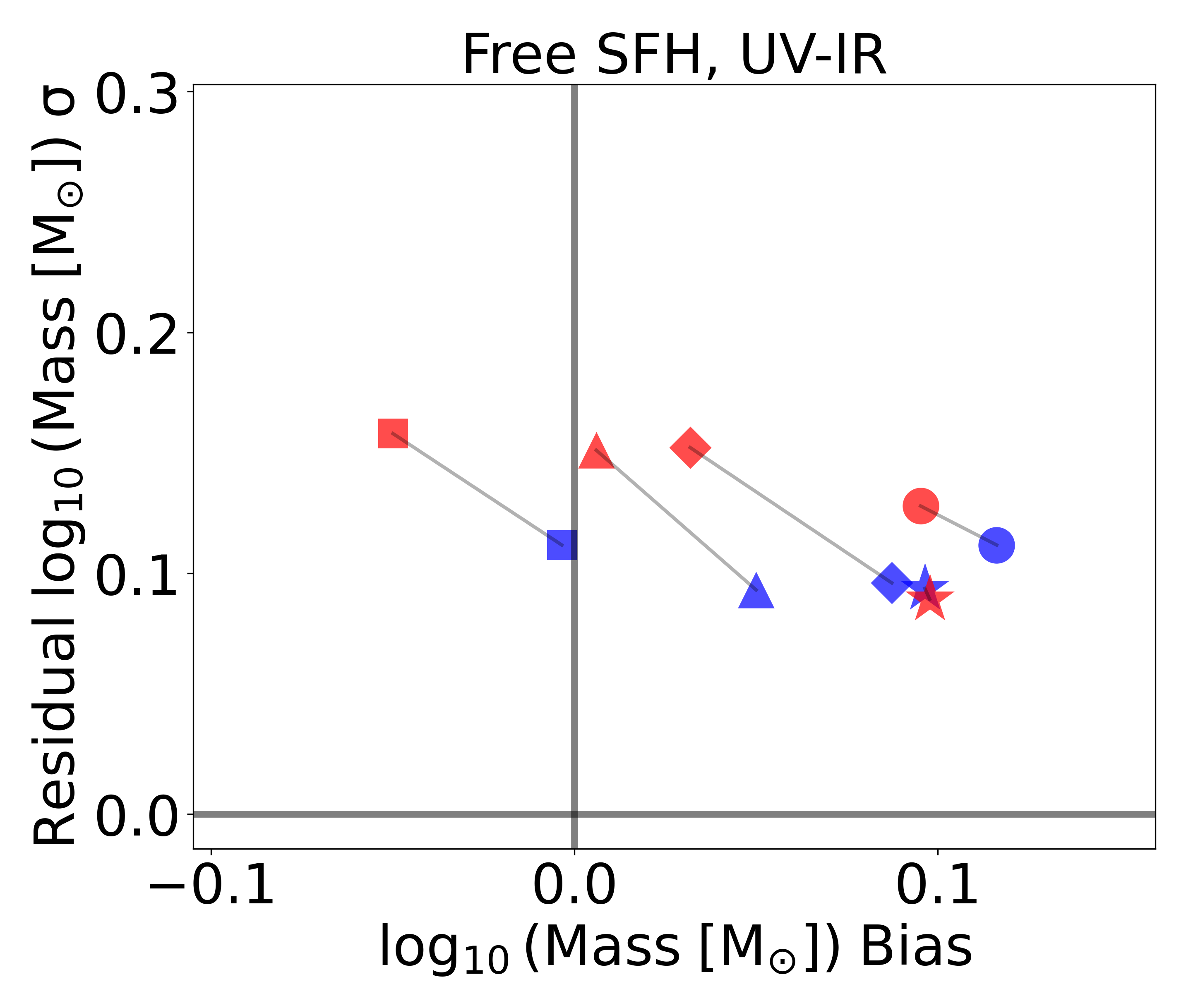}{0.25\textwidth}{}
\fig{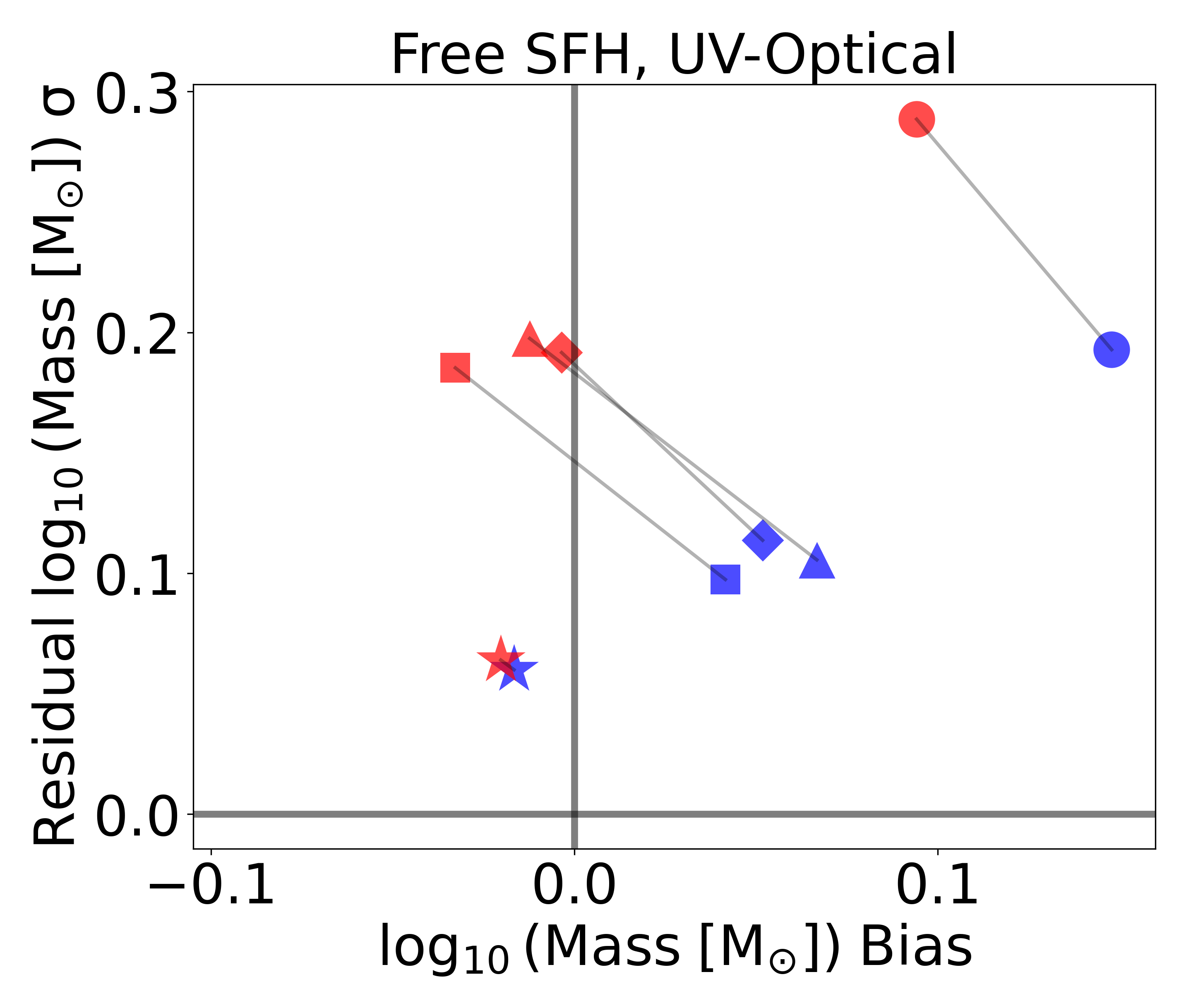}{0.25\textwidth}{}
}
\gridline{\fig{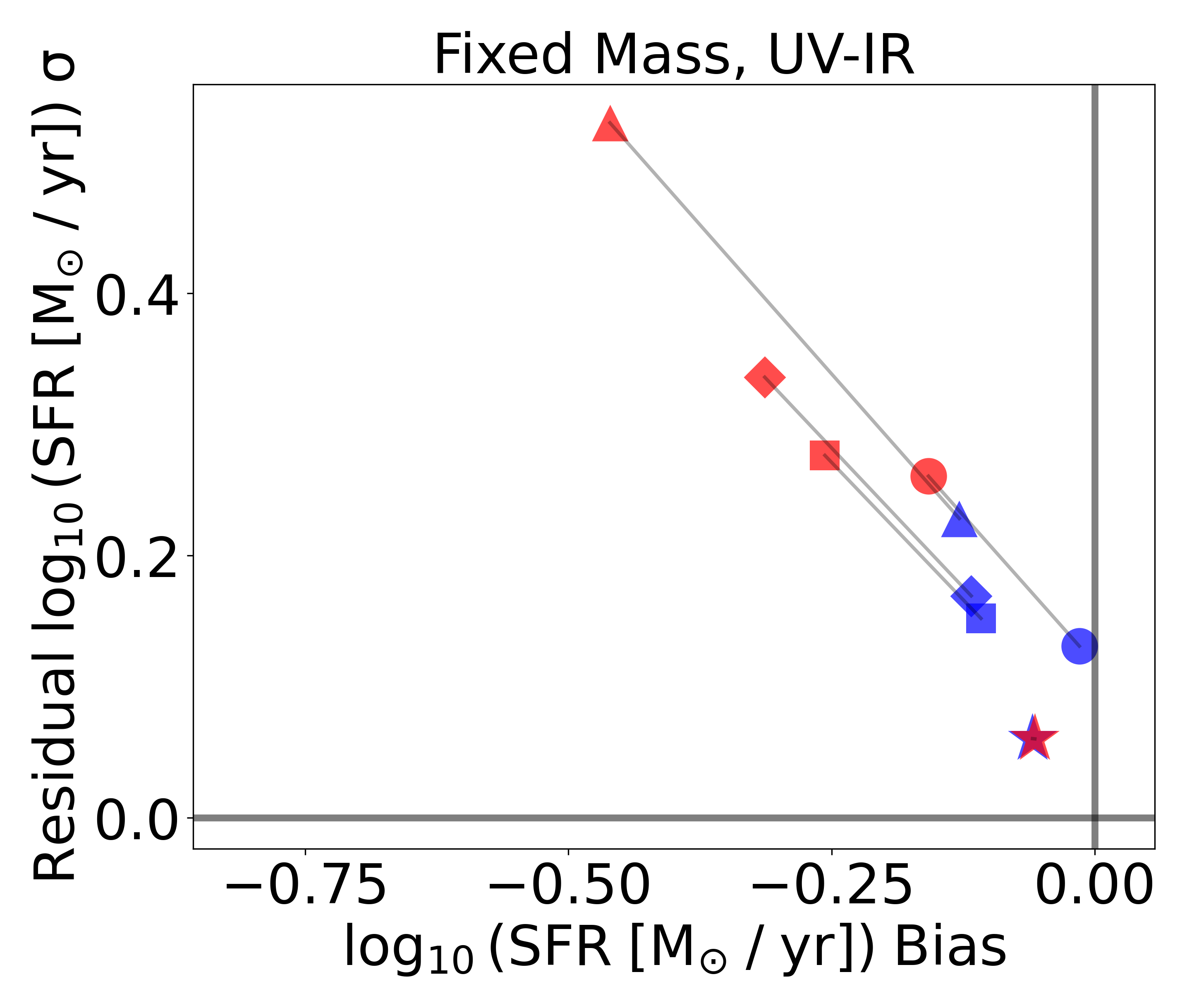}{0.25\textwidth}{}    \fig{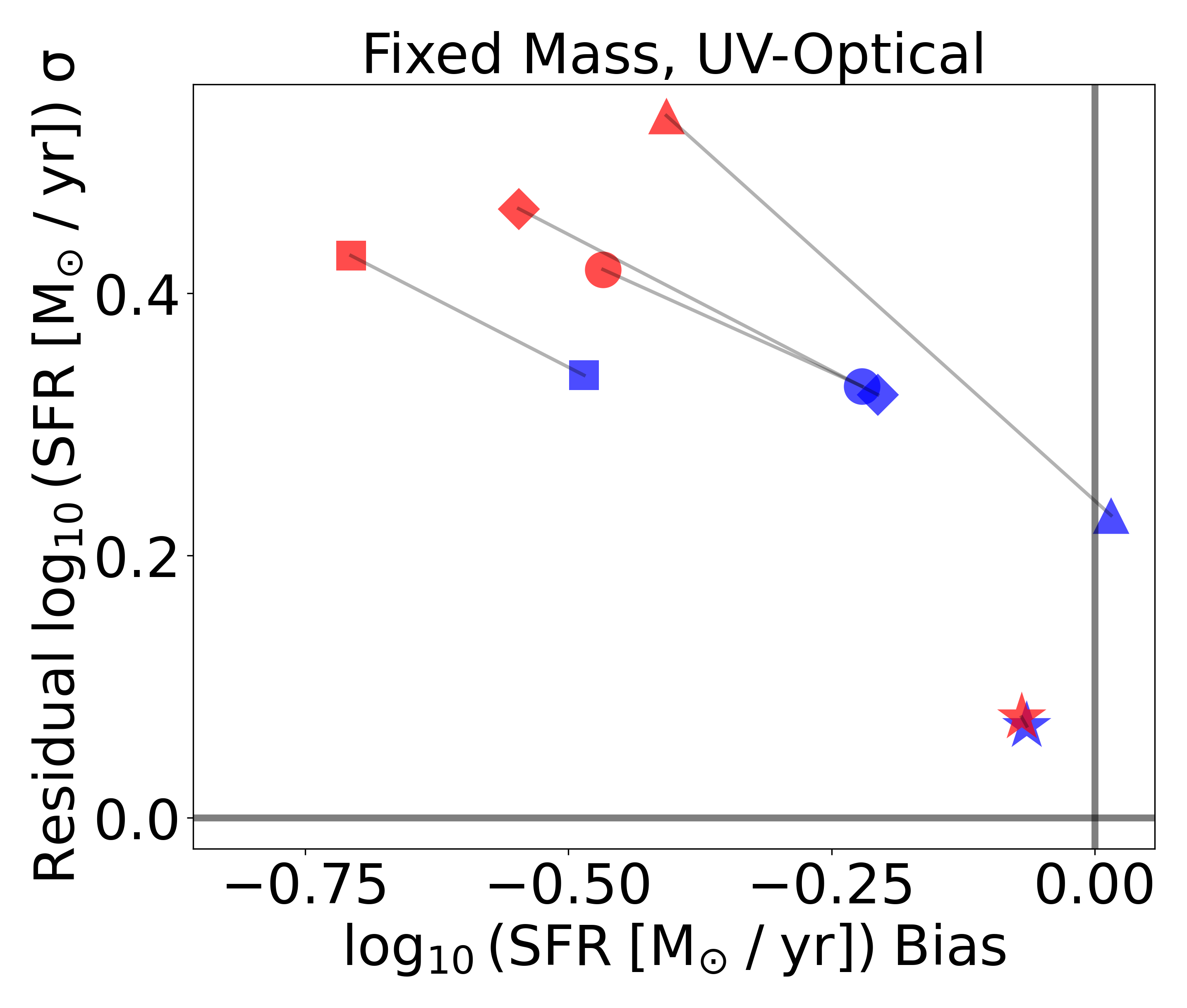}{0.25\textwidth}{}
\fig{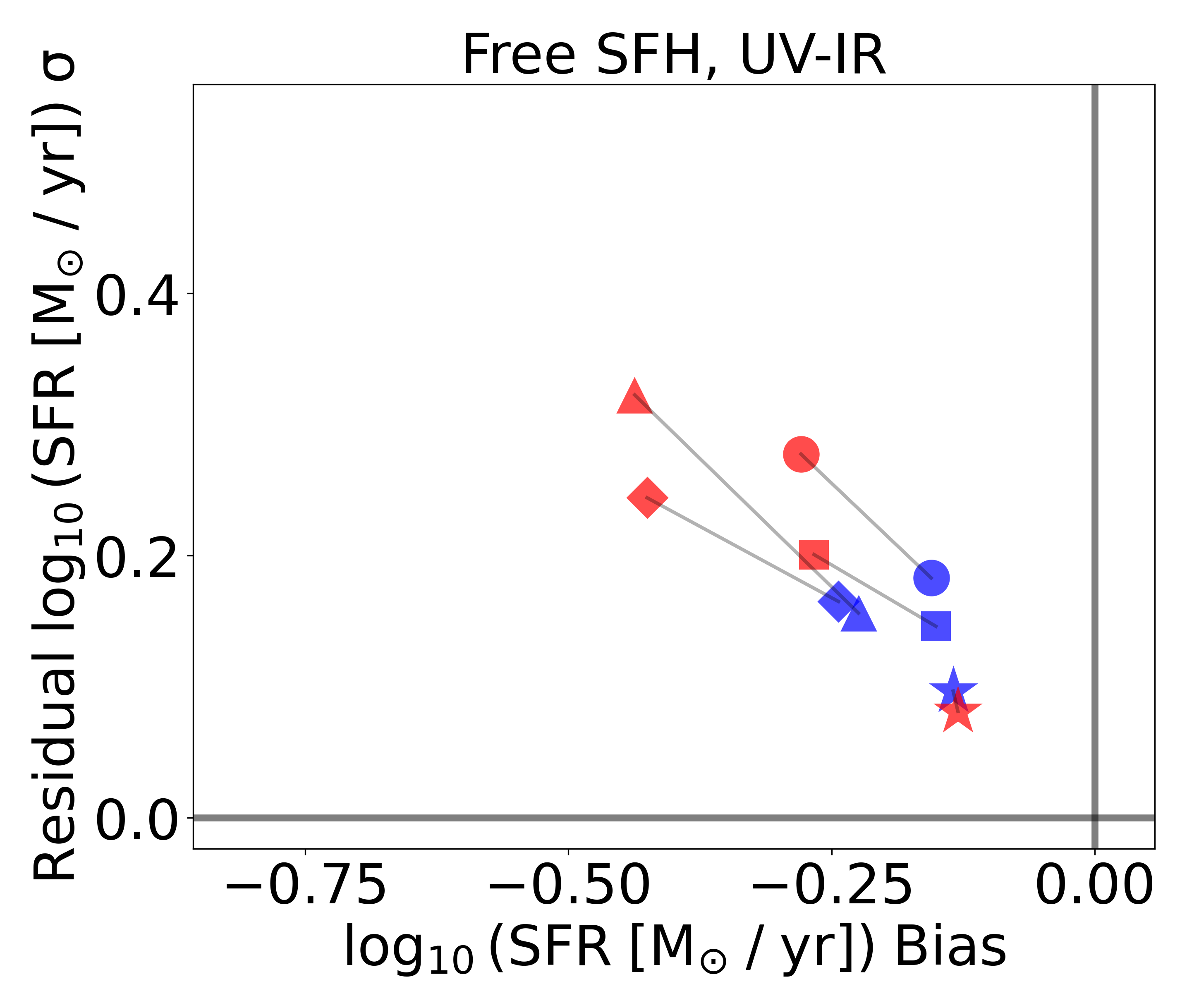}{0.25\textwidth}{}
\fig{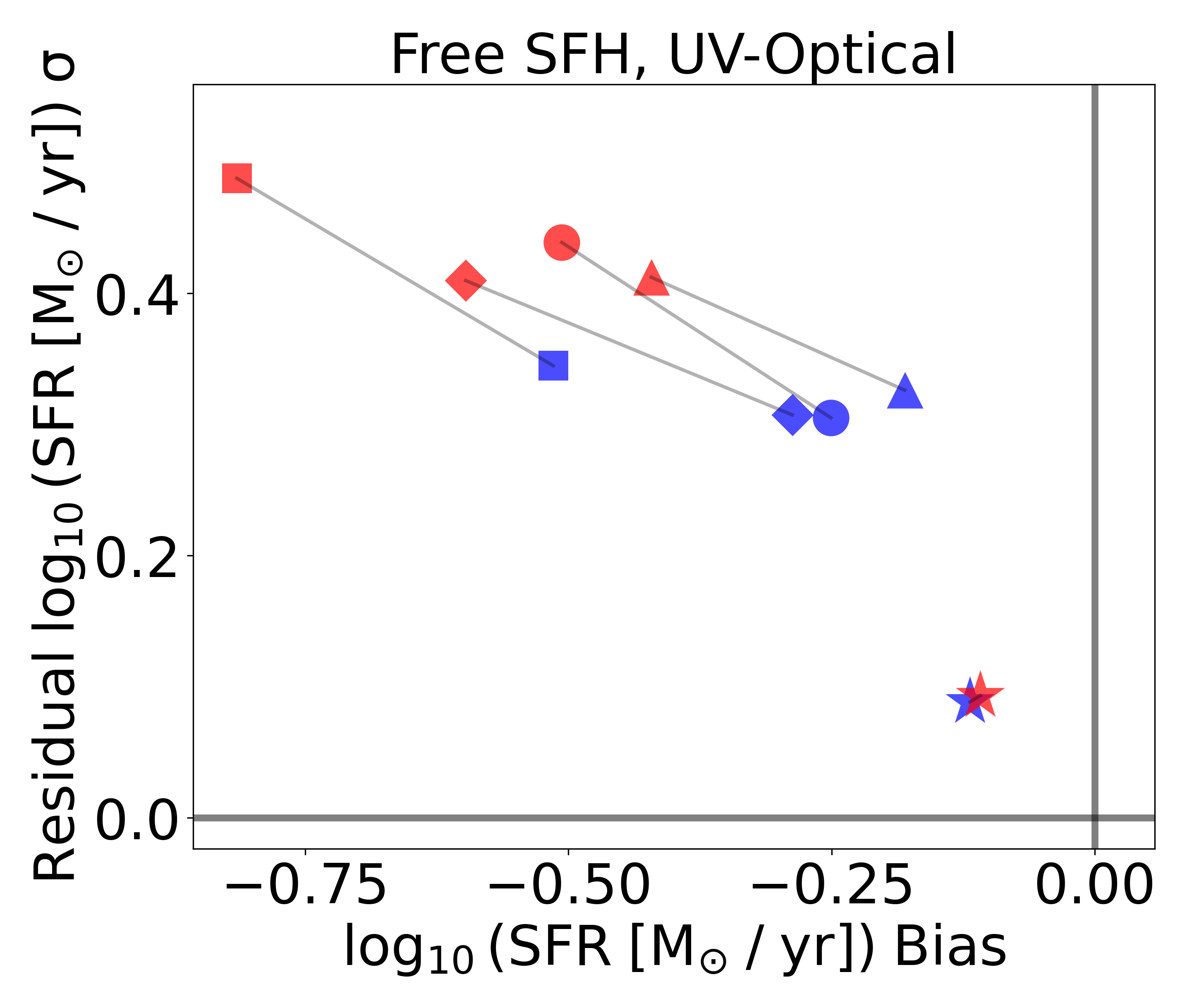}{0.25\textwidth}{}
}
\caption{\label{fig:moments}Distributions of inferred Mass residual standard deviations as a function of inferred Mass Biases (top) and inferred SFR residual standard deviations as a function of inferred SFR Biases (bottom) for face-on (blue) and edge-on (red) orientations of high mass ($> 10^{9.5} M_{\odot}$) galaxies.}
\end{figure}

\begin{figure}
\gridline{
\fig{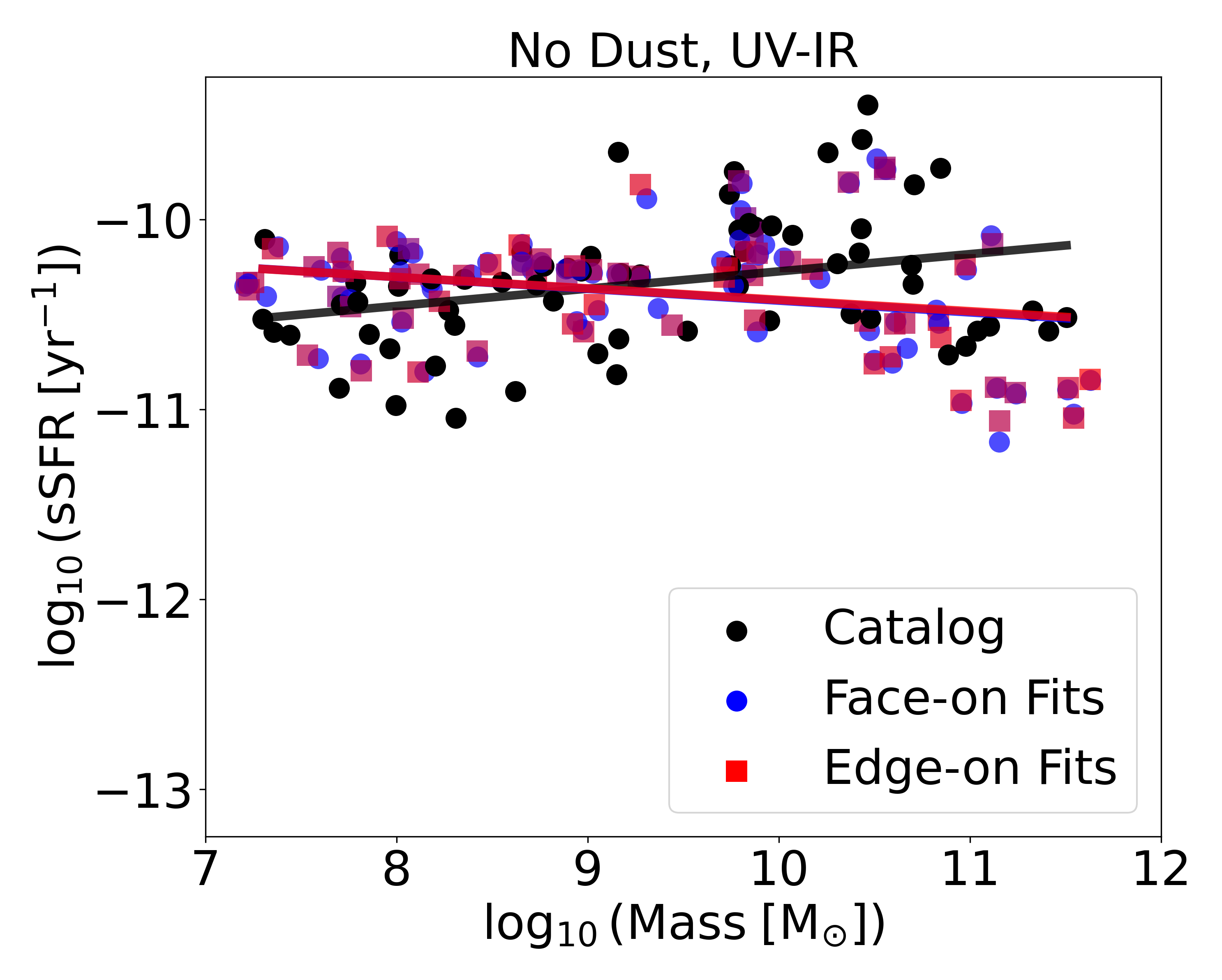}{0.33\textwidth}{}    
\fig{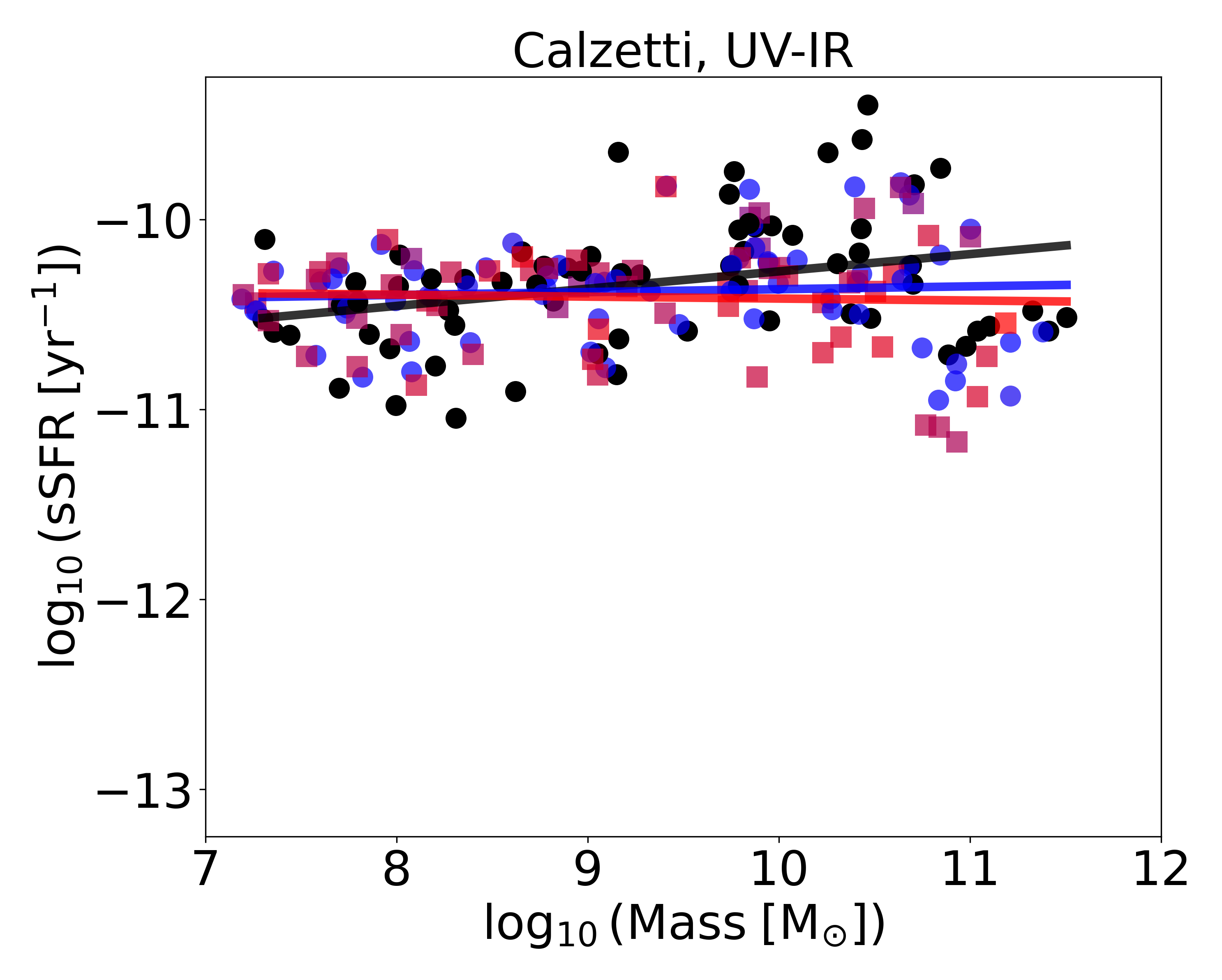}{0.33\textwidth}{}
\fig{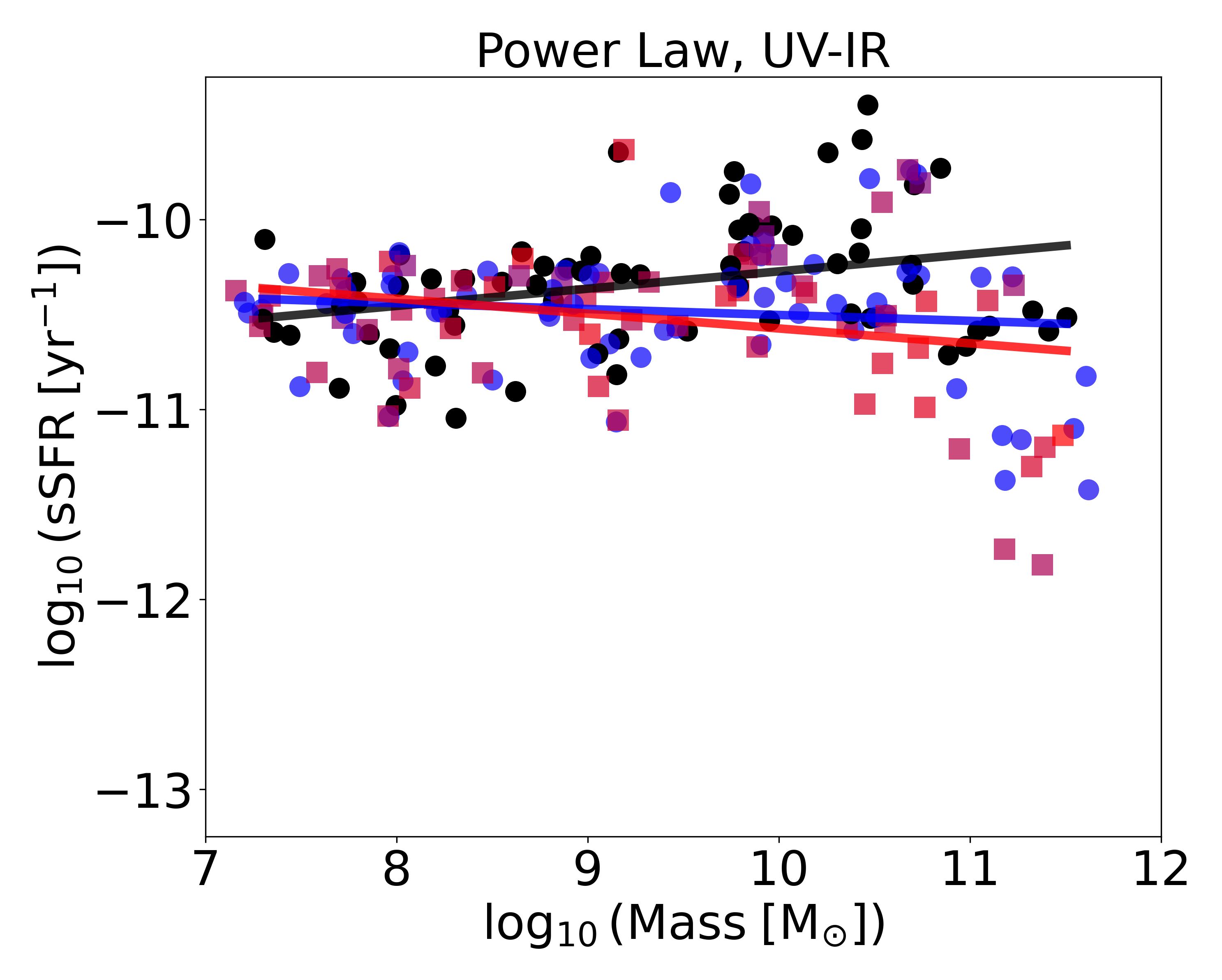}{0.33\textwidth}{}
}
\gridline{
\fig{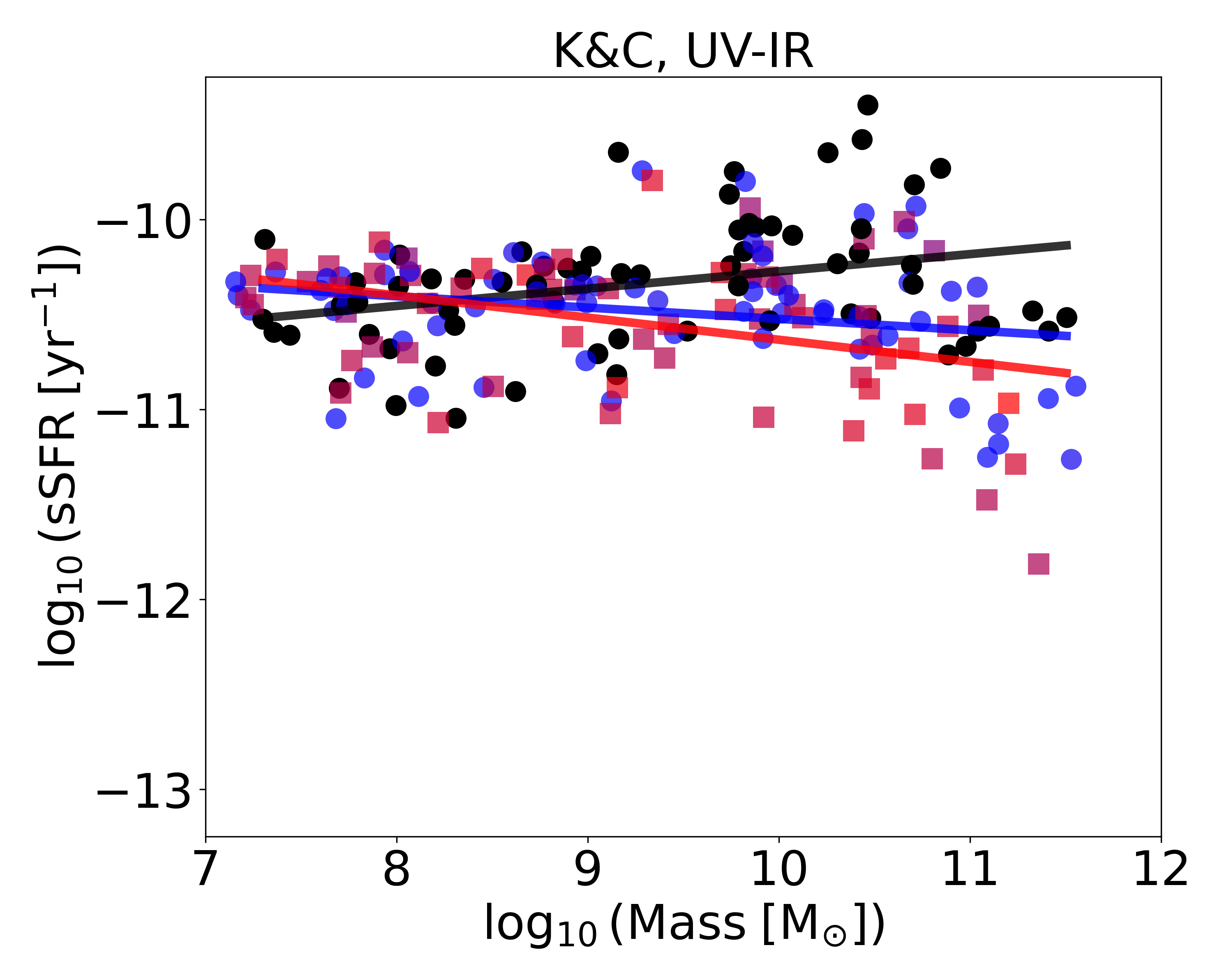}{0.33\textwidth}{}
\fig{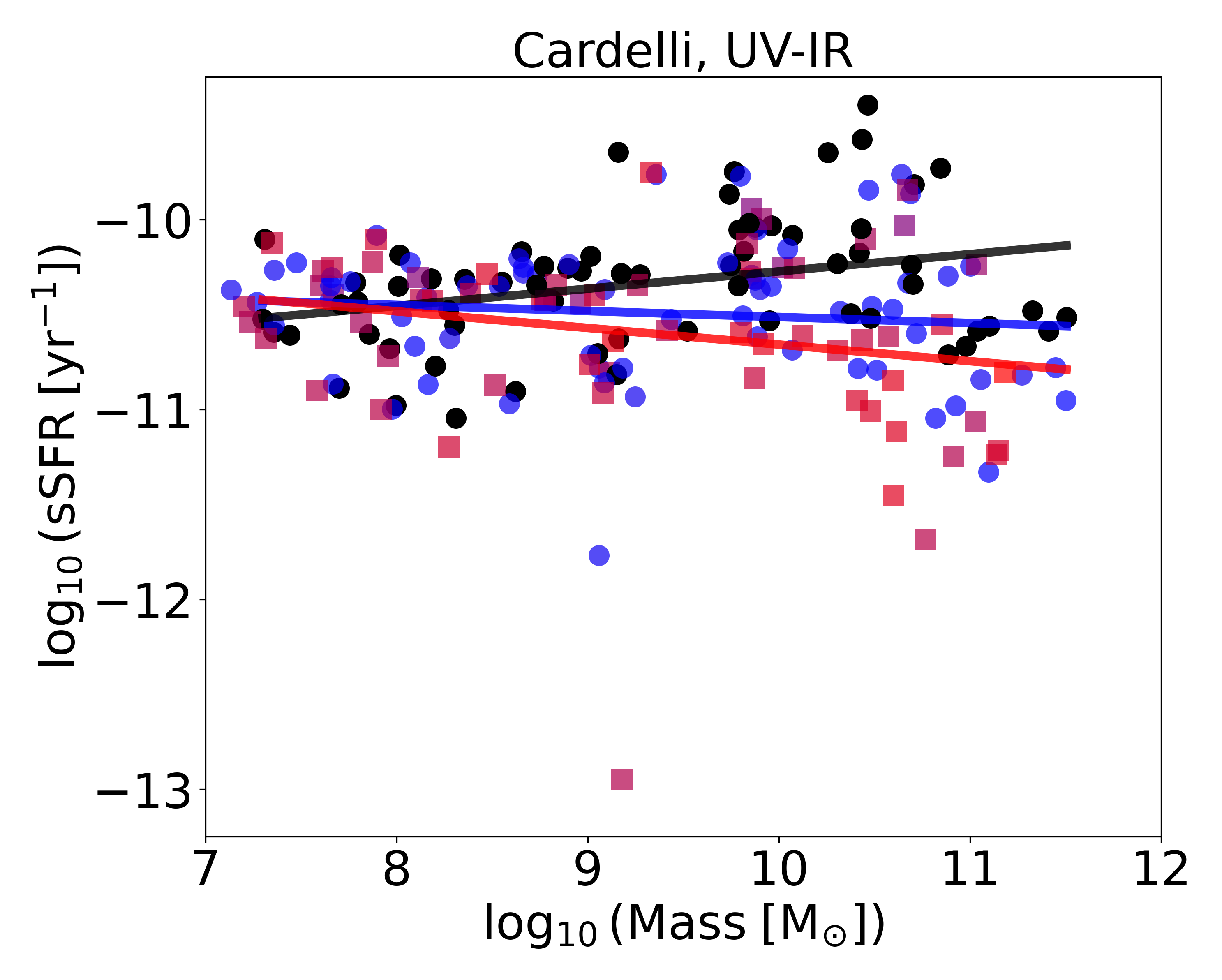}{0.33\textwidth}{}
}
\gridline{
\fig{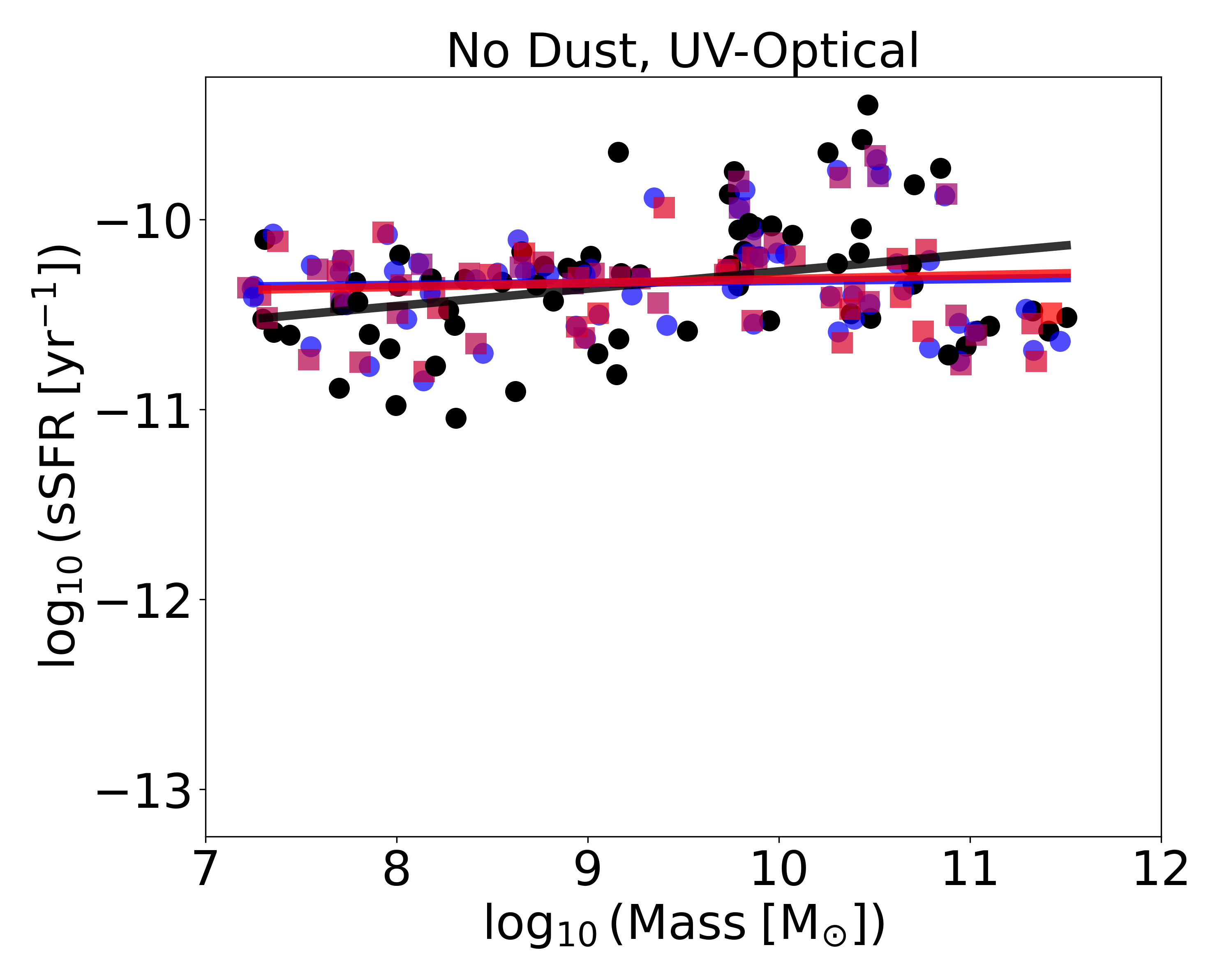}{0.33\textwidth}{}    
\fig{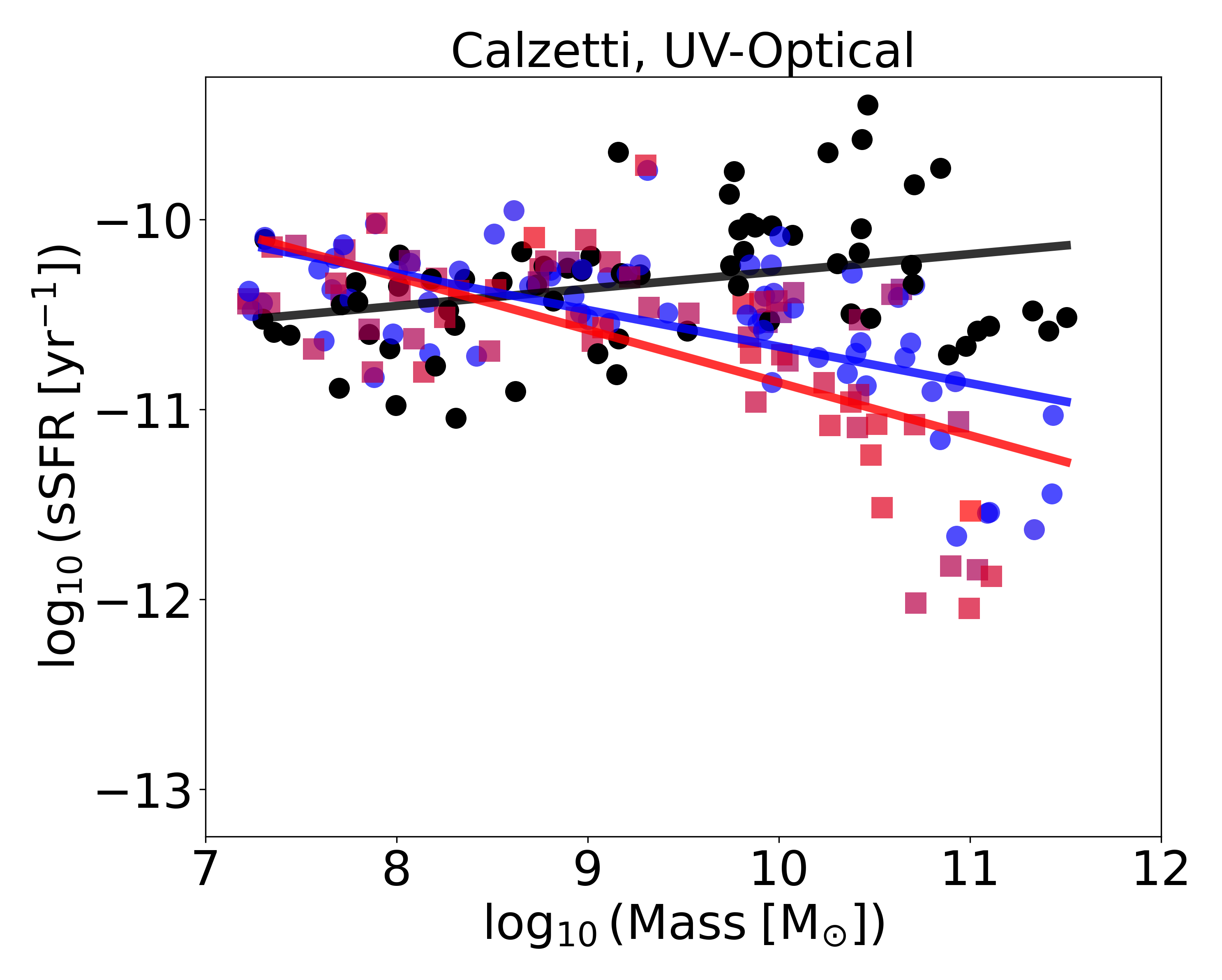}{0.33\textwidth}{}
\fig{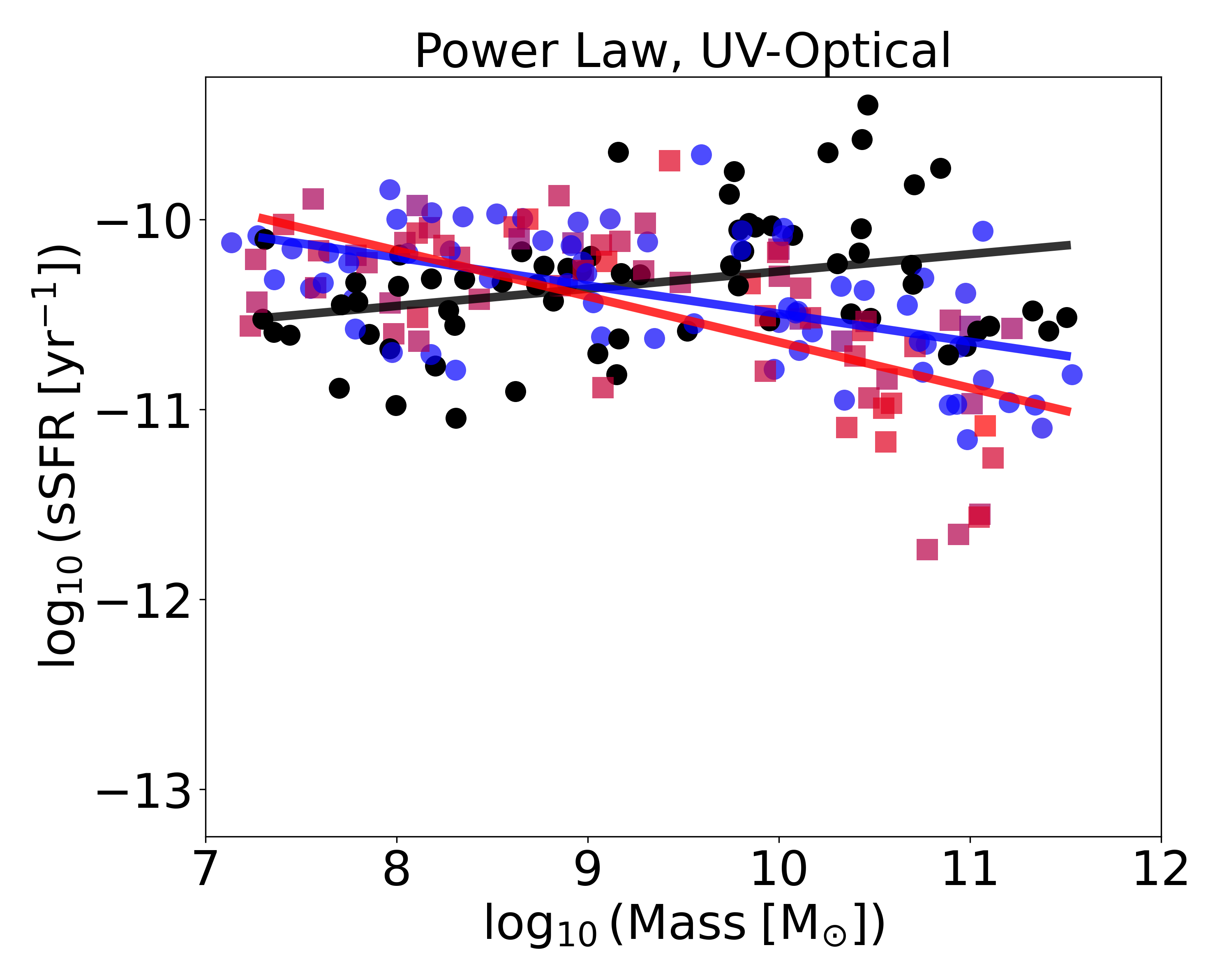}{0.33\textwidth}{}
}
\gridline{
\fig{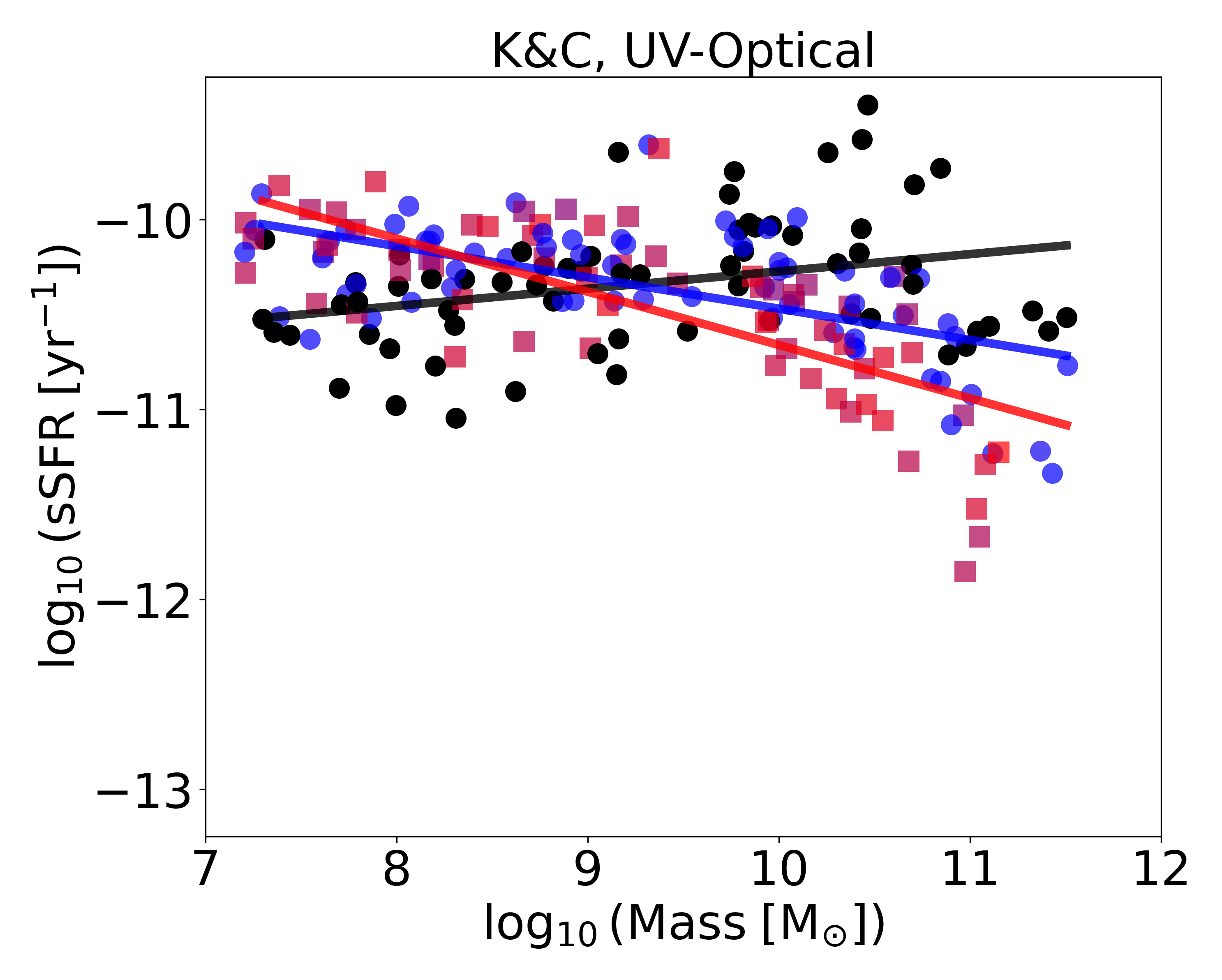}{0.33\textwidth}{}
\fig{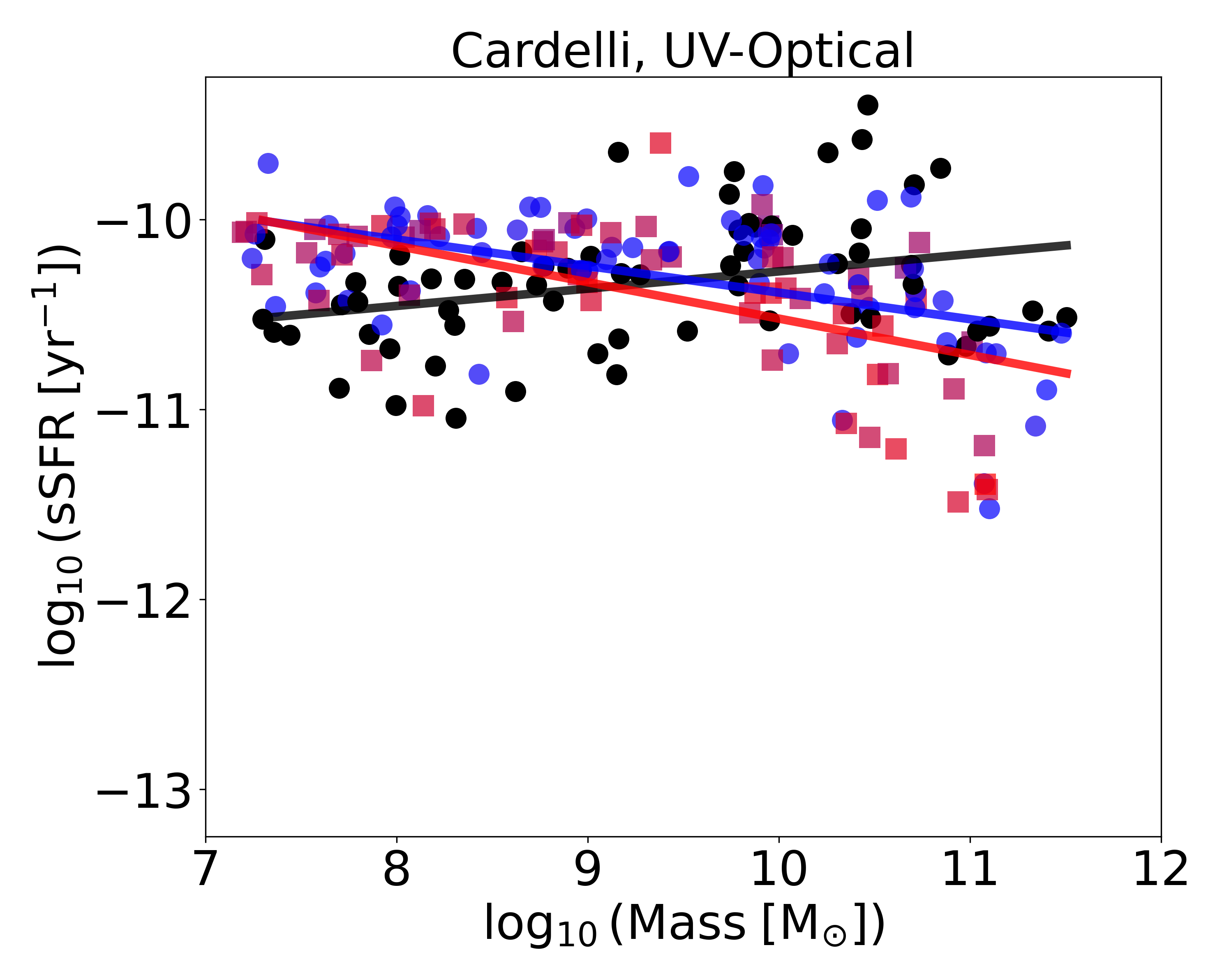}{0.33\textwidth}{}
}
\caption{\label{fig:mass_sSFR}Distributions of true (black circles) and inferred face-on (blueish circles) and edge-on (reddish squares) sSFRs as functions of true and inferred total stellar masses with all SFH parameters free in the fits. The inferred colors indicate their axis ratios as shown in Figure \ref{fig:axisRatioColors}. Best fitting lines for the true values are shown in black, for face-on orientations in blue, and for edge-on orientations in red.}
\end{figure}

\noindent As a quick visual comparison of the results from this work, Figure \ref{fig:moments} shows mass and SFR biases and residual standard deviations for high mass ($> 10^{9.5} M_{\odot}$) galaxies for face-on (blue) and edge-on (red) orientations separately for each dust model, data selection type, and modeling method. If we denote residuals of the $\log_{10}$ parameter values by $\delta$, then biases are given by $\langle \delta \rangle$ and residual standard deviations are given by $\sqrt{\langle \delta^2 \rangle - \langle \delta \rangle^2}$ with averages taken over the set of high mass galaxies.

A few general trends are made clear by comparing the plots in this figure. First, we notice that fits to UV-Optical bands tend to give less accurate inferences comparing to UV-IR fits. Second, inferred masses are significantly more accurate than SFR inferences. Third, dust free SFR inferences are the most accurate, followed by face-on orientations and then edge-on orientations in all cases, emphasizing the strong impact that dust has on our ability to infer SFRs.

As a demonstration of how the inaccuracies in mass and SFR inferences affect widely used relations for studying galaxy evolution, Figure \ref{fig:mass_sSFR} shows the sSFR-mass relation for the true values (black) and the inferred values with all SFH parameters free in the fits for face-on (blue) and edge-on (red) orientations. We can see from the fitted lines that the slope of this relation is significantly affected by the systematic inaccuracies caused by model mismatches, with both orientations showing preference for lower sSFRs at higher masses relative to the true values. These discrepancies become even stronger when only fitting to UV-Optical bands. Even in the dust free cases, we see a decrease in the slope of this relation, which is likely related to the tendency of the Prospector-$\alpha$ model to predict older SFHs in the $10^{10}-10^{11}M_{\odot}$ stellar mass range \citep{leja19a}. 

Comparing these results to those in Figure 10 of \cite{salim23a}, we see that our inferred 
sSFR-mass relations show significantly better agreement with their results than the true 
values of the simulated galaxies. 
One possible interpretation is that the NIHAO simulations  themselves produce 
a sSFR-mass relation that is consistent with that of the real Universe, and 
that the process of observing the simulated galaxies and making inferences from 
those observations leads to a sSFR-mass relation that is consistent with our 
observations and subsequent inferences of galaxies in the real Universe. If so,
it would imply that many of the observational inferences that are used to underpin 
our theoretical understanding of galaxy evolution are severely misinterpreted. 
Our point, however, is not whether the NIHAO prediction of the sSFR-mass relation
matches reality or not.
Instead we emphasize that discrepancies caused by making inferences from 
observations in the face of moderate model mismatches and the incorrect 
assumption of strict energy balance imposed by many widely used SED modeling frameworks such as Prospector, CIGALE, and MAGPHYS, should elicit caution in the interpretation of SED modeling inferences. 

\section{Summary} 
\label{Summary}
\noindent In this work, we have used mock observations of simulated galaxies from the NIHAO-SKIRT-Catalog to test the ability of the SED modeling framework Prospector to infer masses and SFRs. By separating these inferences into four different cases in which we fix certain model parameters related to the galaxies' SFHs to their true values and progressively allow them more freedom in the fits, we have shown that moderate model mismatches tend to be compensated by biases in mass and SFR inferences. 

When fitting to UV-Optical broadband photometry, inferred SFRs are on average underestimated by a factor of 3, with edge-on orientations showing stronger biases that extend well beyond factor of 10 discrepancies. Inferred stellar masses are in general less biased and tend to be within a factor of 2 of the true values. 

When fitting to UV-IR bands, inferred SFRs are on average underestimated by a factor of 2, with edge-on orientations showing stronger biases that extend down to factor of 10 discrepancies. Inferred stellar masses are in general less biased and tend to be within a factor of 2 of the true values. We have also found that the assumption of energy balance between dust attenuation and emission can lead to increased negative biases in the SFR inferences of edge-on galaxies. 

The combined effect of these model mismatches lead to significant inaccuracies in the resulting sSFR-mass relation, with UV-Optical fits showing particularly strong deviations from the true relation exhibited by the simulations. 

In contrast to previous works by \cite{hayward15a}, \cite{lower20a}, \cite{lower22a}, and \cite{trcka20a}, the simulated observations used in this work are generated from simulated galaxies with a multiphase ISM, which results in more realistic dust density structures and diverse attenuation curves. We also perform detailed distance and flux uncertainty calibrations that make the SED modeling inferences of our simulated observations more realistic. We speculate that these differences are the primary drivers of the poorer accuracy we find relative to those works in SFR inferences from SED modeling.

Our results suggest that if astronomers want to ensure the accuracy of stellar 
mass and SFR inferences from existing and upcoming photometric surveys, we must
test SED modeling against realistic simulations such as those used here. 
The NIHAO-SKIRT results may not completely accurately reflect the real universe,
since galaxy formation is not a completely solved problem, but they are likely
far closer to reality than the toy models that SED models use and against which 
they are most often tested.  

\clearpage

\bibliography{refs}{}
\bibliographystyle{aasjournal}

\end{document}